# Quantum Engineering of Single-Crystalline Silver Thin Films

Ilya A. Rodionov[*,1,2], Aleksandr S. Baburin[1,2], Aidar R. Gabidullin[1,2], Sergey S. Maklakov[4], Sven Peters[3], Ilya A. Ryzhikov[3,4], and Alexander V. Andriyash[2]

**There is a demand for the manufacture of ultra low-loss metallic films with high-quality single crystals and surface for quantum optics and quantum information processing. Many researches are devoted to alternative materials, but silver is by far the most preferred low-loss material at optical and near-IR frequencies. Usually, epitaxial growth is used to deposit single-crystalline silver films, but they still suffer from losses and well-known deweting effect. Here we report the two-step approach for e-beam evaporation of atomically smooth single-crystalline metallic films. The proposed method is self-controlled by quantum size effects and is based on the step switch of film growth kinetics between two deposition steps, which allow to overcome the film-surface dewetting. Here we have used it to deposit 35-100 nm thick single-crystalline silver films with sub-100 pm surface roughness and extremely low losses. We anticipate that the proposed approach could be readily adopted for the synthesis of other low-loss single-crystalline metallic thin films.**

[1]Research and Educational Center Functional Micro/Nanosystems, Bauman Moscow State Technical University, Moscow, Russian Federation
[2]Dukhov Research Institute of Automatics, Moscow, Russian Federation, [3]SENTECH Instruments GmbH, Berlin, Germany, [4]Institute for Theoretical and Applied Electromagnetics RAS, Moscow, Russian Federation *e-mail: irodionov@bmstu.ru

Unique large-scale optoelectronic devices utilizing plasmonic effects for near-field manipulation, amplification and sub-wavelength integration open new frontiers in quantum optics and quantum information science[1-15]. However, ohmic losses in metals are still the Chinese wall on the way towards a variety of useful plasmonic devices[14-21]. Many researchers have devoted extensive efforts in clarifying the comprehensive influence of metallic film properties on overall losses to develop a high performance material platform[14-19]. Single-crystalline platform has the potential to alleviate this problem by eliminating material-induced scattering losses[3,4] and nanoscale structure definition impact[15-18]. Silver (Ag) is by far potentially the best plasmonic metal at optical and near-IR frequencies[4,15-19], its properties are even superior to graphene[17]. On the other hand, silver is one of the most challenging metal for single-crystalline growth[18-22]. Moreover, because of silver nature, an epitaxial growth of sub-50 nm thick films – a significant challenge, especially without using loosy wetting underlayers[23-26]. Most of the reported single-crystalline silver growth methods rely on molecular beam epitaxy (MBE) [19,27] or physical vapor deposition (PVD) [18]. For example, it was shown that Ag epitaxial films can be engineered to have almost atomic smoothness and significantly lower optical losses in the 1.8-2.5 eV range[19] than widely cited Johnson and Christy (JC) data[29]. Here we report on a two-step PVD growth approach to obtain atomically smooth single-crystalline metallic films, which is the result of a detailed study on the growth mechanism and a scrupulous analysis of more than 2000 samples[30-32]. Our approach provides the single-crystalline metallic films growth on non-ideally lattice-matched substrates without underlayers using a high vacuum electron-beam evaporator. Unlike colloidal chemistry[4,5], PVD[18,33] and UHV MBE methods[19,27], the proposed two-step process guarantees an accurately controlled thickness (down to 35 nm), atomically smooth surface over a centimetre-scale area, high crystallinity and purity, the unique optical properties and SPP propagation length, high deposition rate, and thermodynamic stability. The process is facile, inexpensive, compatible with lithography and etch nanoscale features definition, and reproducible in a standard cleanroom environment. Moreover, it can be effectively applied to various metals such as silver, gold, and aluminum, which are the most extensively used metals in quantum optics and quantum information science.

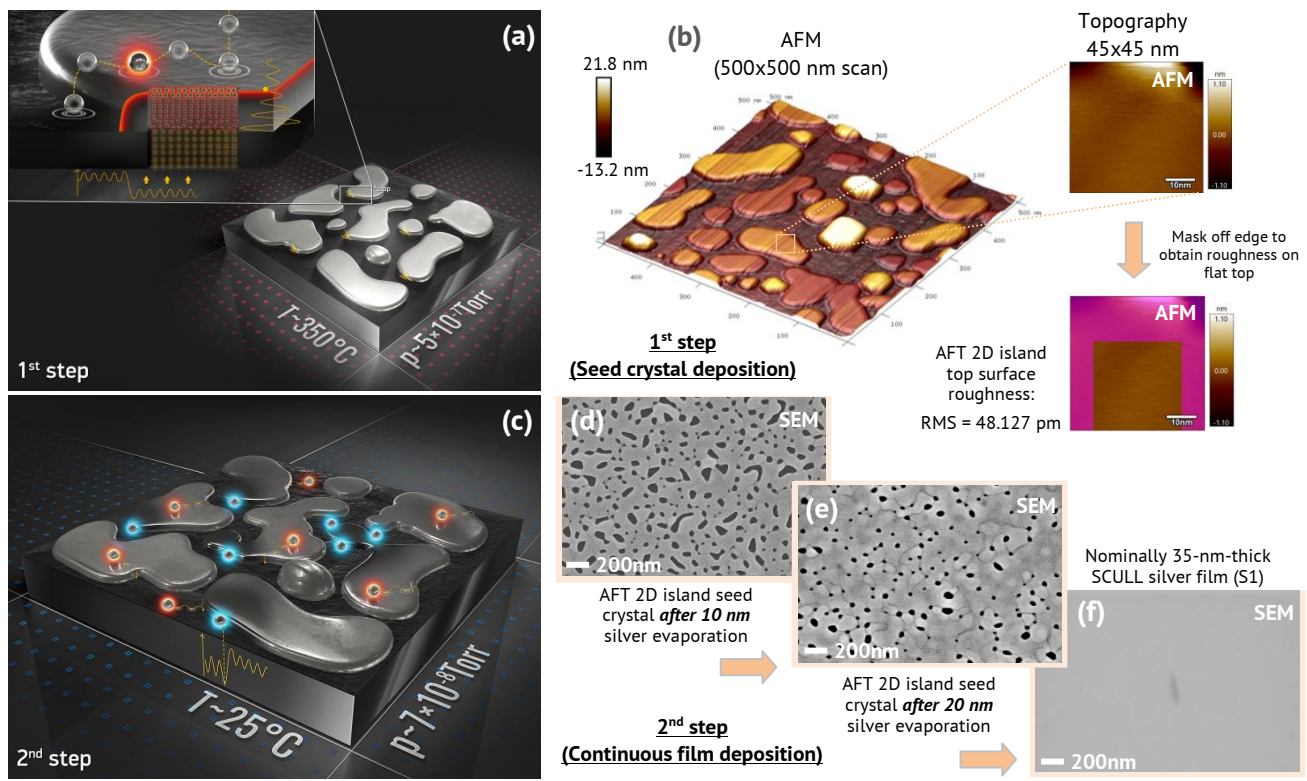

**Figure 1 | Two-step deposition of single-crystalline silver films.** (**a**) In the first step, an AFT 2D Ag (111) seed crystal is deposited under 350 ºC temperature. (**b**) Atomic force microscopy (AFM)

scan of AFT 2D Ag (111) islands deposited on a Si (111) substrate. Most of the AFT 2D Ag(111) islands have atomically flat top surface with an RMS roughness less than 50 pm. In the second step, the process is stopped and substrate is cooled down to 25 ºC followed by additional silver evaporation until continuous silver film is formed (**c**) SEM images illustrate a film morphology evolution during the second step after nominally 10 nm (**d**) and 20 nm (**e**) silver evaporation on a AFT 2D seed crystal at 25 ºC. (**f**) SEM image of nominally 35-nm-thick single-crystalline film. The defect on the film surface is purposefully created (by electron beam burning) to facilitate focusing on the atomically smooth surface.

Figure 1 illustrates the basic principles of atomically smooth single-crystalline silver films two-step deposition process. In the first step, a seed crystal consisting of strained two-dimensional islands with atomically flat top surface (AFT 2D islands) is grown on a substrate under 350 ºC. In the second step, the deposition is stopped and the substrate is cooled down to 25 ºC in the same vacuum cycle to prevent a well-known deweting effect and three-dimensional growth leading to subsequent film imperfections. Then, more silver is deposited on AFT 2D seed at 25 ºC until a continuous film is formed. Following film annealing under elevated temperature (higher than first step) can reduce growth defects density improving crystalline structure and surface roughness. We believe that the dramatic improvement in film characteristics relies on the combination of two mixed evaporation modes combined with the AFT 2D islands growth self-controlled by quantum size effects. With this dual-phase experimental nature in mind as well as improved film parameters, we name our deposition process the «SCULL» (Single-crystalline Continuous Ultra-smooth Low-loss Low-cost).

**The SCULL process**

Consider a silver (111) film on silicon (111), which is a well-known substrate for epitaxial Ag growth. To quantitatively estimate an influence of substrate crystalline structure on films quality, we additionally deposit the SCULL silver films on widely used silicon (100), silicon (110), and mica substrates. Nominally 35-nm-thick single-crystalline silver films were evaporated using the SCULL process (base pressure $3 \times 10^{-8}$ Torr, see Supplementary Information) on different substrates (Table 1), which are the thinnest PVD single-crystalline silver films than those reported previously[18,33]. On the other hand, the SCULL process has no fundamental limitations in a thicker film synthesis, which is crucial for applications sensitive to SPP substrate absorption. To demonstrate this, we deposit nominally 70-nm-thick (S4) and 100-nm-thick Ag (111) films (S5) on Si (111) substrates. All the films are continuous, without voids and pits, and have an atomically smooth surface over $15 \times 15$ mm$^2$ sample area.

According to the epitaxy theory and long-term experience, microstructure, growth mode and morphology are mostly governed by kinetic effects at high deposition rates (non-equilibrium experimental conditions). In this case single-crystalline films can be deposited in the smooth Frank-van der Merwe (layer-by-layer) growth mode[34], when the surface free energy of the substrate ($E_{sub}$) is higher than or equal to the sum of the surface free energies, $E_{sub} \geq E_f + E_{int}$, for the film ($E_f$) and the interface ($E_{int}$). Then, it is energetically favorable the film to cover the substrate completely to eliminate the contribution of the high substrate surface energy. The fundamental idea of our process involves quantum engineering of the AFT 2D seed crystal of a given metal (1st step), which is «frozen» at sweet point of the growth process, followed by dramatic shift of film growth kinetics (2nd step) allowing the lateral spreading of the crystal seed until the perfect continuous film is formed. Three key features (Fig. 1a, 1b) have to be provided at the first process step: the 2D growth of islands with atomically flat top surface, the macroscopic control of thickness and microstructure of AFT 2D islands and the well-defined strain accumulated in islands at the sweet point. 2D growth can be guaranteed by interlayer mass transport control, which is the delicate balance between the adatoms surface diffusion ($D \sim temperature$) and the flux ($F \sim deposition\ rate$), the growing islands state (density, size, shape and strain), the surface diffusion ($E_D$) and the step-edge Ehrlich-Schwoebel ($E_S$) barriers for adatoms to descend (downward transport) or ascend (upward transport) the edges of the growing islands. In order to ascend (descend) an island, the adatoms arriving on a substrate (island) surface may try several times (with the hopping rate $v=v_0\,exp(–E_S/k_BT\,)$) to move over the edge barrier $\Delta E=E_S–E_D$. Based on previously reported data[35,36], we experimentally determine the ranges of adatoms surface diffusion (280-420 ºC temperature), adatoms flux (0.5-10 Å s$^{-1}$ deposition rates) and islands state (1-25 nm thickness) for the layer-by-layer AFT 2D Ag (111) islands growth at the first process step. Above the certain islands state (size, interface area, strain) the sum of the islands surface free energies $E_f + E_{int}$ becomes higher than the substrate surface free energy $E_{sub}$, leading to three-dimensional growth. Moreover, increasing the islands dimensions without reducing the adatom surface diffusion results in an enhanced hopping rate $v$ of adatoms visiting the step edges and, thus, an increased upward mass transport. Thus, there is the sweet point of the process, when the AFT 2D Ag (111) islands seed deposition have to be stopped.

**Table 1 | Thickness, surface roughness and microstructure of SCULL Ag films as a function of substrate type.** AFM RMS roughness was determined from scans over a 2.5 × 2.5 µm$^2$ area. SP RMS roughness was determined from scans over a 20-µm length.

| Substrate | Measured thickness [nm] | Crystalline structure | Average grain size, [nm] | Rocking curve Ag peak, FWHM [º] | AFM RMS roughness [nm] | Sample |
|---|---|---|---|---|---|---|
| Mica | 35 | Single-crystalline | no grains | not measured | 0.35 | M1 |
| Si (111) | 37 | Single-crystalline | no grains | 0.325 | 0.09 | S1 |
| Si (100) | 39 | Single-crystalline | no grains | 0.829 | 0.28 | S2 |
| Si (110) | 39 | Single-crystalline | no grains | 0.831 | 0.37 | S3 |
| Si (111) | 68 | Single-crystalline | no grains | 0.221 | 0.36 | S4 |
| Si (111) | 99 | Single-crystalline | no grains | 0.368 | 0.43 | S5 |
| Quartz | 107 | Nanocrystalline | less than 20 | not measured | 2.18 | NC |
| Quartz | 100 | Polycrystalline | 50 | not measured | 2.34 | PC |
| Quartz | 98 | Polycrystalline | more than 500 | not measured | 2.22 | PCBG |

An electronic growth model[36,37] based on the quantum size effects can explain the second key feature of AFT 2D islands self-controlled thickness and crystalline structure over a large area. According to the electronic growth model, growing AFT 2D Ag (111) islands are considered as an electron gas, which is confined to a 2D quantum well that is as wide as the thickness of the silver islands[38,39]. The energy oscillates as a function of the island thickness (quantum well width), resulting in island thickness quantization (Fig. 1a, inset). Top silver layers of the islands grow under a homoepitaxial regime in the presence of these small energy oscillations, resulting in a quantized island thickness with the silver monolayer accuracy[36,37,40]. Together with the layer-by-layer 2D growth mode it enables formation of the two-dimensional Ag (111) islands with atomically flat top surface and provides the perfect control of islands thickness over macro scale area, even in the presence of typical PVD process deviations.

The third key feature of AFT 2D Ag (111) seed crystal is an energy accumulation in islands which is induced by strained growth under elevated temperature on the substrate with different lattice constants. The accumulated strain affects the growth kinetics at the second process step by lowering the $E_S$ barrier for the atomic surface processes. Strained growth is induced by the onset of defects: a screw dislocation influence and spiral growth which becomes stronger with thickness increase. That is why the AFT 2D seed thickness has to be optimized to provide a dislocation-free crystalline lattice growth of Ag (111) islands, on the one hand, and the ultimate initial strain accumulation, on the other. As the result of the first step the AFT 2D islands seed is formed (Fig. 1b) consisting of the uniform thickness islands (more than 90% substrate area) with the atomically flat top surface (RMS roughness < 50 pm), flat irregular form and the average lateral size from 100 nm to 250 nm.

At the second step, the deposition is stopped and the substrate is cooled down to 25 ºC. Then, more silver is evaporated on the AFT 2D seed (Fig. 1c-1e) until a continuous single-crystalline film is formed (Fig.1f). At room temperature, the reduced surface diffusion length $D^{33}$ and the hopping rate $v$ of adatoms arriving on the substrate lead to decreased upwards mass transport. On the other hand, Ag adatoms arriving on islands hop along the atomically flat top surfaces of the AFT 2D islands with almost no energy dissipation and easily get islands edges. Moreover, the strain relaxation results in the reduced step-edge $E_S$ barrier for the adatoms on the islands surface[39,41] and, thus, increased downward mass transport. Therefore, at the second step almost all the adatoms coming to the substrate are adsorbed at the edges of the islands spreading of dominant Ag(111) 2D islands and, eventually, coalescing the islands with each other, thus, completing the single-crystalline film. In the end, the strain that was accumulated in the first growing step relaxes primarily into interactions with the incoming adatoms improving the crystalline structure of the AFT 2D islands. Upon subsequent annealing at 320-480 ºC, a silver film crystalline structure and surface roughness are improved, along with defect density reduction. Using the SCULL process, the well-known problem of the silver dewetting under elevated temperatures for the continuous sub-50 nm thick single-crystalline silver film deposition is solved.

**Results and discussion**

Utilizing the SCULL silver films as golden samples we demonstrate that grain boundaries, material purity (hence, grain boundaries purity), surface roughness (and associated surface chemical reactivity), and crystallinity imperfection contribute to optical properties of metallic films in descending order of priority. We compare the results for six representative films: three SCULL single-crystalline films of 35 nm (S1), 70 nm (S4) and 100 nm (S5) nominal thickness, and three nominally 100-nm-thick polycrystalline films (PC, PCBG[31], NC) with different grain size and purity (Table 1). Since dielectric permittivity is thickness independent[18,29], optical properties of the Ag (111) / Si (111) films (S1, S4, S5) without grain boundaries and the same material purity properties were compared to estimate a surface roughness and crystallinity impact. High-resolution wide-angle X-ray diffraction (XRD) rocking curves (Fig. 2b, S4, S5) with a full width at half maximum (FWHM) of 0.325 °, 0.221°, 0.368 ° for the film thicknesses of 37 nm (S1), 68 nm (S4) and 99 nm (S5) indicate thickness independent film high quality with minimal level of defects (see Supplementary Information for details). It is important to note the SCULL films deposited on a non lattice-matched Si (100) and Si (110) have predictably worse crystallinity, but also demonstrate atomically smooth surfaces with RMS roughness less that 4 Å (Table 1). High-resolution transmission electron microscopy (HRTEM) image (Fig. 2e) demonstrates the single-crystalline nature of the S1 silver film. Electron backscatter diffraction (EBSD) is used to analyze the domain structures and extract average grain size (Table 1) of single-crystalline (Fig. 2h) and polycrystalline films (Fig. 2f, 2g, S6c).

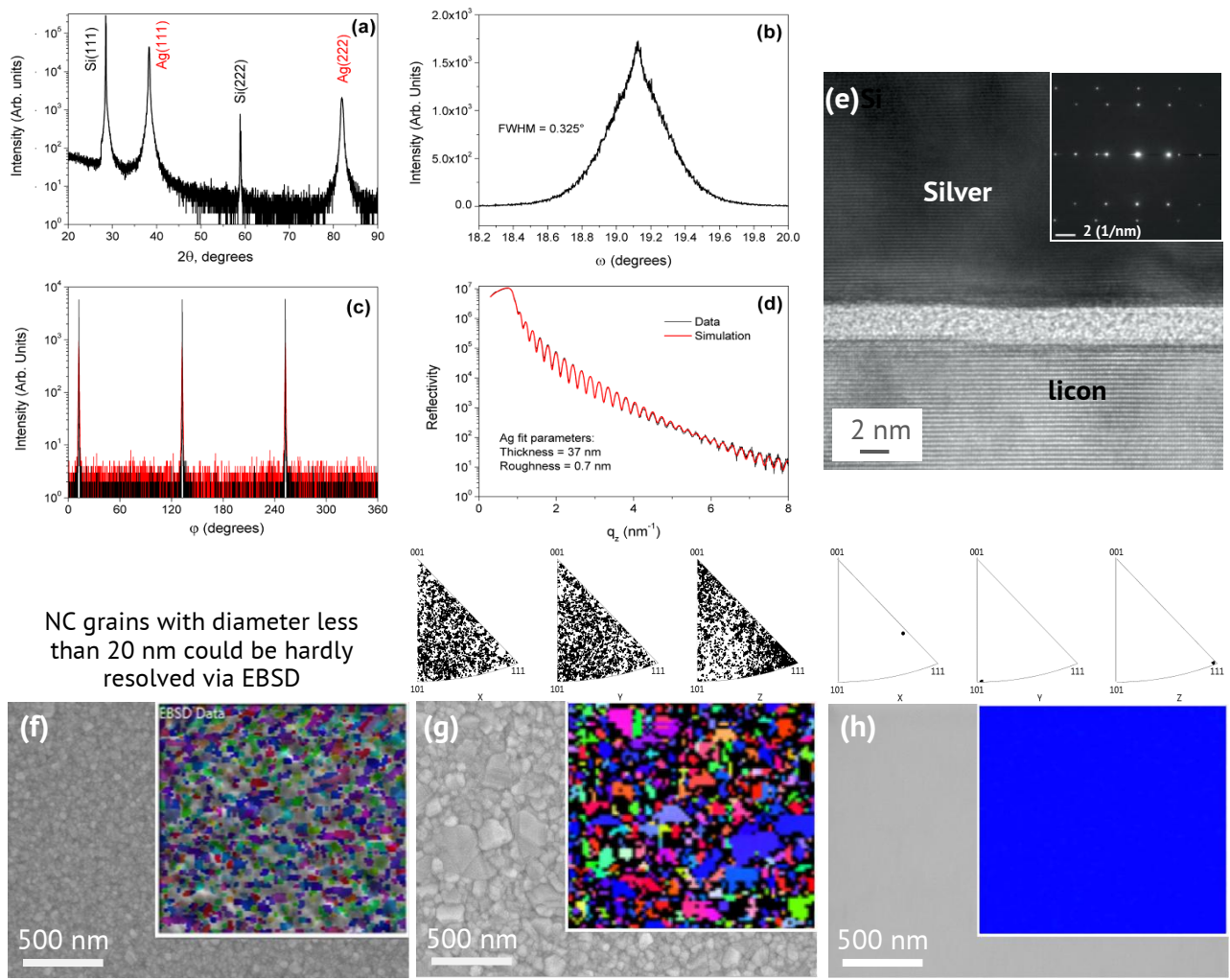

**Figure 2. | Microstructure characterization of a 37-nm-thick Si (111) / Ag (111) film (S1) and SEM images with EBSD insets of (NC), (PC) and (S1) films.** (**a**) XRD (θ-2θ) pattern indicating only Ag (111) and Si (111) substrate peaks. (**b**) Measured transverse scan (rocking curve, ω-scan) through the Ag (111) diffraction peak. (**c**) Grazing incidence of the in-plane X-ray diffraction scan (phi-scans) of the Ag(111) plane. (**d**) X-ray reflectivity curve. (**e**) HRTEM image and the electron diffraction pattern (inset in the right corner), the growth direction is bottom-up. SEM images with EBSD insets of NC (**f**), PC (**g**) and S1 (**h**) silver films highlighting film grains. EBSD inverse pole figures are shown above the SEM images, demonstrating very tight crystal orientation density of the

S1 film (h) along all the normal directions. Only a single domain is observed for S1 film in both small-scale 2 μm (h) and large-scale 400 μm scans (Fig. S6).

To estimate losses and rank the films parameters contribution to optical properties a multi-angle spectroscopic ellipsometry is used (see Supplementary Information for details). We focus on the most practically useful NIR and visible wavelength region for silver lays above the interband transitions (λ > 325 nm), where the contribution to $\varepsilon_1$ mainly comes from Drude terms (dc conductivity), but $\varepsilon_2$ is defined by both intraband and interband components. We observe the dominating contribution of grain boundaries to dielectric permittivity (Fig. 3a-d), the real part becomes more negative with increasing grains size (Fig. 3c) indicating higher conductivity. The NC film with a great number of small grains has the worst $\varepsilon_1$ even compared to PC film deposited in a poor vacuum. In contrast, all the single-crystalline films have larger negative $\varepsilon_1$ compared to JC and polycrystalline films. Observed decrease in negative $\varepsilon_1$ (conductivity) is primarily due to increased number of structural defects (including grain boundaries) in the films leading to the increased electron-phonon interactions, which make the films less metallic. In general the same influence of the films grain size on $\varepsilon_2$ is observed (Fig. 3d), except the PC film in 600-1000 nm wavelength range, which has larger losses than NC in spite of bigger grain size. Indeed, it can be explained by poor PC film purity, which leads to increased Drude term of the imaginary part of the dielectric permittivity[42], elevating losses at longer wavelengths (λ > 500 nm).

Material purity and surface roughness are the factors of the second priority in terms of silver dielectric permittivity in the 600-1000 nm and 325-600 nm wavelengths respectively. To demonstrate material purity effect, we compare the dielectric permittivity of relatively clean (NC, PCBG) and durty (PC) polycrystalline films with JC data. JC data was acquired from the thin films deposited near 170 times faster than the PC film (at very high evaporation rate of 60 Å s$^{-1}$), leading to much more purer silver film. Our measurements (Fig. 3c, 3d) indeed show larger negative $\varepsilon_1$ and lower $\varepsilon_2$ of JC data compared to all the polycrystalline films in the 600-1000 nm wavelengths. However, the above JC permittivity supremacy is almost neglected compared to the PCBG film, because of the very big grains, which, in contrast, improving the film optical quality. These material purity dependencies can be attributed to an increase in the electron-phonon interaction as described above.

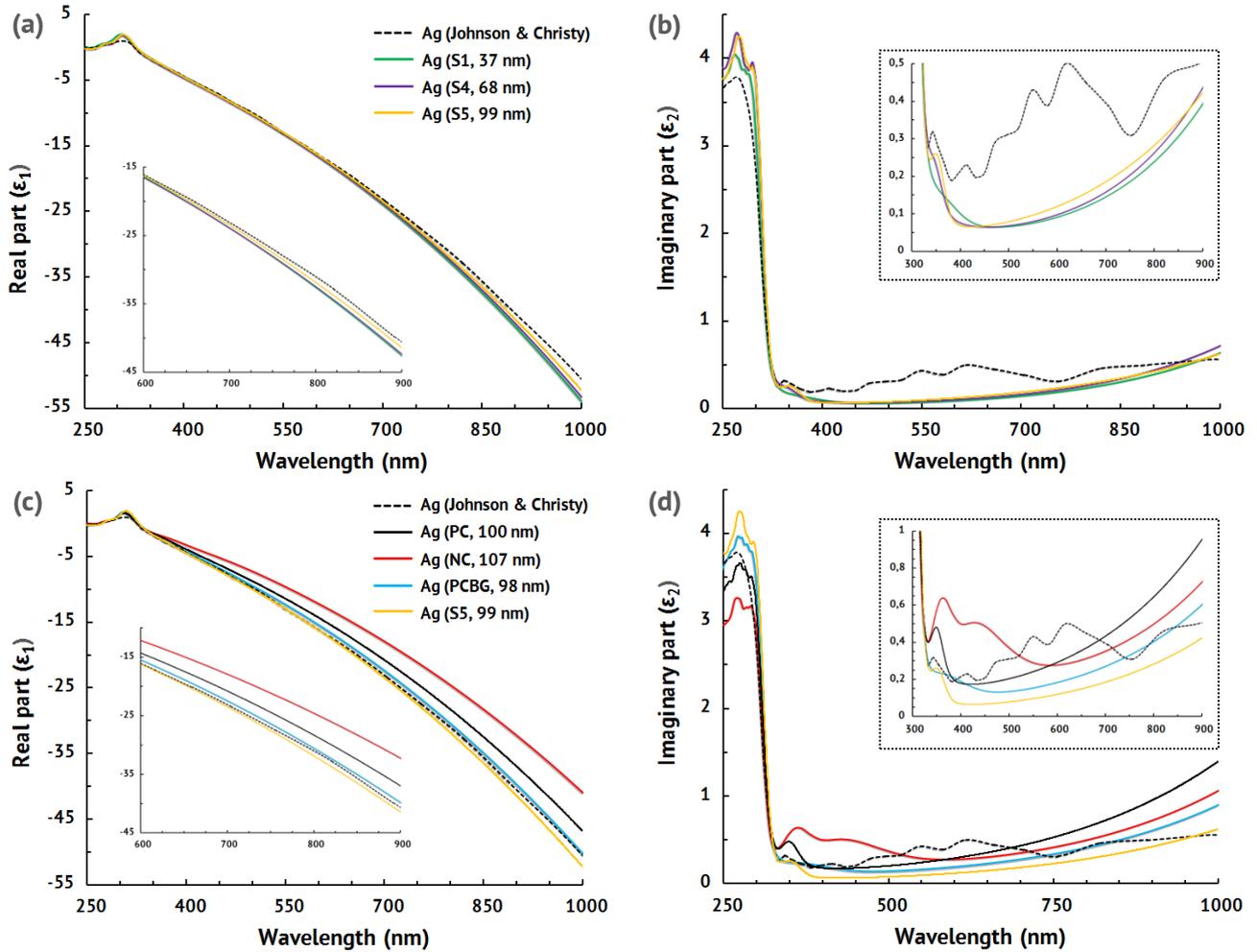

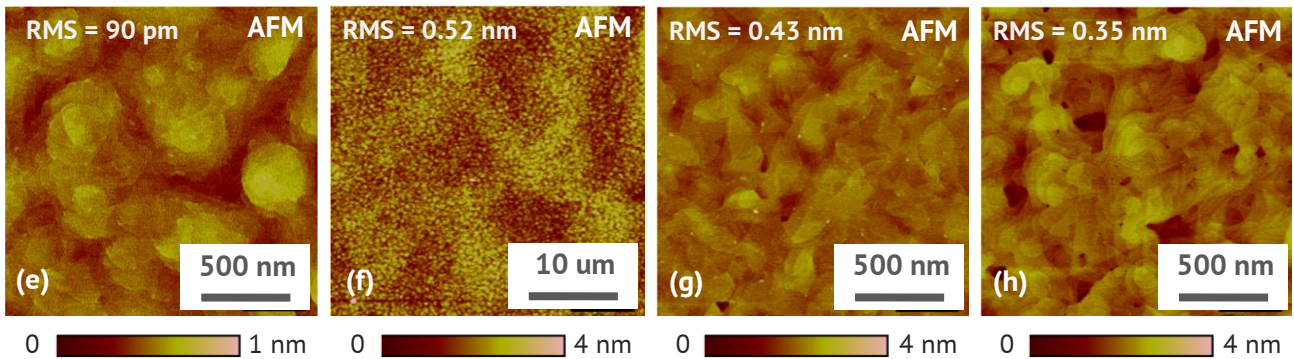

**Figure 3. | Optical properties and surface characterization.** Real (**a**) and imaginary (**b**) part of the dielectric permittivity of the single-crystalline films (S1, S4, S5). Dielectric permittivity (**c, d**) of nominally 100-nm-thick single-crystalline (S5) and polycrystalline (PC, NC, PCBG) films. AFM scans of S1 (**e**), S4 (**g**) and M1 (**h**) films measured over a 2.5×2.5 μm2 area, and S1 (**f**) film, measured over a 50×50 μm$^2$ area. All the films surfaces are continuous without pinholes and we observe no grain boundaries for single-crystalline films (e-h). The S1 film is extremely smooth with an atomic level of root mean square (RMS) roughness equal to 90 pm (e), which is the smoothest reported single-crystalline silver film. The RMS roughnesses of thicker films S4 and M1 are slightly larger, but still extremely smooth of 0.43 nm (c) and 0.35 nm (d).

At the 325-600 nm wavelengths, with increasing surface roughness (averaged value) and surface morphology singularities (absolute number of surface non uniformities) the $\varepsilon_2$ is dramatically increased, and for the NC film it becomes more than five times and more than twice larger (Fig. 3d) compared to the S5 and PCBG films respectively. Furthermore, there are typical peaks in $\varepsilon_2$ between 340 nm and 400 nm wavelengths for all the samples, and it is important to note that the S1 film peak amplitude is four times lower than the NC film peak amplitude. These $\varepsilon_2$ spectrum features can be explained by internal interfaces effects[43], that is, with the surface roughness and morphology increase a silver surface oxidation and chemical reactivity is boosting. The observed typical peaks in $\varepsilon_2$ are primarily due to the surface reaction with adsorbed sulphur[32,43], which transforms the silver into a non-metal silver-sulphide (by transfer of S-ions through the interface). In case of polycrystalline films the surface topography (active surface area) plays the key role in the increased silver surface chemical reactivity leading to $\varepsilon_2$ spectrum degradation close to interband transition threshold. For the single-crystalline films with improving the surface roughness (to sub-100 pm level) the typical peak associated with the interband transitions is almost eliminated (but is still present) due to silver surface perfect thermodynamic stability and weaker sorptivity to chemical elements from ambient.

As atomically smooth single-crystalline silver films with extremely low optical absorption and high conductivity can result in enhanced SPP propagation length, SCULL films should be of a great interest for high performance plasmonic applications. The extremely low losses and improved plasmonic properties of SCULL silver films were confirmed by demonstrating the longest SPP propagation length above two hundred microns[44]. We believe that it is the result of the SCULL films unique characteristics and synergistic effect from the dedicated single-crystalline nature, atomically smooth surface, process-induced high purity and thermodynamic stability.

In summary, we have achieved the growth of continuous atomically smooth single-crystalline metallic films in wide range of thicknesses down to 35 nm using PVD SCULL process. The process is based on the two-step approach involving quantum engineering of AFT 2D islands crystal seed, known for 3D bulk materials, but not previously realized for thin films. Unlike the colloidal and MBE methods, it provides extremely low optical losses and improved plasmonic performance, allowing the demonstration of theoretically limited SPP propagation length above two hundred microns for silver thin films[44]. The key feature of our approach is the combination of two mixed evaporation modes together with AFT 2D crystal seed growth self-controlled by quantum size effects, which enables deposition of perfect single-crystalline metallic films on non-ideally lattice-matched substrates, even with imperfect standard deposition tools and typical process deviations. The proposed process has been approved for silver, gold and aluminum single-crystalline films growth on silicon, sapphire and mica substrates. We believe that it could be used for deposition of various atomically smooth single-crystalline metallic thin films. The unique physical and optical properties of SCULL films may open fundamentally new possibilities in nanophotonics[1,15], biotechnology[6,10] and quantum technologies[9]. The SCULL process, as well, could be easily integrated and is compatible with planar top-down device fabrication technology.

**References**

1. Melikyan, A. *et al.* High-speed plasmonic phase modulators. *Nat. Photon.* **8**, 229-233 (2014).

2. Bozhevolnyi, S.I. *et al* Channel plasmon subwavelength waveguide components including interferometers and ring resonators. *Nature* **440**, 508 (2006).

3. Huang, J.-S. *et al.* Atomically flat single-crystalline gold nanostructures for plasmonic nanocircuitry. *Nat. Commun.* **1**, 150 (2010).

4. Wang, C. Y. *et al.* Giant colloidal silver crystals for low-loss linear and nonlinear plasmonics. *Nat. Commun.* **6**, 7734 (2015).

5. Pyayt, A.L. *et al.* Integration of photonic and silver nanowire plasmonic waveguides. *Nat. Nanotech.* **3**, 660 (2008).

6. Melentiev, P. *et al.* Plasmonic nanolaser for intracavity spectroscopy and sensorics. *Appl. Phys. Lett.* **111**, 213104 (2017).

7. Oulton, R.F. *et al.* Plasmon lasers at deep subwavelength scale. *Nature* **461**, 629 (2009).

8. Lu, Y.J. *et al.* Plasmonic nanolaser using epitaxially grown silver film. *Science* **337**, 450-453 (2012).

9. Bogdanov, S. *et al.* Ultrabright room-temperature sub-nanosecond emission from single nitrogen-vacancy centers coupled to nano-patch antennas. *Nano Lett.* Article ASAP (2018).

10. Anker, J.N. *et al.* Biosensing with plasmonic nanosensors. *Nat. Mater.* **7**, 442–453 (2008).

11. Durmanov, N.N. *et al.* Non-labeled selective virus detection with novel SERS-active porous silver nanofilms fabricated by Electron Beam Physical Vapor Deposition. *Sensors Act. B: Chem.* **257**, 37-47 (2018).

12. Atwater, H.A. and Polman, A. Plasmonics for improved photovoltaic devices. *Nat. Mater.* **9**, 205 (2010).

13. Pendry, J. B. Negative refraction makes a perfect lens. *Phys. Rev. Lett.* **85**, 3966–3969 (2000).

14. Boltasseva, A. and Atwater, H.A. Low-loss plasmonic metamaterials. *Science* **331**, 290-291 (2011).

15. High, A.A. *et al.* Visible-frequency hyperbolic metasurface. *Nature* **522**, 192 (2015).

16. West, P.R. *et al.* A. Searching for better plasmonic materials. *Laser & Photon. Rev.* **4**, 795-808 (2010).

17. Dastmalchi, B. *et al.* A new perspective on plasmonics: confinement and propagation length of surface plasmons for different materials and geometries. *Advanced Opt. Mater.* **4**, 177-184 (2016).

18. Park, J.H. *et al.* Single-Crystalline Silver Films for Plasmonics. *Adv. Mater.* **24**, 3988-3992 (2012).

19. Wu, Y. *et al.* Intrinsic optical properties and enhanced plasmonic response of epitaxial silver. *Adv. Mater.* **26**, 6106-6110 (2014).

20. Malureanu, R. and Lavrinenko, A. Ultra-thin films for plasmonics: a technology overview. *Nanotech. Rev.* **4**, 259-275 (2015).

21. McPeak, K.M. *et al.* Plasmonic films can easily be better: rules and recipes. *ACS Photon.* **2**, 326-333 (2015).

22. Kunwar, S. *et al.* Various silver nanostructures on sapphire using plasmon self-assembly and dewetting of thin films. *Nano-Micro Lett.* **9**, 17 (2017).

23. Gather, M.C. *et al.* Net optical gain in a plasmonic waveguide embedded in a fluorescent polymer. *Nat. Photon.* **4**, 457 (2010).

24. Ciesielski, A., *et al.* Controlling the optical parameters of self-assembled silver films with wetting layers and annealing. *Appl. Surf. Sci.* **421**, 349-356 (2017).

25. Logeeswaran, V.J. *et al.* Ultrasmooth silver thin films deposited with a germanium nucleation layer. *Nano Lett.* **9**, 178-182 (2008).

26. Formica, N. *et al.* Ultrastable and atomically smooth ultrathin silver films grown on a copper seed layer. *ACS Appl. Mater. & Interfaces* **5**, 3048-3053 (2013).

27. Hanawa, T. and Oura, K. Deposition of Ag on Si (100) Surfaces as Studied by LEED-AES. *Japanese Journal of Appl. Phys.* **16**, 519 (1977).

28. Ditlbacher, H. *et al.* Silver Nanowires as Surface Plasmon Resonators. *Phys. Rev. Lett.* **95**, 257403 (2005).

29. Johnson, P.B. and Christy, R.W. Optical constants of the noble metals. *Phys. Rev. B* **6**, 4370 (1972).

30. Baburin, A.S. *et al.* Silver Films Deposited by Electron-Beam Evaporation for Application in Nanoplasmonics. *Herald of the Bauman Moscow State Tech. Univ. Instrum. Eng*, **6**, 4-14 (2016).

31. Baburin, A.S. *et al.* Highly directional plasmonic nanolaser based on highperformance noble metal film photonic crystal. *Nanophotonics VII Proc. SPIE* **10672**, 106724D (2018).

32. Rodionov, I.A. *et al.* Crystalline structure dependence on optical properties of silver thin film over time. *2017 Progress In Electromagnetics Research Symposium — Spring (PIERS)*, 1497-1502 (2017).

33. Baski, A.A. and Fuchs, H. Epitaxial growth of silver on mica as studied by AFM and STM. *Surf. Sci.* **313**, 275-288 (1994).

34. Kern, R., Le Lay, G. and J.J. Metois. *Current Topics in Materials Science* **3**, 131, ed. by E. Kaldis (1979).

35. Goswami, D.K. *et al.* Preferential heights in the growth of Ag islands on Si (1 1 1)-(7× 7) surfaces. *Surf. Sci.* **601**, 603-608 (2007).

36. Fokin, D.A. *et al.* Electronic growth of Pb on the vicinal Si surface. *Phys. Status Solidi C* **7**, 165-168 (2010).

37. Su, W.B. *et al.* Correlation between quantized electronic states and oscillatory thickness relaxations of 2D Pb islands on Si (111)-(7× 7) surfaces. *Phys. Rev. Lett.* **86**, 5116 (2001).

38. Czoschke, P. *et al.* Quantum size effects in the surface energy of Pb∕Si (111) film nanostructures studied by surface x-ray diffraction and model calculations. *Phys. Rev. B*, **72**, 075402 (2005).

39. Bromann, K. *et al.* Interlayer mass transport in homoepitaxial and heteroepitaxial metal growth. *Phys. Rev. Lett.* **75**, 677 (1995).

40. Altfeder, I.B., Matveev, K.A. and Chen, D.M. Electron fringes on a quantum wedge. *Phys. Rev. Lett*. **78**, 2815 (1997).

41. Sette, F. *et al.* Coverage and chemical dependence of adsorbate-induced bond weakening in metal substrate surfaces. *Phys. Rev. Lett.* **61**, 1384 (1988).

42. Pinchuk, A., Kreibig, U., Hilger A. Optical properties of metallic nanoparticles: influence of interface effects and interband transitions. *Surface Science* **557**, 269–280 (2004).

43. Kreibig, U. *et al.* Interfaces in nanostructures: optical investigations on cluster-matter. *Nanostructured Materials*, **11**, 1335-1342 (1999).

44. Baburin, A.S. *et al.* Toward theoretically limited SPP propagation length above two hundred microns on ultra-smooth silver surface. *eprint arXiv:1806.07606* (2018)

**Acknowledgements**

We would like to thank Alexey P. Vinogradov, Vladimir M. Shalaev, Alexandra Boltasseva, Denis A. Fokin, Alexander M. Merzlikin, Alexander V. Baryshev and Alexander S. Dorofeenko for the helpful discussions. The SCULL process was developed and the samples were prepared at the BMSTU Nanofabrication Facility (Functional Micro/Nanosystems, FMNS REC, ID 74300).

**Author contributions**

The SCULL metal growth process was developed by I. A. R., I. A. R. and A. S. B. A. R. G. analyzed the data and developed samples cleaning procedure, S. S. M. performed the XRD measurements, S. P. performed the ellipsometry measurements and fitting. A. V. A. supervised the study. All authors analyzed the data and contributed to writing the manuscript.

**Additional information**

**Competing financial interests**

The authors declare no competing financial interests.

# Methods

## Preparation of Epitaxial Films

Epitaxial films of silver were deposited on prime-grade degenerately doped Si(111), Si(100), Si(110) wafers (0.0015-0.005 Ω-cm) and muscovite mica substrates using 10 kW e-beam evaporator (Angstrom Engineering) with a base pressure lower than $3 \times 10^{-8}$ Torr. We first cleaned the wafers in a 2:1 sulfuric acid: hydrogen peroxide solution (80°C), followed by further cleaning in isopropanol to eliminate organics. Finally, we placed the wafers in 49% hydrofluoric acid for approximately 20 s to remove the native oxide layer. After oxide removal, we immediately transferred the wafers into the evaporation tool and pumped the system down to limit native oxide growth. Mica substrates were cleaved perpendicular to the c-axis to reveal fresh surfaces, prior to deposition. All films were grown using 5N (99.999%) pure silver. Films were deposited with rate of 0.5-10 Å·s$^{-1}$ measured with quartz monitor at approximate source to substrate distance of 30 cm. Deposition is done in two steps using SCULL process. First, seed island layer is deposited on the elevated temperature substrate. At the second step, evaporation is stopped and the substrate is cooled to room temperature. Then, silver deposition is started on seed layer until continuous single-crystalline film is formed. Different temperature in a range from room temperature to 450°C were used.

The NC film was e-beam evaporated onto liquid nitrogen cooled quartz substrate, the conditions were adjusted so that the film had an average grains size around 20 nm. The PCBG film was e-beam evaporated on a quartz substrate using optimized two-step process to obtain an average grains size lager than 500 nm. A 100-nm-thick polycrystalline film (PC) was deposited under $10^{-6}$ Torr pressure (dry-pumped) at room temperature and 2 Å·s$^{-1}$ deposition rate. In order to emulate highly cited JC silver films the 100-nm-thick polycrystalline film (PC) was deposited under poor vacuum conditions ($10^{-6}$ Torr).

## X-Ray Diffraction (XRD)

X-ray diffraction was studied by means of the Rigaku SmartLab diffractometer. Parallel beam geometry and Cu Kα1 radiation was applied. X-ray diffraction pattern was measured in 2θ/ω mode from 20° to 90° (2θ) with 0.02° step. To study texture and azimuthal orientation of silver crystals in relation to silicon monocrystal axes, φ-scans for both Ag and Si layers were measured from 0° to 360° with 0.052° step. Rocking curves (or ω-scans, 0.001° step) were applied to characterize in-plane perfection of silver crystals. Position of X-ray reflections and full width at half maximum values of the rocking curves were determined through experimental data approximation with a pseudo-Voigt function by means of a CSD software package.[46]

In each sample, observed reflections were caused by sets of crystallographic planes with divisible Miller indices. This means that crystallographic planes of silver crystals were parallel to planes of the substrate with the same Miller indices. Ag(111) was parallel to Si(111), Ag(110)//Si(110) and Ag(100)//Si(100). A difference in value of lattice parameter of silver on substrate with various Si orientations, if any, was lower than uncertainty of the method of investigation. Averaged out of all samples lattice parameter was 4.076 ± 0.004 Å, which was in good agreement with known value for pure Ag atomic weight.[47] Curves of φ-scans consisted of sharp reflections for all the samples (See Supplementary Fig. S1-S5). The number and position of these reflections coincided with standard (111), (110) and (001) FCC crystal projections. In addition, the position of reflections from the silver film matched in each case to a position of those of a silicon substrate. This result meant that each studied silver film possessed biaxial texture. Analysis of 2θ/ω and φ-scans showed that each studied sample was biaxially textured silver film on mono-crystalline silicon substrate with the following epitaxial relationships: (111)Ag//(111)Si:[111]Ag//[111]Si, (110)Ag//(110)Si:[110]Ag//[110]Si and (001)Ag//(001)Si:[001]Ag//[001]Si. Additionally, φ-scans showed that misorientation of these films and a substrate did not exceed 0.1°. Appearance of a Si(222) forbidden reflection is in accordance with recently published investigations.[48]

Since full-width-at-half-maximum (FWHM) of a rocking curve profile serves as a practical numerical characteristic of mosaic spread[47] in thin crystalline films for plasmonics[49], precise determination of this value is of importance. The values derived were: 0.325 for S1 sample, 0.829 for S2 and 0.831 for S3. The FWHM value for S1 sample was comparable to that of a previously reported single-crystalline films for modern plasmonic applications but of larger thicknesses: 0.116° for 1-μm thick Ag film[15] and 0.22° for 200-nm thick Ag film[18] S4 and S5 samples showed same diffractometric character as S1 and a rocking curve FWHM value of 0.221° and 0.368°.

**Atomic Force Microscopy (AFM)**

The atomic force microscope Bruker Dimension Icon with SCANASYST-AIR-HR probe (with nominal tip radius of 2 nm) was used. All AFM images were obtained by using PeakForce Tapping mode with ScanAsyst imaging and the scanned area was 2.5×2.5 μm$^2$ and 50×50 μm$^2$. Nanoscope software was utilized to analyze the images and extract root mean square roughness.

**Ellipsometry**

Dielectric functions of the silver films were measured using a multi-angle spectroscopic ellipsometer (SER 800, Sentech GmbH). Additionally, ellipsometers in three different laboratories have been crosschecked (for single-crystalline silver films on Si(111)) to eliminate the possibility of systematic errors. Modeling and analysis were performed with the ellipsometer SENresearch 4.0 software. The models were developed in cooperation with Sentech GmbH application department. Measurement spectral wavelength range was from 240 to 1000 nm, with an interval of 1 nm, and the reflected light was analyzed at incidence angles of 50˚, 60˚, 70˚. For all samples, the mean square errors, representing the quality of the match between the measured and theoretically calculated dielectric functions, were less than 1.3˚, which implies good consistency between them.

**Profilometry**

The stylus profiler KLA Tencor P17 (with Durasharp 38-nm tip radius stylus) was used. All measurements were done by using 0.5 mg taping strength, scan rate was 2 $\mu m \cdot s^{-1}$ and the scanned line length was 20 μm.

**Scanning Electron Microscopy.**

In order to check the quality and uniformity of the deposited layers silver films surfaces after deposition were investigated by means of a scanning electron microscope Zeiss Merlin with a Gemini II column. All SEM images were obtained using in-lens detector and the accelerating voltage 5 kV and working distance from the sample to detector from 1 to 4 mm. Magnifications 3000, 7000, 15000 and 50000 were used to fully analyze samples.

**Back-scattered electron detection.**

EBSD scans were performed in a Zeiss Merlin SEM coupled with an Oxford Instrument NordlysNano EBSD detector and Aztec 3.0 software. EBSD data shown in Supplementary Fig. 2 and Supplementory were generated under 5-10 kV.

# Quantum Engineering of Single-Crystalline Silver Thin Films

*Ilya A. Rodionov*[*,1,2], *Aleksandr S. Baburin*[1,2], *Aidar R. Gabidullin*[1,2], *Sergey S. Maklakov*[4], *Swen Peters*[3], *Ilya A. Ryzhikov*[3,4], *and Alexander V. Andriyash*[2]

[1]Research and Educational Center Functional Micro/Nanosystems, Bauman Moscow State Technical University, Moscow, Russian Federation

[2]Dukhov Research Institute of Automatics, Moscow, Russian Federation,

[3]SENTECH Instruments GmbH, Berlin, Germany

[4]Institute for Theoretical and Applied Electromagnetics RAS, 125412, Moscow, Russian Federation

*e-mail: irodionov@bmstu.ru

**Table of content**

1. **Results and discussion**

Grain boundaries, material purity (hence, grain boundaries purity), surface roughness (and associated surface chemical reactivity), and crystalline imperfection contribute to optical properties of metallic films in descending order of priority. To demonstrate this, we compare the results for six representative films: three SCULL single-crystalline films of 35 nm (S1), 70 nm (S4) and 100 nm (S5) nominal thickness, and three nominally 100-nm-thick polycrystalline films (PC, PCBG, NC) with different grain size and purity. To estimate grain boundaries and surface roughness impact, we compare nominally 100-nm-thick polycrystalline films (PC, PCBG, NC) with various grain size and identical surface roughness (Table 1) with the single-crystalline (S5) film. The NC film was e-beam evaporated onto a liquid nitrogen cooled quartz substrate, the conditions were adjusted so that the film had an average grains size around 20 nm. The PCBG film was e-beam evaporated using optimized two-step process[6] to obtain an average grains size lager than 500 nm. In order to highlight material purity effect and emulate highly cited JC silver films, which was obtained under poor vacuum conditions ($4 \cdot 10^{-6}$ Torr, oil-pumped), a 100-nm-thick polycrystalline film (PC) was deposited under $10^{-6}$ Torr preasure (dry-pumped). Since dielectric permittivity is thickness independent[11,13], optical properties of the Ag(111)/Si(111) films (S1, S4, S5) without grain boundaries and the same material purity properties were compared to estimate a surface roughness and crystallinity impact. High-resolution wide-angle X-ray diffraction (XRD) rocking curves (see the XRD section for details) with a full width at half maximum (FWHM) of 0.325 °, 0.221°, 0.368 ° for the film thicknesses 37 nm (S1), 68 nm (S4) and 99 nm (S5) indicate thickness independent film high quality with minimal level of defects, which is comparable to previously best reported even thicker single-crystalline silver PVD films[18]. It is important to note the SCULL films deposited on a non lattice-matched Si (100) and Si (110) have predictably worse crystallinity, but also demonstrate atomically smooth surfaces with RMS roughness less that 4 Å (Table 1). The high-resolution transmission electron microscopy (HRTEM) image (Fig. 9) demonstrates the single-crystalline nature of the S1 silver film. Electron backscatter diffraction (EBSD) is used to analyse the domain structures and extract average grain size (Table 1) of single-crystalline (Fig. 7c, 8d, 8e) and polycrystalline films (Fig. 7a, 7b, 8a, 8b, 8c). Only a single domain is observed in both small-scale 2 μm (Fig. 8d) and large-scale 400 μm scans (Fig. 8f).

To estimate losses and rank the films parameters contribution to optical properties a multi-angle spectroscopic ellipsometry is used (see the Ellipsometry section for details). We focus on the most practically useful NIR and visible wavelength region for silver lays above the interband transitions ($\lambda > 325$ nm), where the contribution to $\varepsilon_1$ mainly comes from Drude terms (dc conductivity), but $\varepsilon_2$ is defined by both intraband and interband components. We observe the dominating contribution of grain boundaries to dielectric permittivity (Fig. 10), the real part becomes more negative with increasing grains size (Fig. 10c) indicating higher conductivity. The NC film with a great number of small grains has the worst $\varepsilon_1$ even compared to PC film deposited in a poor vacuum. Opposite, all the single-crystalline films have larger negative $\varepsilon_1$ compared to JC and polycrystalline films. The observed decrease in negative $\varepsilon_1$ (conductivity) is primarily due to increased number of structural defects (including grain boundaries) in the films leading in the the increased electron-phonon interactions, which make the films less metallic. In general the same influence of the films grain size on $\varepsilon_2$ is observed (Fig. 10d), except the PC film in 600-1000 nm wavelength range, which has lager losses than NC in spite of bigger grain size. Indeed, it can be explained by poor PC film purity, which leads to increased Drude term of the imaginary part of the dielectric permittivity[14], elevating losses at longer wavelengths ($\lambda > 500$ nm).

Material purity and surface roughness are the factors of the second priority in term of silver dielectric permittivity in the 600-1000 nm and 325-600 nm wavelengths respectively. To demonstrate material purity effect, we compare the dielectric permittivity of relatively clean (NC, PCBG) and duty (PC) polycrystalline films with widely cited JC data. One should note that JC dielectric permittivity was averaged from the measurements of the films with 30.4 nm and 37.5 nm thickness deposited at very high evaporation rate (60 Å·s$^{-1}$). A dielectric permittity is thickness independent, however, the JC films were deposited near 170 times faster than the PC film (see the Deposition section), leading to much more pure silver film. Our measurements (Fig. 10c, 10d) indeed show larger negative $\varepsilon_1$ and lower $\varepsilon_2$ of JC data compared to all the polycrystalline films in the 600-1000 nm wavelengths. However, the above JC permittivity supremacy is almost neglected compared to the PCBG film, because of the very big grains, which, in contrast, improving the film optical quality. These material purity dependencies can be attributed to an increase in the electron-phonon interaction as described above.

The surface morphology of the films is characterized by AFM. All the films surfaces are continuous without pinholes and we observe no grain boundaries for single-crystalline films (Fig. 9). The S1 film is extremely smooth with an atomical level of root mean square (RMS) roughness equal to 90 pm (Fig. 9a), which is the smoothest reported single-crystalline silver film. The RMS roughness of thicker films S4 and M1 are slightly larger, but still extremely smooth of 0.43 nm (Fig. 9b) and 0.35 nm (Fig. 9c). At the 325-600 nm wavelengths, with increasing surface roughness (averaged value) and surface morphology singularites (absolute number of surface nonuniformities) the $\varepsilon_2$ is dramatically increased, and for the NC film it becomes more than five times and more than twice larger (Fig. 10d) compared to the S5 and PCBG films respectively. Futhermore, there are typical peaks in $\varepsilon_2$ between 340 nm and 400 nm wavelengths for all the samples, and it is important to note that the S1 film peak amplitude is four times lower than the NC film peak amplitude. These $\varepsilon_2$ spectrum features can be explained by internal interfaces effects[43], that is, with the surface roughness and morphology increase a silver surface oxidation and chemical reactivity is boosting. The observed typical peaks in $\varepsilon_2$ are primarily due to the surface reaction with adsorbed sulphur[15,16] which transforms, by transfer of S-ions through the interface, the silver into a non-metal silver-sulphide. In case of polycrystalline films the surface topography (active suface area) plays the key role in the increased silver surface chemical reactivity leading to $\varepsilon_2$ spectrum degradation close to interband transition threshold. For the single-crystalline films with improving the surface roughness (to sub-100-pm level) the typical peak associated with the interband transitions is almost eliminated (but is still present) due to silver surface perfect thermodynamic stability and weaker sorbtivity to chemical elements from ambient.

As atomically smooth single-crystalline silver films with extremely low optical absorption and high conductivity can result in an enhanced SPP propagation length, SCULL films should be of a great interest for high performance plasmonic applications. The extremely low losses and improved plasmonic properties of SCULL films was confirmed by demonstrating SPP propagation length above two hundred microns over the 100 nm single-crystalline silver film[17], which is twice longer than previously reported experimental results[10,11,18]. We believe that it is the result of the SCULL films unique characteristics and synergistic effect from the dedicated single-crystalline nature, atomically smooth surface, process-induced high purity and thermodynamic stability.

In this paper, we have demonstrates the two-step PVD SCULL process to obtain a continuous atomically smooth single-crystalline metallic films deposition over a wide range of thicknesses. The fundamental idea of the process involves quantum engineering of an effective underlying layer (1$^{st}$ step), which thermodinamically emulate a lattice-matched substrate for the targeted metal, followed by a single-crystalline film growth (2$^{nd}$ step) on the just syntesied lattice-matched substrate. The process provides the single-crystalline metallic films growth on non-ideally lattice-matched substrates without underlayers using a high vacuum electron-beam evaporator. The extremely low optical losses and improved plasmonic perfomance have been confirmed by demonstrating the longest reported SPP propagation length above two hundred microns[8].

## 2. Methods

### 2.1 Deposition

Silver thin films were deposited on prime-grade degenerately doped Si(111), Si(100), Si(110) wafers (0.0015-0.005 Ω-cm) and muscovite mica substrates using 10 kW e-beam evaporator (Angstrom Engineering) with a base pressure lower than $3 \times 10^{-8}$ Torr. We first cleaned the wafers in a 2:1 sulfuric acid: hydrogen peroxide solution (80°C), followed by further cleaning in isopropanol to eliminate organics. Finally, we placed the wafers in 49% hydrofluoric acid for approximately 20 s to remove the native oxide layer. After oxide removal, we immediately transferred the wafers into the evaporation tool and pumped the system down to limit native oxide growth. Mica substrates were cleaved perpendicular to the c-axis to reveal fresh surfaces, prior to deposition. All films were grown using 5N (99.999%) pure silver. Films were deposited with rate of 0.5-10 Å·s$^{-1}$ measured with quartz monitor at approximate source to substrate distance of 30 cm. Deposition is done in two steps using SCULL process. First, seed island layer is deposited on the elevated temperature substrate. At the second step, evaporation is stopped and the substrate is cooled to room temperature. Then, silver deposition is started on seed layer until continuous single-crystalline film is formed. Different temperature in a range from room temperature to 450˚C were used.

**2.2 X-Ray Diffraction (XRD)**

X-ray diffraction was studied by means of the Rigaku SmartLab diffractometer . PaTo study texture and azimuthal orientation of silver crystals in relation to silicon monocrystal axes, φ-scans for both Ag and Si layers were measured from 0° to 360° with 0.052° step. Rocking curves (or ω-scans, 0.001° step) were applied to characterize in-plane perfection of silver crystals. In each sample, observed reflections were caused by sets of crystallographic planes with divisible Miller indices. This means that crystallographic planes of silver crystals were parallel to planes of the substrate with the same Miller indices. Ag(111) was parallel to Si(111), Ag(110)//Si(110) and Ag(100)//Si(100). A difference in value of lattice parameter of silver on substrate with various Si orientations, if any, was lower than uncertainty of the method of investigation. Averaged out of all samples lattice parameter was 4.076 ± 0.004 Å, which was in good agreement with known value for pure Ag atomic weight[7]. Curves of φ-scans consisted of sharp reflections for all the samples (Fig. 1-5). The number and position of these reflections coincided with standard (111), (110) and (001) FCC crystal projections. Also, the position of reflections from the silver film matched in each case to a position of those of a silicon substrate. This result meant that each studied silver film possessed biaxial texture. Analysis of 2θ/ω and φ-scans showed that each studied sample was biaxially textured silver film on mono-crystalline silicon substrate with the following epitaxial relationships: (111)Ag//(111)Si:[111]Ag//[111]Si, (110)Ag//(110)Si:[110]Ag//[110]Si and (001)Ag//(001)Si:[001]Ag//[001]Si. Additionally, φ-scans showed that misorientation of these films and a substrate did not exceed 0.1°. Appearance of a Si(222) forbidden reflection is in accordance with recently published investigations[8]. Since full-width-at-half-maximum (FWHM) of a rocking curve profile serves as a practical numerical characteristic of mosaic spread[1] in thin crystalline films for plasmonics[9], precise determination of this value is of importance. The values derived were: 0.325 for S1 sample, 0.829 for S2 and 0.831 for S3. The FWHM value for S1 sample was comparable to that of a previously reported single-crystalline films for modern plasmonic applications but of larger thicknesses: 0.116° for 1-μm thick Ag film[10] and 0.22° for 200-nm thick

Ag film[11]. S7 sample showed same diffractometric character as S1 and a rocking curve FWHM value of 0.368°.

## 2.3. Atomic Force Microscopy (AFM)

The atomic force microscope Bruker Dimension Icon with SCANASYST-AIR-HR probe (with nominal tip radius of 2 nm) was used. All AFM images were obtained by using PeakForce Tapping mode with ScanAsyst imaging and the scanned area was 2.5×2.5 µm$^2$ and 50×50 µm$^2$. Nanoscope software was utilized to analyze the images and extract root mean square roughness.

## 2.4. Ellipsometry

Dielectric functions of the silver films were measured using a multi-angle spectroscopic ellipsometer (SER 800, Sentech GmbH). Additionally, ellipsometers in three different laboratories have been crosschecked (for single-crystalline silver films on Si(111)) to eliminate the possibility of systematic errors. We specifically measured the optical constants of the HF treated silicon substrate used in the deposition process to eliminate any discrepancy and uncertainty introduced by the substrate. These measured silicon optical constants and silver thickness are fixed in the subsequent data fitting for all samples, and only the silver parameters are allowed to change.

Modeling and analysis were performed with the ellipsometer SENresearch 4.0 software. The models were developed in cooperation with Sentech GmbH application department. Measurement spectral wavelength range was from 240 to 1000 nm, with an interval of approx. 2 nm, and the reflected light was analyzed at incidence angles of 50˚, 60˚, 70˚. To characterize the optical losses, the real ($\varepsilon_1$) and imaginary ($\varepsilon_2$) parts of the dielectric permittivity were extracted by fitting the measured raw ellipsometric data (Ψ and Δ). In our fitting, we used a bilayer Ag/Si structural model and a simple phenomenological Brendel-Bormann (BB) oscillator model[12] to interpret both the free electron and the interband parts of the dielectric response of our samples:

$$\hat{\varepsilon}(\omega) = \varepsilon_\infty - \frac{\omega_p^2}{\omega^2 + i\Gamma_D \omega} + \sum_{j=1}^{k} \chi_j(\omega), \qquad (1)$$

where $\omega_p$ is the plasma frequency, $\varepsilon_\infty$ is the background dielectric constant, $\Gamma_D$ is Drude damping, $\chi_j(\omega)$ is BB oscillators interband part of dielectric function, and k is the number of BB oscillators used to interpret the interband part of the spectrum.

Modern ellipsometers are capable of obtaining a presice optical data. Using flexible and sophisticated models and analysis software, it is possible accurately determine the optical constants of materials. While data obtaining creates no difficulties for high quality samples, analysis is not trivial. Extracting reliable permittivity is challenging because it is an inverse problem. Polarization ratio of reflected light are measured and the optical constants of the structure under investigation and layer thicknesses are retrieved.

The mean square error (MSE) is a crucial parameter to quantify the quality of fitted permittivity parameters. However, a small MSE alone is not a conclusive proof that the model is totally reliable. A model contains highly correlated parameters, so it is possible to have multiple solutions with similarly low MSE values. A strong correlation exist between thickness and optical constants in absorptive metal thin films and lead to unreliable permittivity values. To verify that the final fit solution is truly unique, we need to do a test showing that there is indeed a best fit at a singular value of a chosen parameter. The parameter we chose to perform the uniqueness test on is the independelty-measured thickness of the film. By fixing the thickness of the film at a measured value with tolerance 2 nm, while letting the other parameters vary during the fitting process, we calculated the MSE of each final fit result. For all samples, the mean square errors, representing the quality of the match between the measured and theoretically calculated dielectric functions, were the best for measured thickness less than 1.3°(Table S1-5). From these uniqueness tests, we conclude that our model is indeed reliable and the retrieved optical constants are valid.

**2.5. Thickness measurement**

The film thickness was measured independently by SEM crossection, profilometer. Zeiss Merlin with a Gemini II column was used for measurement, obtained image resolution is better than 1 nm. Fabricated film step was measured by stylus profiler KLA Tencor P17 (with Durasharp 38-nm tip radius stylus), repetability of vertical size measurement by profilometer is 0.4 nm. The measurments for nominally 65-nm thick film (S4) are presented on fig. 6.

**2.6. Profilometry**

The stylus profiler KLA Tencor P17 (with Durasharp 38-nm tip radius stylus) was used. All measurements were done by using 0.5 mg taping strength, scan rate was 2 $\mu m \cdot s^{-1}$ and the scanned line length was 20 $\mu m$.

**2.7. Scanning electron microscopy**

In order to check the quality and uniformity of the deposited layers silver films surfaces after deposition were investigated by means of a scanning electron microscope Zeiss Merlin with a Gemini II column. All SEM images were obtained using in-lens detector and the accelerating voltage 5 kV and working distance from the sample to detector from 1 to 4 mm. Magnifications 3000, 7000, 15000 and 50000 were used to fully analyze samples.

**2.8. Electron back scattered diffraction (EBSD) characterisation**

The Ag films were observed and structurally characterized by field emission scanning electron microscopy (FE-SEM: Zeiss Merlin Gemini II). The crystal orientation maps of the Ag films were obtained by FE-SEM equipped with an EBSD system (NordlysNano from Oxford Instrument, Oxford Instruments Corp., UK). EBSD patterns were acquired at the following shooting modes: tilt angle – 70°, accelerating voltage – 10 keV, probe current – 1.7 nA and scan sizes 2x2 um$^2$ and 20 x 20 um$^2$ for SC film. EBSD has proven to be a useful tool for characterizing the crystallographic orientation aspects of microstructures at length scales ranging from dozens of nanometers to millimeters in the scanning electron microscope. Detector provides single grains detection by means of orientation measurement based on acquired Kikuchi patterns. Colored image represents grains orientation map, correlation between colors and orientations is shown on triangle diagram at the bottom left corner. We extract average grain size for our polycrystalline films using the embedded software package for an EBSD image processing AZtecHKL software package .

The growth direction is bottom-up along the [111] direction. EBSD images of NC (Fig. 7a), PC (Fig. 7b) PCBG(Fig. 7c) and S1 (Fig. 7d,e) silver films demonstrating single-crystalline nature without grain boundaries.

We extract average grain size for our polycrystalline films using the embedded software package for an EBSD image processing. The NC film was e-beam evaporated onto a liquid nitrogen cooled quartz substrate, the conditions were adjusted so that the film had an average grains size around 20 nm. The PCBG film (Fig. 7c) was e-beam evaporated using optimized two-step process[31] to obtain an average grains size lager than 500 nm. Only a single domain is observed in both small-scale 2 μm (Fig. 7d) and large-scale 400 μm scans (Fig. 7e), confirming the quality of our single-crystalline silver films.

**2.9. Transmission electron microscopy (TEM)**

In order to check the quality and crystallinity of the nominally 35-nm deposited silver film on Si(111) its crossection made by ion milling was investigated by means of a transmission electron microscope TITAN³ 300. TEM image (Fig. 11) was obtained using in-lens detector and the accelerating voltage 100 kV, spot size 3.

## 3. Supplementary Information Figures.

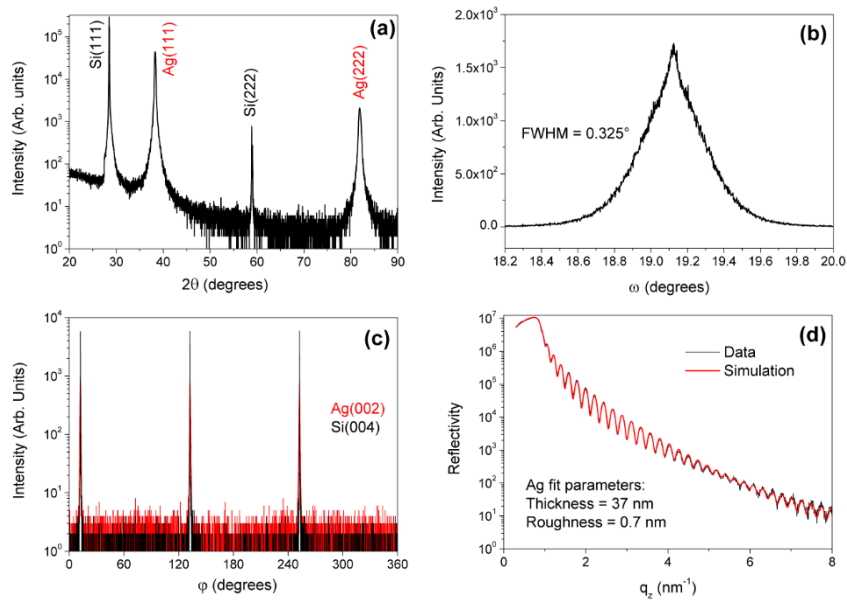

**Supplementary Figure 1. | XRD characterization of a nominally 35-nm-thick Si(111)/Ag(111) film (S1).** (**a**) High-resolution X-ray diffraction (θ-2θ) pattern. (**b**) Measured transverse scan (rocking curve, ω-scan) through the Ag(111) diffraction peak. (**c**) Grazing incidence of the in-plane X-ray diffraction scan (phi-scans) of the Ag(111) plane. (**d**) X-ray reflectivity curve.

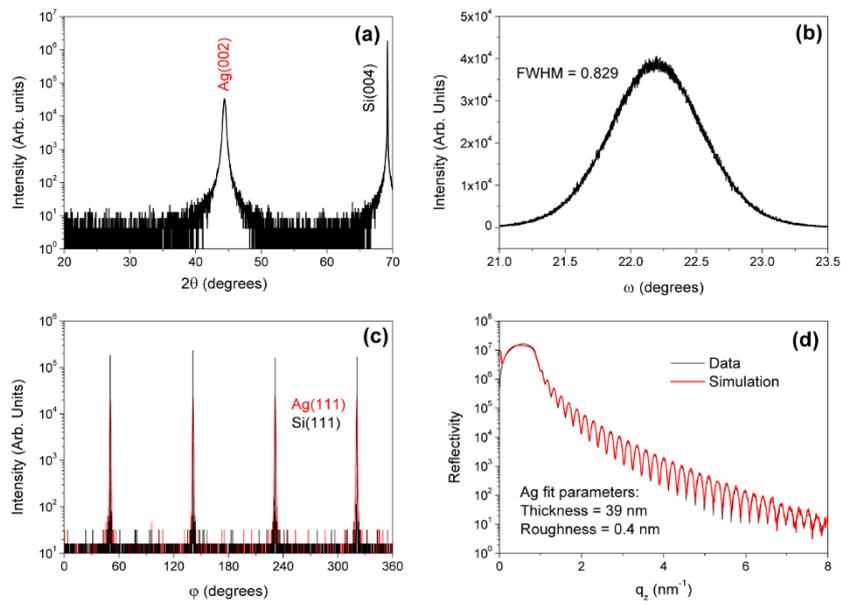

**Supplementary Figure 2. | XRD characterization of a nominally 35-nm-thick Si(100)/Ag(100) film (S2).** (**a**) High-resolution X-ray diffraction (θ-2θ) pattern. (**b**) Measured transverse scan (rocking curve, ω-scan) through the Ag(100) diffraction peak. (**c**) Grazing incidence of the in-plane X-ray diffraction scan (phi-scans) of the Ag(100) plane. (**d**) X-ray reflectivity curve.

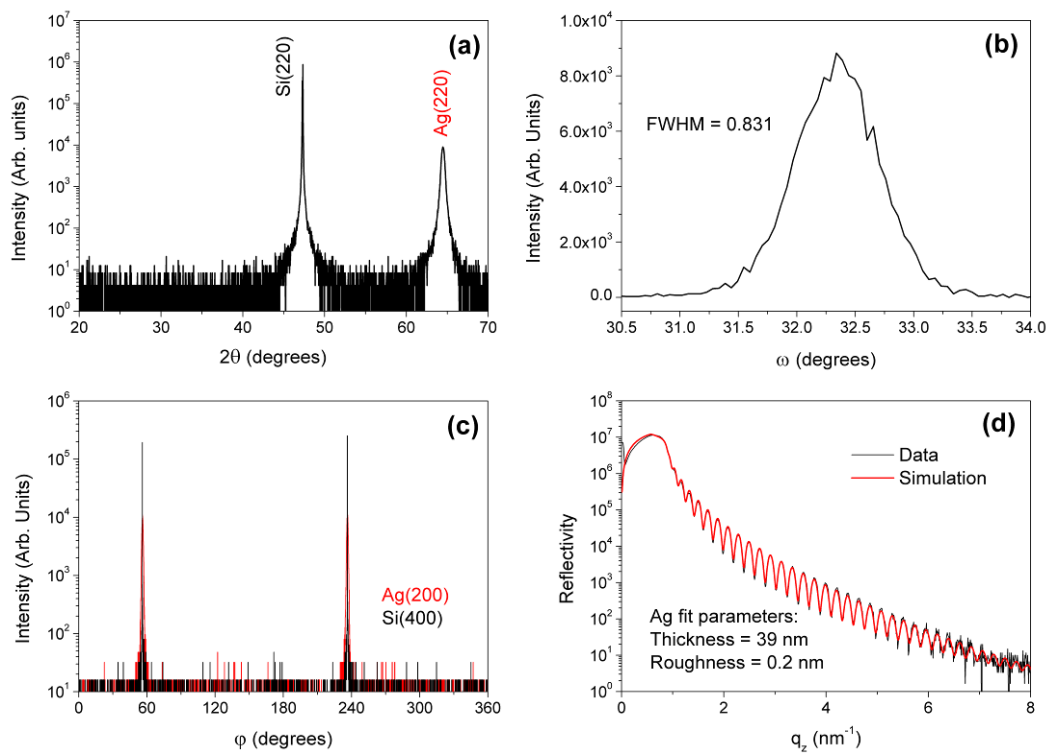

**Supplementary Figure 3. | XRD characterization of a nominally 35-nm-thick Si(110)/Ag(110) film (S3).** (**a**) High-resolution X-ray diffraction (θ-2θ) pattern. (**b**) Measured transverse scan (rocking curve, ω-scan) through the Ag(110) diffraction peak. (**c**) Grazing incidence of the in-plane X-ray diffraction scan (phi-scans) of the Ag(110) plane. (**d**) X-ray reflectivity curve.

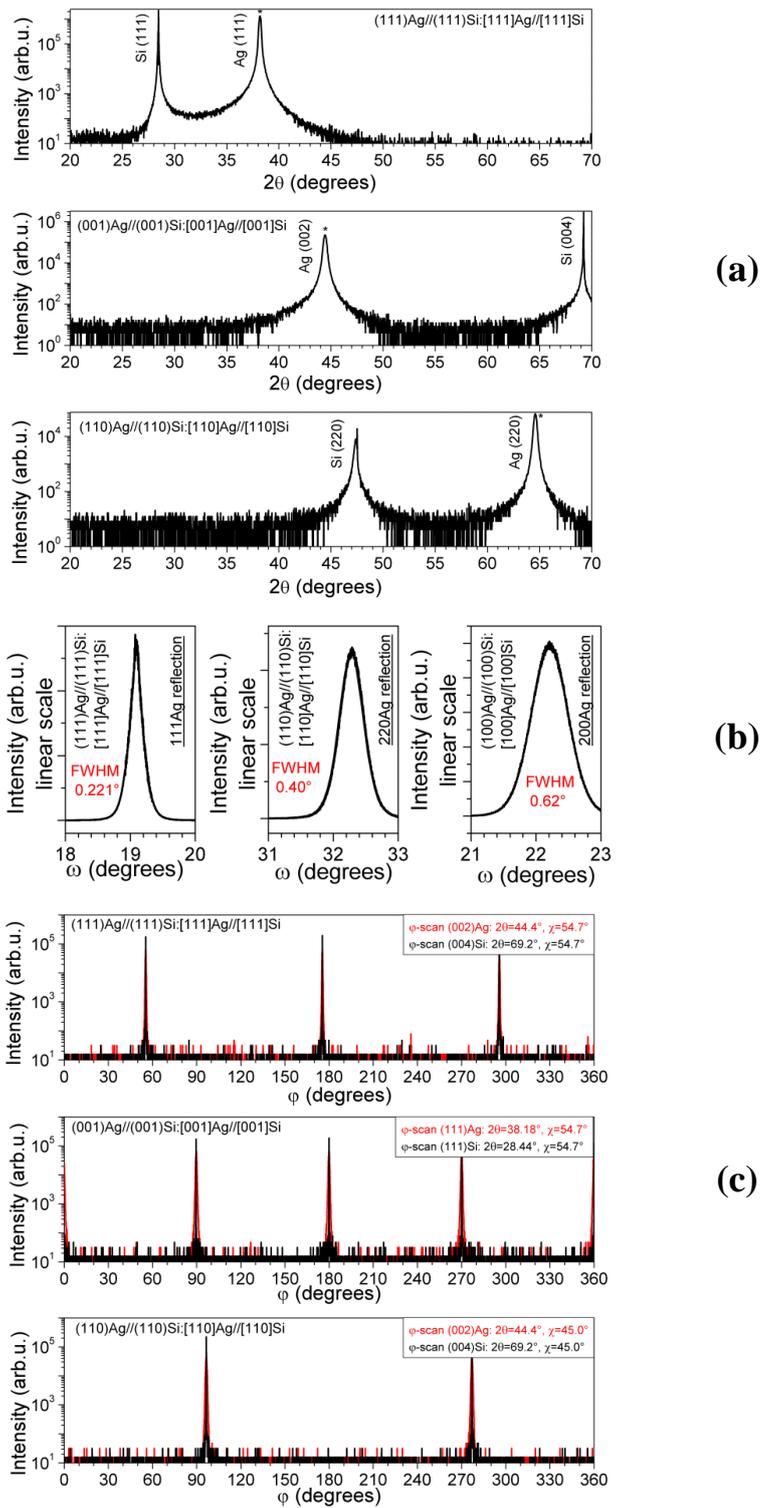

**Supplementary Figure 4. | XRD characterization of a nominally 65-nm-thick Ag on Si(111), Si(100), Si(110) substrates respectively.** (a) High-resolution X-ray diffraction (θ-2θ) pattern. (b) Measured transverse scan (rocking curve, ω-scan) through the Ag diffraction peak. (c) Grazing incidence of the in-plane X-ray diffraction scan (phi-scans) of the Ag plane.

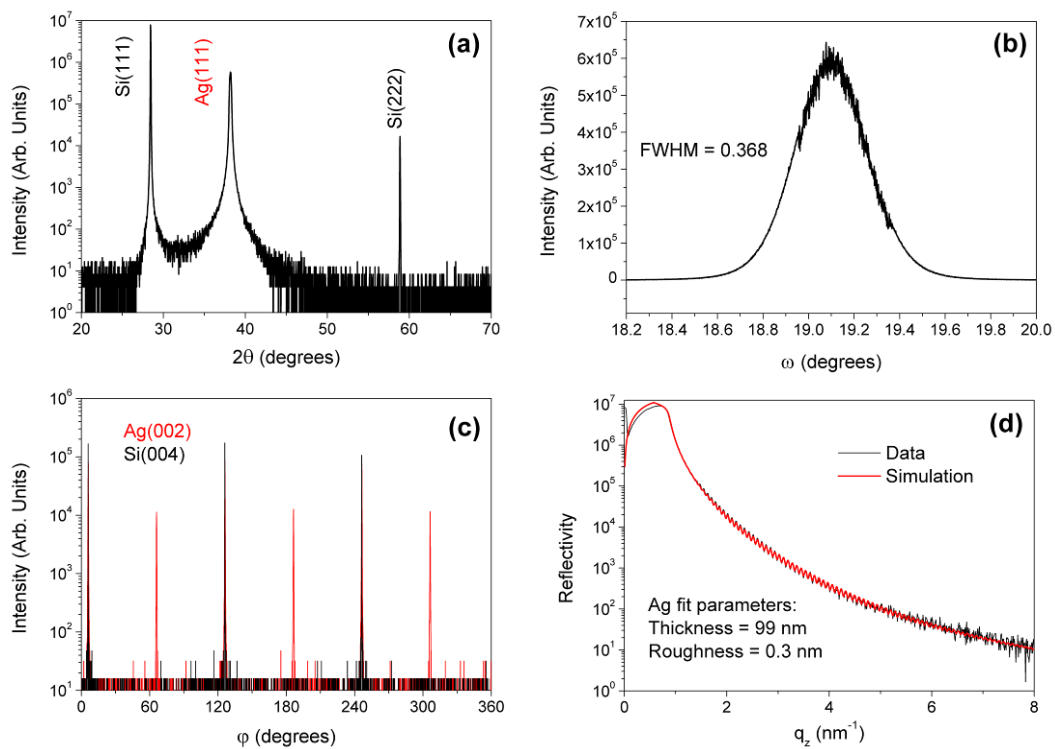

**Supplementary Figure 5. | XRD characterization of a nominally 100-nm-thick Si(111)/Ag(111) film (S5).** (**a**) High-resolution X-ray diffraction (θ-2θ) pattern. (**b**) Measured transverse scan (rocking curve, ω-scan) through the Ag(111) diffraction peak. (**c**) Grazing incidence of the in-plane X-ray diffraction scan (phi-scans) of the Ag(111) plane. (**d**) X-ray reflectivity curve.

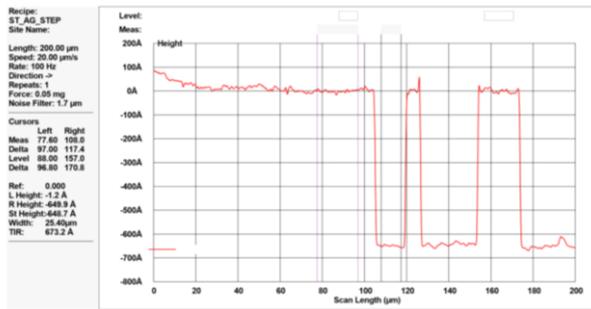 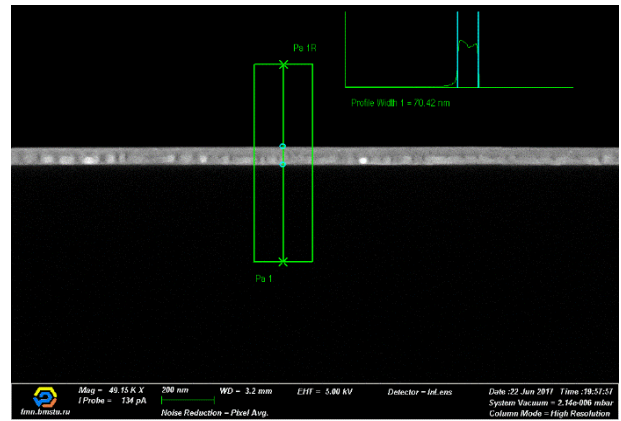

**Supplementary Figure 6. | Thickness measurments of a nominally 70-nm-thick Si(111)/Ag(111) film (S4).** (**a**) Profilometer step height measurment (**b**) SEM cross-section.

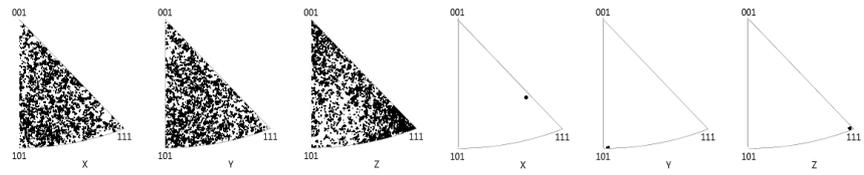

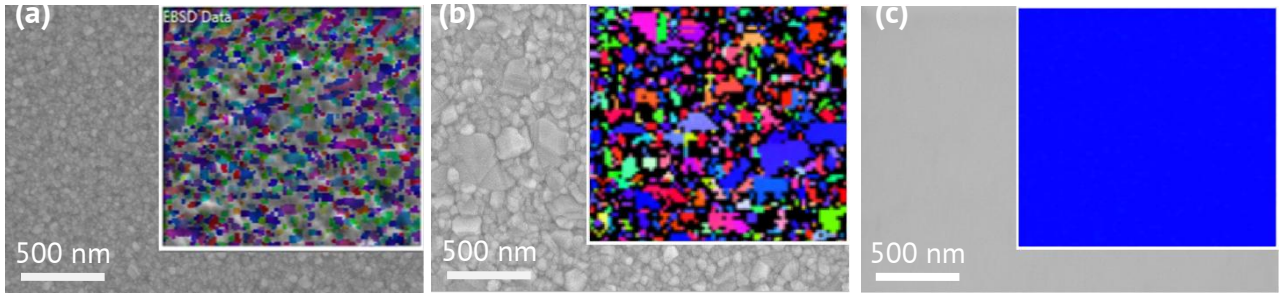

**Supplementary Figure 7. | SEM images with EBSD insets.** NC (**a**), PC (**b**) and S1 (**c**) silver films highlighting film grains. EBSD inverse pole figures are shown above the SEM images, demostrating very tight crystal orientation density of the S1 film (c) along all the normal directions.

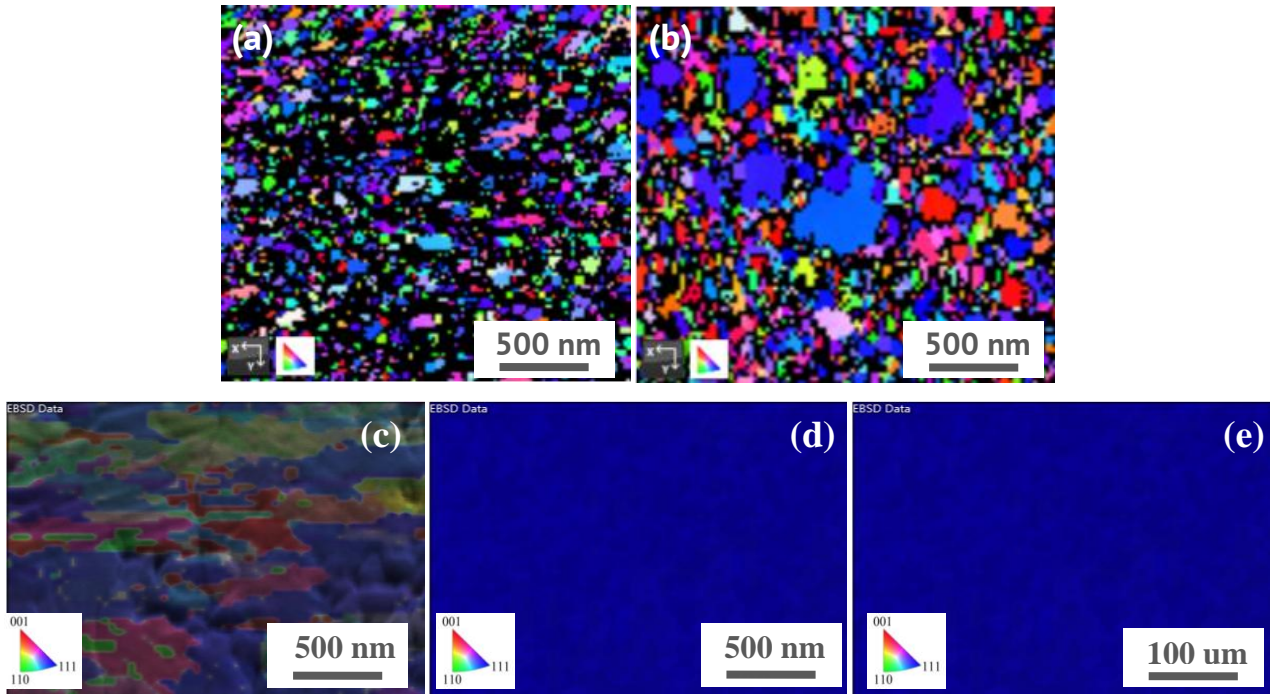

**Supplementary Figure 8. | EBSD images of silver thin films.** Nanocrystalline NC (a), polycrystalline PC (b), polycrystalline PCBG (c) and single-crystalline S1 (d, e) silver films. Only a single domain is observed in both small-scale 2 μm (d) and large-scale 400 μm scans (e), confirming the high quality and single-crystalline nature without grain boundaries over a large length scale.

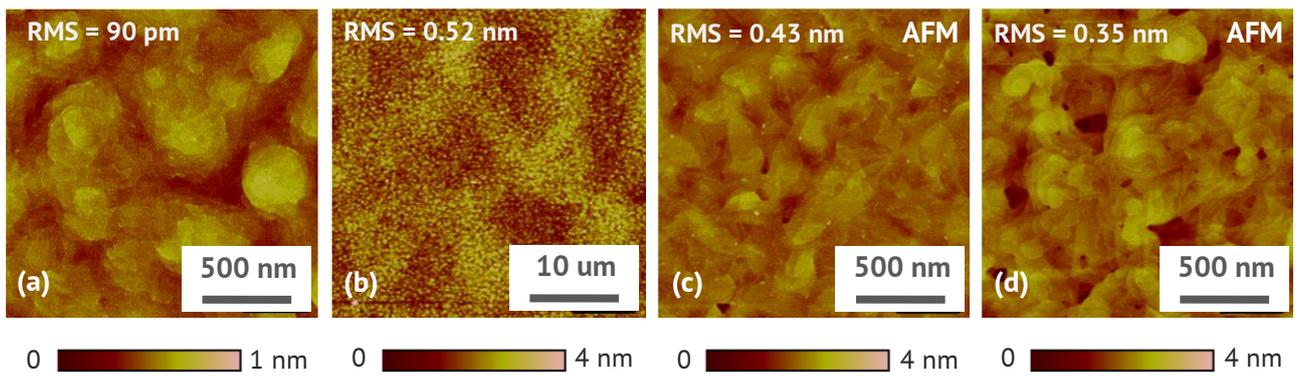

**Supplementary Figure 9. | AFM scans**. S1 (**a**), S4 (**b**) and M1 (**c**) films measured over a 2.5×2.5 μm² area, and S1 (**d**) film, measured over a 50×50 μm² area.

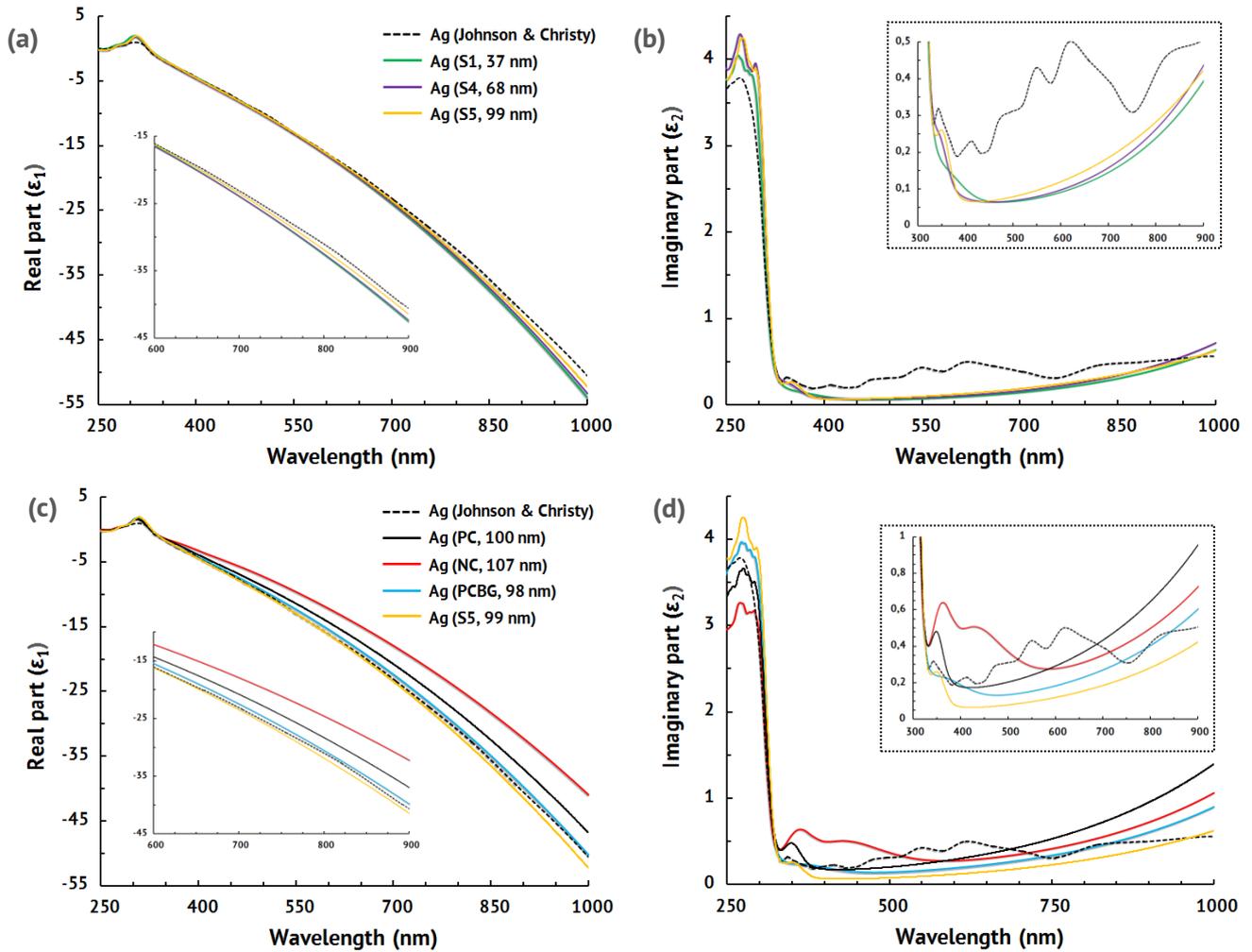

**Supplementary Figure 10. | Silver films dielectric permittivity.** Real (**a**) and imaginary (**b**) part of the dielectric permittivity of the single-crystalline films (S1, S4, S5). Dielectric permittivity (**c, d**) of nominally 100-nm-thick single-crystalline (S5) and polycrystalline (PC, NC, PCBG) films.

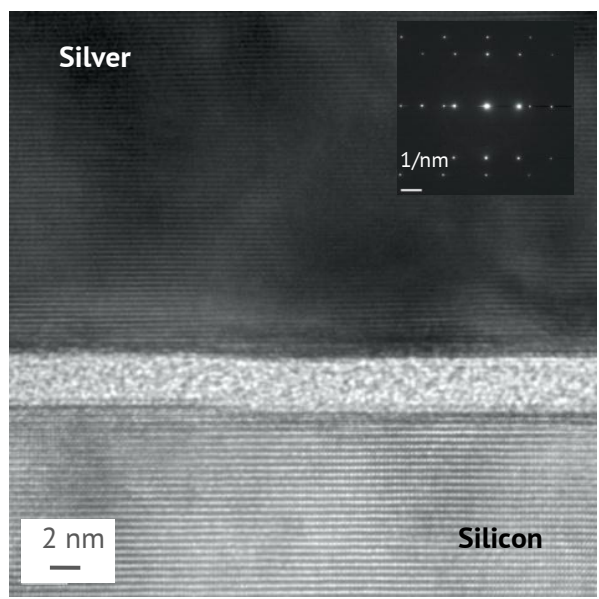

**Supplementary Figure 11. | HRTEM image and the electron diffraction pattern (inset in the right corner).** The growth direction is bottom-up.

**Supplementary Table 1. | Thickness, surface roughness and microstructure of SCULL Ag films as a function of substrate type.** AFM RMS roughness was determined from scans over a 2.5 × 2.5 μm$^2$. area. SP RMS roughness was determined from scans over a 20-μm length.

| Substrate | Measured thickness [nm] | Crystalline structure | Average grain size, [nm] | Rocking curve Ag peak, FWHM [°] | AFM RMS roughness [nm] | Sample |
|---|---|---|---|---|---|---|
| Mica | 35 | Single-crystalline | no grains | not measured | 0.35 | M1 |
| Si (111) | 37 | Single-crystalline | no grains | 0.325 | 0.09 | S1 |
| Si (100) | 39 | Single-crystalline | no grains | 0.829 | 0.28 | S2 |
| Si (110) | 39 | Single-crystalline | no grains | 0.831 | 0.37 | S3 |
| Si (111) | 68 | Single-crystalline | no grains | 0.221 | 0.36 | S4 |
| Si (111) | 99 | Single-crystalline | no grains | 0.368 | 0.43 | S5 |
| Quartz | 107 | Nanocrystalline | less than 20 | not measured | 2.18 | NC |
| Quartz | 100 | Polycrystalline | 50 | not measured | 2.34 | PC |
| Quartz | 98 | Polycrystalline | more than 500 | not measured | 2.22 | PCBG |

# Supplementary Table 2. | The silver thin films dielectric permittivity comparison

| | Sample S1 | | Sample S2 | | Sample S3 | | Sample S4 | | Sample S5 | | Sample NC | | Sample PC | | Sample PCBG | |
|---|---|---|---|---|---|---|---|---|---|---|---|---|---|---|---|---|
| λ, nm | $\varepsilon_1$ | $\varepsilon_2$ | $\varepsilon_1$ | $\varepsilon_2$ | $\varepsilon_1$ | $\varepsilon_2$ | $\varepsilon_1$ | $\varepsilon_2$ | $\varepsilon_1$ | $\varepsilon_2$ | $\varepsilon_1$ | $\varepsilon_2$ | $\varepsilon_1$ | $\varepsilon_2$ | $\varepsilon_1$ | $\varepsilon_2$ |
| 240.0 | -0.440 | 3.8356 | -0.259 | 3.5933 | -0.425 | 3.7566 | -0.175 | 4.0373 | -0.474 | 3.4993 | -0.095 | 2.9829 | -0.207 | 3.1800 | -0.333 | 3.4440 |
| 241.9 | -0.363 | 3.8381 | -0.251 | 3.6428 | -0.354 | 3.7547 | -0.172 | 3.9977 | -0.456 | 3.6096 | -0.053 | 2.9673 | -0.227 | 3.1983 | -0.347 | 3.4640 |
| 243.8 | -0.300 | 3.8379 | -0.236 | 3.6937 | -0.302 | 3.7509 | -0.177 | 3.9648 | -0.409 | 3.7073 | -0.028 | 2.9548 | -0.243 | 3.2208 | -0.357 | 3.4966 |
| 245.7 | -0.246 | 3.8351 | -0.214 | 3.7420 | -0.263 | 3.7503 | -0.187 | 3.9380 | -0.336 | 3.7694 | -0.013 | 2.9473 | -0.261 | 3.2550 | -0.358 | 3.5356 |
| 247.6 | -0.200 | 3.8286 | -0.187 | 3.7839 | -0.232 | 3.7534 | -0.201 | 3.9161 | -0.268 | 3.7858 | -0.004 | 2.9441 | -0.261 | 3.3022 | -0.350 | 3.5747 |
| 249.5 | -0.164 | 3.8191 | -0.157 | 3.8173 | -0.206 | 3.7599 | -0.221 | 3.8991 | -0.230 | 3.7754 | -0.001 | 2.9450 | -0.244 | 3.3377 | -0.336 | 3.6114 |
| 251.4 | -0.141 | 3.8106 | -0.130 | 3.8430 | -0.184 | 3.7707 | -0.246 | 3.8891 | -0.223 | 3.7651 | -0.001 | 2.9509 | -0.233 | 3.3661 | -0.317 | 3.6441 |
| 253.3 | -0.128 | 3.8084 | -0.109 | 3.8656 | -0.164 | 3.7864 | -0.276 | 3.8896 | -0.234 | 3.7696 | -0.003 | 2.9626 | -0.211 | 3.3995 | -0.296 | 3.6708 |
| 255.2 | -0.121 | 3.8165 | -0.092 | 3.8909 | -0.143 | 3.8059 | -0.307 | 3.9046 | -0.249 | 3.7917 | -0.004 | 2.9789 | -0.189 | 3.4178 | -0.278 | 3.6938 |
| 257.1 | -0.114 | 3.8350 | -0.074 | 3.9220 | -0.120 | 3.8266 | -0.333 | 3.9362 | -0.259 | 3.8285 | -0.005 | 2.9979 | -0.168 | 3.4414 | -0.263 | 3.7195 |
| 259.0 | -0.105 | 3.8629 | -0.048 | 3.9554 | -0.100 | 3.8493 | -0.346 | 3.9807 | -0.256 | 3.8734 | -0.009 | 3.0211 | -0.137 | 3.4526 | -0.245 | 3.7509 |
| 260.9 | -0.091 | 3.9011 | -0.016 | 3.9835 | -0.081 | 3.8803 | -0.343 | 4.0297 | -0.241 | 3.9167 | -0.014 | 3.0547 | -0.124 | 3.4473 | -0.217 | 3.7841 |
| 262.8 | -0.067 | 3.9515 | 0.014 | 4.0055 | -0.055 | 3.9257 | -0.328 | 4.0776 | -0.222 | 3.9526 | -0.012 | 3.1044 | -0.137 | 3.4594 | -0.181 | 3.8079 |
| 264.7 | -0.020 | 4.0095 | 0.038 | 4.0326 | -0.008 | 3.9805 | -0.306 | 4.1289 | -0.209 | 3.9882 | 0.011 | 3.1658 | -0.143 | 3.5055 | -0.150 | 3.8197 |
| 266.6 | 0.054 | 4.0604 | 0.068 | 4.0751 | 0.066 | 4.0271 | -0.269 | 4.1905 | -0.199 | 4.0410 | 0.063 | 3.2221 | -0.115 | 3.5673 | -0.135 | 3.8382 |
| 268.5 | 0.147 | 4.0892 | 0.125 | 4.1234 | 0.155 | 4.0500 | -0.200 | 4.2539 | -0.168 | 4.1175 | 0.133 | 3.2562 | -0.055 | 3.6126 | -0.117 | 3.8827 |
| 270.4 | 0.243 | 4.0928 | 0.209 | 4.1510 | 0.243 | 4.0506 | -0.098 | 4.2945 | -0.094 | 4.1963 | 0.205 | 3.2661 | 0.009 | 3.6301 | -0.062 | 3.9400 |
| 272.3 | 0.337 | 4.0775 | 0.292 | 4.1469 | 0.326 | 4.0403 | 0.015 | 4.2963 | 0.015 | 4.2433 | 0.269 | 3.2637 | 0.062 | 3.6410 | 0.026 | 3.9701 |
| 274.2 | 0.425 | 4.0446 | 0.357 | 4.1312 | 0.409 | 4.0192 | 0.117 | 4.2704 | 0.127 | 4.2512 | 0.333 | 3.2553 | 0.125 | 3.6606 | 0.108 | 3.9622 |
| 276.1 | 0.497 | 3.9951 | 0.420 | 4.1232 | 0.484 | 3.9799 | 0.213 | 4.2319 | 0.229 | 4.2419 | 0.399 | 3.2285 | 0.213 | 3.6560 | 0.167 | 3.9515 |
| 278.0 | 0.546 | 3.9459 | 0.494 | 4.1045 | 0.537 | 3.9378 | 0.309 | 4.1674 | 0.340 | 4.2214 | 0.444 | 3.1808 | 0.280 | 3.6139 | 0.237 | 3.9549 |
| 279.9 | 0.585 | 3.9145 | 0.554 | 4.0677 | 0.581 | 3.9147 | 0.373 | 4.0696 | 0.452 | 4.1573 | 0.458 | 3.1420 | 0.313 | 3.5846 | 0.324 | 3.9265 |
| 281.8 | 0.637 | 3.8948 | 0.598 | 4.0471 | 0.643 | 3.9025 | 0.388 | 3.9792 | 0.514 | 4.0644 | 0.464 | 3.1336 | 0.354 | 3.5823 | 0.371 | 3.8803 |
| 283.7 | 0.700 | 3.8681 | 0.662 | 4.0426 | 0.719 | 3.8781 | 0.382 | 3.9280 | 0.535 | 4.0002 | 0.491 | 3.1441 | 0.423 | 3.5759 | 0.406 | 3.8729 |
| 285.6 | 0.757 | 3.8375 | 0.749 | 4.0226 | 0.789 | 3.8432 | 0.391 | 3.9075 | 0.564 | 3.9724 | 0.539 | 3.1486 | 0.496 | 3.5426 | 0.479 | 3.8810 |
| 287.5 | 0.815 | 3.8265 | 0.831 | 3.9824 | 0.857 | 3.8226 | 0.421 | 3.8897 | 0.618 | 3.9476 | 0.586 | 3.1446 | 0.546 | 3.5024 | 0.575 | 3.8585 |
| 289.4 | 0.913 | 3.8316 | 0.906 | 3.9495 | 0.959 | 3.8147 | 0.448 | 3.8698 | 0.674 | 3.9057 | 0.637 | 3.1557 | 0.589 | 3.4921 | 0.654 | 3.8154 |
| 291.3 | 1.059 | 3.8061 | 1.008 | 3.9283 | 1.103 | 3.7768 | 0.472 | 3.8818 | 0.711 | 3.8746 | 0.724 | 3.1802 | 0.673 | 3.5055 | 0.724 | 3.7971 |
| 293.2 | 1.215 | 3.7299 | 1.147 | 3.8802 | 1.251 | 3.6924 | 0.549 | 3.9353 | 0.767 | 3.8867 | 0.856 | 3.1744 | 0.810 | 3.4890 | 0.841 | 3.7948 |
| 295.1 | 1.368 | 3.6320 | 1.290 | 3.7886 | 1.395 | 3.5914 | 0.712 | 3.9627 | 0.895 | 3.9042 | 0.994 | 3.1249 | 0.956 | 3.4210 | 1.005 | 3.7499 |
| 297.0 | 1.548 | 3.4917 | 1.429 | 3.6805 | 1.569 | 3.4534 | 0.905 | 3.9163 | 1.070 | 3.8631 | 1.139 | 3.0602 | 1.098 | 3.3305 | 1.168 | 3.6575 |
| 298.9 | 1.702 | 3.2718 | 1.597 | 3.5366 | 1.721 | 3.2330 | 1.100 | 3.8354 | 1.245 | 3.7765 | 1.316 | 2.9424 | 1.263 | 3.1991 | 1.344 | 3.5434 |
| 300.8 | 1.801 | 3.0304 | 1.744 | 3.3090 | 1.812 | 2.9903 | 1.339 | 3.6904 | 1.450 | 3.6560 | 1.454 | 2.7498 | 1.399 | 2.9933 | 1.539 | 3.3482 |
| 302.7 | 1.879 | 2.7726 | 1.826 | 3.0578 | 1.883 | 2.7376 | 1.533 | 3.4378 | 1.654 | 3.4319 | 1.549 | 2.5425 | 1.480 | 2.7727 | 1.664 | 3.0871 |
| 304.6 | 1.921 | 2.4758 | 1.883 | 2.8032 | 1.922 | 2.4492 | 1.664 | 3.1641 | 1.783 | 3.1601 | 1.625 | 2.3084 | 1.543 | 2.5436 | 1.743 | 2.8253 |
| 306.5 | 1.895 | 2.1609 | 1.914 | 2.5186 | 1.899 | 2.1393 | 1.772 | 2.8548 | 1.881 | 2.8744 | 1.652 | 2.0363 | 1.578 | 2.2803 | 1.801 | 2.5358 |
| 308.4 | 1.808 | 1.8665 | 1.888 | 2.2106 | 1.811 | 1.8470 | 1.815 | 2.4910 | 1.940 | 2.5411 | 1.609 | 1.7608 | 1.554 | 1.9976 | 1.800 | 2.2115 |
| 310.3 | 1.686 | 1.6039 | 1.798 | 1.9163 | 1.688 | 1.5883 | 1.763 | 2.1241 | 1.917 | 2.1812 | 1.516 | 1.5143 | 1.471 | 1.7323 | 1.723 | 1.8956 |
| 312.2 | 1.545 | 1.3633 | 1.669 | 1.6560 | 1.546 | 1.3546 | 1.646 | 1.7970 | 1.816 | 1.8468 | 1.398 | 1.2963 | 1.356 | 1.4974 | 1.598 | 1.6176 |
| 314.1 | 1.380 | 1.1388 | 1.523 | 1.4232 | 1.387 | 1.1369 | 1.499 | 1.5046 | 1.675 | 1.5525 | 1.262 | 1.0963 | 1.225 | 1.2840 | 1.453 | 1.3707 |
| 316.0 | 1.190 | 0.9367 | 1.361 | 1.2067 | 1.204 | 0.9378 | 1.323 | 1.2329 | 1.510 | 1.2841 | 1.101 | 0.9141 | 1.074 | 1.0846 | 1.289 | 1.1417 |
| 317.9 | 0.979 | 0.7663 | 1.179 | 1.0063 | 1.000 | 0.7664 | 1.112 | 0.9873 | 1.317 | 1.0357 | 0.918 | 0.7590 | 0.899 | 0.9035 | 1.098 | 0.9324 |
| 319.8 | 0.760 | 0.6309 | 0.975 | 0.8300 | 0.784 | 0.6279 | 0.870 | 0.7829 | 1.091 | 0.8179 | 0.721 | 0.6374 | 0.703 | 0.7503 | 0.884 | 0.7542 |

| | | | | | | | | | | | | | | | |
|---|---|---|---|---|---|---|---|---|---|---|---|---|---|---|---|
| 321.7 | 0.542 | 0.5272 | 0.757 | 0.6849 | 0.568 | 0.5210 | 0.616 | 0.6264 | 0.845 | 0.6423 | 0.523 | 0.5485 | 0.496 | 0.6302 | 0.657 | 0.6140 |
| 323.6 | 0.333 | 0.4488 | 0.536 | 0.5715 | 0.359 | 0.4402 | 0.365 | 0.5134 | 0.592 | 0.5104 | 0.331 | 0.4870 | 0.289 | 0.5417 | 0.431 | 0.5100 |
| 325.5 | 0.134 | 0.3895 | 0.320 | 0.4856 | 0.160 | 0.3796 | 0.126 | 0.4341 | 0.346 | 0.4162 | 0.148 | 0.4467 | 0.088 | 0.4798 | 0.214 | 0.4350 |
| 327.4 | -0.054 | 0.3440 | 0.112 | 0.4212 | -0.029 | 0.3337 | -0.096 | 0.3791 | 0.113 | 0.3508 | -0.023 | 0.4225 | -0.103 | 0.4390 | 0.010 | 0.3816 |
| 329.3 | -0.232 | 0.3086 | -0.084 | 0.3730 | -0.207 | 0.2984 | -0.302 | 0.3408 | -0.105 | 0.3069 | -0.184 | 0.4104 | -0.282 | 0.4145 | -0.181 | 0.3435 |
| 331.2 | -0.401 | 0.2806 | -0.270 | 0.3365 | -0.376 | 0.2710 | -0.494 | 0.3141 | -0.308 | 0.2784 | -0.334 | 0.4075 | -0.449 | 0.4027 | -0.360 | 0.3161 |
| 333.1 | -0.561 | 0.2581 | -0.447 | 0.3087 | -0.536 | 0.2493 | -0.672 | 0.2956 | -0.497 | 0.2607 | -0.475 | 0.4114 | -0.605 | 0.4005 | -0.528 | 0.2962 |
| 335.0 | -0.715 | 0.2399 | -0.614 | 0.2873 | -0.689 | 0.2318 | -0.839 | 0.2830 | -0.673 | 0.2504 | -0.606 | 0.4205 | -0.749 | 0.4055 | -0.686 | 0.2816 |
| 336.9 | -0.862 | 0.2250 | -0.774 | 0.2708 | -0.835 | 0.2176 | -0.995 | 0.2743 | -0.838 | 0.2454 | -0.730 | 0.4334 | -0.882 | 0.4156 | -0.836 | 0.2709 |
| 338.8 | -1.005 | 0.2128 | -0.927 | 0.2580 | -0.976 | 0.2059 | -1.142 | 0.2676 | -0.994 | 0.2447 | -0.846 | 0.4495 | -1.004 | 0.4288 | -0.978 | 0.2628 |
| 340.8 | -1.142 | 0.2029 | -1.074 | 0.2480 | -1.112 | 0.1961 | -1.282 | 0.2619 | -1.140 | 0.2474 | -0.955 | 0.4681 | -1.116 | 0.4433 | -1.114 | 0.2567 |
| 342.7 | -1.275 | 0.1949 | -1.216 | 0.2404 | -1.244 | 0.1879 | -1.415 | 0.2565 | -1.276 | 0.2520 | -1.058 | 0.4889 | -1.219 | 0.4574 | -1.245 | 0.2521 |
| 344.6 | -1.405 | 0.1884 | -1.353 | 0.2345 | -1.371 | 0.1808 | -1.543 | 0.2512 | -1.404 | 0.2561 | -1.154 | 0.5111 | -1.312 | 0.4693 | -1.372 | 0.2485 |
| 346.5 | -1.531 | 0.1832 | -1.485 | 0.2301 | -1.496 | 0.1745 | -1.666 | 0.2455 | -1.526 | 0.2587 | -1.243 | 0.5335 | -1.399 | 0.4775 | -1.494 | 0.2457 |
| 348.4 | -1.653 | 0.1790 | -1.614 | 0.2268 | -1.617 | 0.1688 | -1.785 | 0.2386 | -1.643 | 0.2600 | -1.326 | 0.5549 | -1.480 | 0.4811 | -1.613 | 0.2435 |
| 350.3 | -1.773 | 0.1757 | -1.740 | 0.2245 | -1.736 | 0.1637 | -1.902 | 0.2307 | -1.756 | 0.2604 | -1.403 | 0.5745 | -1.558 | 0.4795 | -1.728 | 0.2416 |
| 352.2 | -1.890 | 0.1730 | -1.863 | 0.2230 | -1.852 | 0.1590 | -2.016 | 0.2222 | -1.864 | 0.2585 | -1.477 | 0.5919 | -1.633 | 0.4725 | -1.841 | 0.2401 |
| 354.1 | -2.005 | 0.1709 | -1.983 | 0.2222 | -1.966 | 0.1545 | -2.129 | 0.2130 | -1.969 | 0.2530 | -1.547 | 0.6073 | -1.709 | 0.4606 | -1.952 | 0.2388 |
| 356.0 | -2.118 | 0.1691 | -2.100 | 0.2218 | -2.079 | 0.1502 | -2.240 | 0.2030 | -2.074 | 0.2453 | -1.614 | 0.6207 | -1.785 | 0.4443 | -2.060 | 0.2377 |
| 357.9 | -2.228 | 0.1673 | -2.215 | 0.2219 | -2.190 | 0.1461 | -2.351 | 0.1923 | -2.177 | 0.2365 | -1.678 | 0.6314 | -1.863 | 0.4245 | -2.167 | 0.2367 |
| 359.8 | -2.337 | 0.1655 | -2.328 | 0.2221 | -2.300 | 0.1420 | -2.462 | 0.1814 | -2.280 | 0.2260 | -1.739 | 0.6389 | -1.945 | 0.4023 | -2.271 | 0.2357 |
| 361.7 | -2.444 | 0.1636 | -2.439 | 0.2223 | -2.408 | 0.1379 | -2.572 | 0.1707 | -2.383 | 0.2132 | -1.799 | 0.6428 | -2.029 | 0.3788 | -2.374 | 0.2347 |
| 363.6 | -2.549 | 0.1614 | -2.548 | 0.2225 | -2.516 | 0.1339 | -2.682 | 0.1606 | -2.486 | 0.1989 | -1.860 | 0.6437 | -2.117 | 0.3552 | -2.476 | 0.2336 |
| 365.5 | -2.653 | 0.1589 | -2.655 | 0.2225 | -2.622 | 0.1300 | -2.792 | 0.1510 | -2.592 | 0.1844 | -1.920 | 0.6423 | -2.207 | 0.3321 | -2.576 | 0.2324 |
| 367.4 | -2.757 | 0.1560 | -2.761 | 0.2222 | -2.728 | 0.1261 | -2.901 | 0.1419 | -2.698 | 0.1707 | -1.980 | 0.6388 | -2.300 | 0.3102 | -2.676 | 0.2311 |
| 369.3 | -2.859 | 0.1528 | -2.866 | 0.2217 | -2.833 | 0.1222 | -3.010 | 0.1333 | -2.805 | 0.1578 | -2.041 | 0.6332 | -2.395 | 0.2900 | -2.774 | 0.2296 |
| 371.2 | -2.962 | 0.1492 | -2.969 | 0.2210 | -2.938 | 0.1184 | -3.119 | 0.1253 | -2.912 | 0.1456 | -2.102 | 0.6253 | -2.491 | 0.2718 | -2.872 | 0.2280 |
| 373.1 | -3.063 | 0.1453 | -3.072 | 0.2198 | -3.042 | 0.1146 | -3.228 | 0.1180 | -3.020 | 0.1339 | -2.165 | 0.6154 | -2.589 | 0.2559 | -2.969 | 0.2264 |
| 375.0 | -3.165 | 0.1411 | -3.174 | 0.2183 | -3.146 | 0.1109 | -3.336 | 0.1115 | -3.128 | 0.1228 | -2.230 | 0.6041 | -2.687 | 0.2422 | -3.065 | 0.2246 |
| 376.9 | -3.266 | 0.1366 | -3.275 | 0.2164 | -3.249 | 0.1073 | -3.444 | 0.1059 | -3.236 | 0.1127 | -2.297 | 0.5922 | -2.785 | 0.2306 | -3.161 | 0.2227 |
| 378.8 | -3.367 | 0.1320 | -3.375 | 0.2140 | -3.352 | 0.1039 | -3.551 | 0.1010 | -3.345 | 0.1039 | -2.366 | 0.5802 | -2.882 | 0.2210 | -3.257 | 0.2208 |
| 380.7 | -3.468 | 0.1274 | -3.476 | 0.2112 | -3.455 | 0.1007 | -3.658 | 0.0969 | -3.454 | 0.0964 | -2.437 | 0.5687 | -2.980 | 0.2130 | -3.352 | 0.2187 |
| 382.6 | -3.569 | 0.1227 | -3.576 | 0.2080 | -3.558 | 0.0976 | -3.765 | 0.0934 | -3.563 | 0.0903 | -2.509 | 0.5577 | -3.076 | 0.2063 | -3.446 | 0.2166 |
| 384.5 | -3.670 | 0.1182 | -3.675 | 0.2045 | -3.660 | 0.0947 | -3.870 | 0.0904 | -3.671 | 0.0853 | -2.582 | 0.5474 | -3.172 | 0.2008 | -3.540 | 0.2143 |
| 386.4 | -3.772 | 0.1138 | -3.775 | 0.2007 | -3.763 | 0.0919 | -3.975 | 0.0877 | -3.778 | 0.0814 | -2.655 | 0.5376 | -3.267 | 0.1962 | -3.634 | 0.2120 |
| 388.3 | -3.873 | 0.1096 | -3.875 | 0.1967 | -3.865 | 0.0892 | -4.080 | 0.0855 | -3.885 | 0.0783 | -2.730 | 0.5284 | -3.362 | 0.1923 | -3.728 | 0.2095 |
| 390.2 | -3.975 | 0.1056 | -3.974 | 0.1923 | -3.967 | 0.0867 | -4.184 | 0.0835 | -3.991 | 0.0758 | -2.805 | 0.5198 | -3.456 | 0.1891 | -3.822 | 0.2070 |
| 392.1 | -4.076 | 0.1018 | -4.074 | 0.1877 | -4.069 | 0.0844 | -4.288 | 0.0817 | -4.097 | 0.0737 | -2.881 | 0.5119 | -3.549 | 0.1864 | -3.916 | 0.2044 |
| 394.0 | -4.178 | 0.0982 | -4.174 | 0.1830 | -4.171 | 0.0822 | -4.391 | 0.0801 | -4.202 | 0.0721 | -2.958 | 0.5049 | -3.642 | 0.1841 | -4.009 | 0.2017 |
| 395.9 | -4.280 | 0.0950 | -4.274 | 0.1780 | -4.272 | 0.0802 | -4.494 | 0.0787 | -4.306 | 0.0708 | -3.036 | 0.4990 | -3.734 | 0.1822 | -4.103 | 0.1989 |
| 397.8 | -4.382 | 0.0920 | -4.374 | 0.1731 | -4.374 | 0.0784 | -4.596 | 0.0774 | -4.410 | 0.0697 | -3.113 | 0.4941 | -3.826 | 0.1805 | -4.196 | 0.1962 |
| 399.7 | -4.484 | 0.0893 | -4.475 | 0.1681 | -4.476 | 0.0767 | -4.698 | 0.0763 | -4.513 | 0.0688 | -3.191 | 0.4904 | -3.917 | 0.1791 | -4.290 | 0.1934 |
| 401.6 | -4.585 | 0.0869 | -4.576 | 0.1632 | -4.577 | 0.0752 | -4.800 | 0.0752 | -4.616 | 0.0680 | -3.269 | 0.4876 | -4.009 | 0.1779 | -4.384 | 0.1906 |
| 403.5 | -4.687 | 0.0847 | -4.677 | 0.1584 | -4.679 | 0.0738 | -4.902 | 0.0742 | -4.719 | 0.0674 | -3.347 | 0.4858 | -4.099 | 0.1769 | -4.477 | 0.1879 |
| 405.4 | -4.789 | 0.0828 | -4.778 | 0.1537 | -4.781 | 0.0725 | -5.004 | 0.0733 | -4.821 | 0.0669 | -3.424 | 0.4848 | -4.190 | 0.1761 | -4.571 | 0.1852 |
| 407.3 | -4.891 | 0.0811 | -4.880 | 0.1492 | -4.882 | 0.0714 | -5.105 | 0.0725 | -4.923 | 0.0664 | -3.501 | 0.4845 | -4.280 | 0.1754 | -4.665 | 0.1826 |

| | | | | | | | | | | | | | | | |
|---|---|---|---|---|---|---|---|---|---|---|---|---|---|---|---|
| 409.2 | -4.993 | 0.0796 | -4.982 | 0.1447 | -4.984 | 0.0703 | -5.207 | 0.0718 | -5.025 | 0.0661 | -3.577 | 0.4847 | -4.371 | 0.1748 | -4.759 | 0.1800 |
| 411.1 | -5.094 | 0.0783 | -5.084 | 0.1404 | -5.085 | 0.0694 | -5.308 | 0.0711 | -5.127 | 0.0658 | -3.653 | 0.4853 | -4.461 | 0.1744 | -4.854 | 0.1775 |
| 413.0 | -5.196 | 0.0771 | -5.187 | 0.1362 | -5.187 | 0.0686 | -5.409 | 0.0704 | -5.229 | 0.0656 | -3.728 | 0.4862 | -4.551 | 0.1740 | -4.948 | 0.1751 |
| 414.9 | -5.298 | 0.0761 | -5.289 | 0.1322 | -5.288 | 0.0678 | -5.510 | 0.0698 | -5.331 | 0.0654 | -3.803 | 0.4874 | -4.641 | 0.1738 | -5.042 | 0.1727 |
| 416.8 | -5.400 | 0.0751 | -5.392 | 0.1283 | -5.390 | 0.0671 | -5.611 | 0.0693 | -5.432 | 0.0653 | -3.877 | 0.4886 | -4.730 | 0.1736 | -5.137 | 0.1704 |
| 418.7 | -5.502 | 0.0743 | -5.495 | 0.1247 | -5.492 | 0.0664 | -5.713 | 0.0688 | -5.534 | 0.0652 | -3.951 | 0.4899 | -4.820 | 0.1735 | -5.232 | 0.1681 |
| 420.6 | -5.603 | 0.0735 | -5.599 | 0.1212 | -5.594 | 0.0659 | -5.814 | 0.0683 | -5.635 | 0.0651 | -4.025 | 0.4911 | -4.910 | 0.1734 | -5.327 | 0.1659 |
| 422.5 | -5.705 | 0.0728 | -5.702 | 0.1179 | -5.696 | 0.0653 | -5.915 | 0.0679 | -5.737 | 0.0651 | -4.098 | 0.4922 | -5.000 | 0.1734 | -5.422 | 0.1638 |
| 424.4 | -5.807 | 0.0722 | -5.806 | 0.1148 | -5.798 | 0.0649 | -6.016 | 0.0675 | -5.838 | 0.0651 | -4.171 | 0.4931 | -5.089 | 0.1735 | -5.517 | 0.1617 |
| 426.3 | -5.910 | 0.0717 | -5.910 | 0.1120 | -5.900 | 0.0644 | -6.118 | 0.0671 | -5.939 | 0.0651 | -4.244 | 0.4939 | -5.179 | 0.1736 | -5.612 | 0.1596 |
| 428.2 | -6.012 | 0.0712 | -6.015 | 0.1094 | -6.002 | 0.0640 | -6.219 | 0.0668 | -6.041 | 0.0652 | -4.316 | 0.4944 | -5.269 | 0.1738 | -5.708 | 0.1577 |
| 430.1 | -6.114 | 0.0707 | -6.119 | 0.1070 | -6.104 | 0.0637 | -6.321 | 0.0664 | -6.143 | 0.0653 | -4.389 | 0.4947 | -5.359 | 0.1741 | -5.804 | 0.1558 |
| 432.0 | -6.217 | 0.0703 | -6.224 | 0.1048 | -6.207 | 0.0634 | -6.422 | 0.0662 | -6.244 | 0.0654 | -4.461 | 0.4947 | -5.448 | 0.1743 | -5.900 | 0.1539 |
| 433.9 | -6.319 | 0.0700 | -6.328 | 0.1028 | -6.310 | 0.0631 | -6.524 | 0.0659 | -6.346 | 0.0655 | -4.533 | 0.4944 | -5.538 | 0.1747 | -5.996 | 0.1522 |
| 435.8 | -6.422 | 0.0697 | -6.433 | 0.1009 | -6.413 | 0.0628 | -6.626 | 0.0657 | -6.448 | 0.0657 | -4.606 | 0.4939 | -5.628 | 0.1750 | -6.092 | 0.1505 |
| 437.7 | -6.525 | 0.0694 | -6.538 | 0.0992 | -6.516 | 0.0626 | -6.728 | 0.0655 | -6.550 | 0.0658 | -4.678 | 0.4930 | -5.719 | 0.1754 | -6.189 | 0.1489 |
| 439.6 | -6.628 | 0.0692 | -6.644 | 0.0977 | -6.619 | 0.0624 | -6.830 | 0.0653 | -6.652 | 0.0660 | -4.751 | 0.4919 | -5.809 | 0.1759 | -6.286 | 0.1473 |
| 441.5 | -6.731 | 0.0689 | -6.749 | 0.0963 | -6.722 | 0.0622 | -6.932 | 0.0651 | -6.754 | 0.0662 | -4.824 | 0.4905 | -5.899 | 0.1763 | -6.383 | 0.1459 |
| 443.4 | -6.835 | 0.0688 | -6.855 | 0.0950 | -6.826 | 0.0621 | -7.035 | 0.0650 | -6.856 | 0.0665 | -4.896 | 0.4889 | -5.990 | 0.1768 | -6.480 | 0.1445 |
| 445.3 | -6.939 | 0.0686 | -6.960 | 0.0939 | -6.930 | 0.0620 | -7.137 | 0.0649 | -6.959 | 0.0667 | -4.969 | 0.4870 | -6.080 | 0.1774 | -6.578 | 0.1433 |
| 447.2 | -7.042 | 0.0685 | -7.066 | 0.0928 | -7.034 | 0.0619 | -7.240 | 0.0648 | -7.061 | 0.0670 | -5.043 | 0.4848 | -6.171 | 0.1779 | -6.676 | 0.1421 |
| 449.1 | -7.147 | 0.0684 | -7.172 | 0.0918 | -7.138 | 0.0618 | -7.343 | 0.0647 | -7.164 | 0.0672 | -5.116 | 0.4825 | -6.262 | 0.1785 | -6.774 | 0.1410 |
| 451.0 | -7.251 | 0.0683 | -7.279 | 0.0910 | -7.243 | 0.0617 | -7.446 | 0.0647 | -7.267 | 0.0675 | -5.190 | 0.4798 | -6.353 | 0.1792 | -6.872 | 0.1400 |
| 452.9 | -7.355 | 0.0683 | -7.385 | 0.0902 | -7.347 | 0.0617 | -7.549 | 0.0646 | -7.370 | 0.0678 | -5.264 | 0.4770 | -6.444 | 0.1798 | -6.971 | 0.1391 |
| 454.8 | -7.460 | 0.0683 | -7.492 | 0.0894 | -7.452 | 0.0617 | -7.653 | 0.0646 | -7.473 | 0.0681 | -5.338 | 0.4739 | -6.536 | 0.1805 | -7.069 | 0.1383 |
| 456.7 | -7.565 | 0.0683 | -7.598 | 0.0888 | -7.558 | 0.0617 | -7.757 | 0.0646 | -7.577 | 0.0684 | -5.413 | 0.4706 | -6.627 | 0.1812 | -7.168 | 0.1375 |
| 458.6 | -7.671 | 0.0683 | -7.705 | 0.0882 | -7.663 | 0.0617 | -7.861 | 0.0646 | -7.681 | 0.0688 | -5.488 | 0.4671 | -6.719 | 0.1819 | -7.267 | 0.1369 |
| 460.5 | -7.776 | 0.0683 | -7.813 | 0.0876 | -7.769 | 0.0617 | -7.965 | 0.0646 | -7.785 | 0.0691 | -5.563 | 0.4634 | -6.811 | 0.1827 | -7.367 | 0.1363 |
| 462.4 | -7.882 | 0.0684 | -7.920 | 0.0871 | -7.875 | 0.0618 | -8.070 | 0.0647 | -7.889 | 0.0695 | -5.639 | 0.4595 | -6.903 | 0.1835 | -7.466 | 0.1358 |
| 464.3 | -7.988 | 0.0685 | -8.028 | 0.0866 | -7.981 | 0.0619 | -8.174 | 0.0647 | -7.993 | 0.0699 | -5.715 | 0.4555 | -6.996 | 0.1843 | -7.566 | 0.1353 |
| 466.2 | -8.094 | 0.0686 | -8.135 | 0.0862 | -8.088 | 0.0619 | -8.279 | 0.0648 | -8.098 | 0.0702 | -5.791 | 0.4513 | -7.088 | 0.1851 | -7.666 | 0.1350 |
| 468.1 | -8.201 | 0.0687 | -8.244 | 0.0858 | -8.195 | 0.0620 | -8.385 | 0.0649 | -8.203 | 0.0706 | -5.868 | 0.4470 | -7.181 | 0.1860 | -7.767 | 0.1347 |
| 470.0 | -8.308 | 0.0688 | -8.352 | 0.0855 | -8.302 | 0.0622 | -8.490 | 0.0650 | -8.308 | 0.0710 | -5.946 | 0.4426 | -7.274 | 0.1869 | -7.867 | 0.1344 |
| 471.9 | -8.415 | 0.0690 | -8.461 | 0.0852 | -8.409 | 0.0623 | -8.596 | 0.0651 | -8.413 | 0.0715 | -6.023 | 0.4381 | -7.367 | 0.1878 | -7.968 | 0.1342 |
| 473.8 | -8.522 | 0.0691 | -8.569 | 0.0849 | -8.517 | 0.0624 | -8.702 | 0.0652 | -8.519 | 0.0719 | -6.102 | 0.4335 | -7.461 | 0.1887 | -8.069 | 0.1341 |
| 475.7 | -8.630 | 0.0693 | -8.679 | 0.0846 | -8.625 | 0.0626 | -8.808 | 0.0654 | -8.624 | 0.0723 | -6.180 | 0.4288 | -7.555 | 0.1896 | -8.171 | 0.1340 |
| 477.6 | -8.738 | 0.0695 | -8.788 | 0.0844 | -8.733 | 0.0628 | -8.915 | 0.0655 | -8.730 | 0.0728 | -6.259 | 0.4241 | -7.649 | 0.1906 | -8.272 | 0.1340 |
| 479.5 | -8.847 | 0.0698 | -8.898 | 0.0842 | -8.842 | 0.0630 | -9.021 | 0.0657 | -8.837 | 0.0732 | -6.339 | 0.4194 | -7.743 | 0.1916 | -8.374 | 0.1340 |
| 481.4 | -8.955 | 0.0700 | -9.007 | 0.0840 | -8.951 | 0.0632 | -9.129 | 0.0658 | -8.943 | 0.0737 | -6.419 | 0.4146 | -7.837 | 0.1926 | -8.476 | 0.1341 |
| 483.3 | -9.064 | 0.0702 | -9.118 | 0.0839 | -9.060 | 0.0634 | -9.236 | 0.0660 | -9.050 | 0.0742 | -6.499 | 0.4099 | -7.932 | 0.1936 | -8.579 | 0.1342 |
| 485.2 | -9.173 | 0.0705 | -9.228 | 0.0837 | -9.170 | 0.0636 | -9.344 | 0.0662 | -9.157 | 0.0746 | -6.580 | 0.4051 | -8.027 | 0.1947 | -8.681 | 0.1344 |
| 487.1 | -9.283 | 0.0708 | -9.339 | 0.0836 | -9.279 | 0.0638 | -9.452 | 0.0665 | -9.265 | 0.0751 | -6.661 | 0.4004 | -8.122 | 0.1957 | -8.784 | 0.1346 |
| 489.0 | -9.393 | 0.0711 | -9.450 | 0.0835 | -9.390 | 0.0641 | -9.560 | 0.0667 | -9.372 | 0.0756 | -6.743 | 0.3957 | -8.217 | 0.1968 | -8.887 | 0.1349 |
| 490.9 | -9.503 | 0.0714 | -9.561 | 0.0835 | -9.500 | 0.0643 | -9.668 | 0.0669 | -9.480 | 0.0762 | -6.825 | 0.3910 | -8.313 | 0.1979 | -8.991 | 0.1351 |
| 492.8 | -9.614 | 0.0717 | -9.673 | 0.0834 | -9.611 | 0.0646 | -9.777 | 0.0672 | -9.589 | 0.0767 | -6.908 | 0.3864 | -8.409 | 0.1991 | -9.095 | 0.1355 |
| 494.7 | -9.724 | 0.0720 | -9.785 | 0.0834 | -9.722 | 0.0649 | -9.886 | 0.0674 | -9.697 | 0.0772 | -6.991 | 0.3818 | -8.505 | 0.2002 | -9.198 | 0.1358 |

| | | | | | | | | | | | | | | | |
|---|---|---|---|---|---|---|---|---|---|---|---|---|---|---|---|
| 496.6 | -9.836 | 0.0724 | -9.897 | 0.0833 | -9.833 | 0.0652 | -9.996 | 0.0677 | -9.806 | 0.0777 | -7.074 | 0.3773 | -8.601 | 0.2014 | -9.303 | 0.1362 |
| 498.5 | -9.947 | 0.0728 | -10.010 | 0.0833 | -9.945 | 0.0655 | -10.106 | 0.0680 | -9.915 | 0.0783 | -7.158 | 0.3728 | -8.698 | 0.2026 | -9.407 | 0.1366 |
| 500.4 | -10.059 | 0.0731 | -10.122 | 0.0834 | -10.057 | 0.0658 | -10.216 | 0.0683 | -10.024 | 0.0788 | -7.242 | 0.3684 | -8.795 | 0.2038 | -9.512 | 0.1370 |
| 502.3 | -10.171 | 0.0735 | -10.236 | 0.0834 | -10.170 | 0.0662 | -10.326 | 0.0686 | -10.134 | 0.0794 | -7.327 | 0.3640 | -8.892 | 0.2050 | -9.617 | 0.1375 |
| 504.2 | -10.284 | 0.0739 | -10.349 | 0.0834 | -10.283 | 0.0665 | -10.437 | 0.0689 | -10.244 | 0.0800 | -7.412 | 0.3597 | -8.990 | 0.2062 | -9.723 | 0.1380 |
| 506.1 | -10.396 | 0.0744 | -10.463 | 0.0835 | -10.396 | 0.0669 | -10.548 | 0.0692 | -10.354 | 0.0806 | -7.497 | 0.3554 | -9.088 | 0.2075 | -9.828 | 0.1385 |
| 508.0 | -10.510 | 0.0748 | -10.577 | 0.0836 | -10.509 | 0.0672 | -10.659 | 0.0695 | -10.465 | 0.0812 | -7.583 | 0.3513 | -9.186 | 0.2088 | -9.934 | 0.1391 |
| 509.9 | -10.623 | 0.0752 | -10.691 | 0.0837 | -10.623 | 0.0676 | -10.771 | 0.0699 | -10.576 | 0.0818 | -7.669 | 0.3471 | -9.284 | 0.2100 | -10.041 | 0.1396 |
| 511.8 | -10.737 | 0.0757 | -10.806 | 0.0838 | -10.737 | 0.0680 | -10.883 | 0.0702 | -10.687 | 0.0824 | -7.756 | 0.3431 | -9.383 | 0.2114 | -10.147 | 0.1402 |
| 513.7 | -10.851 | 0.0762 | -10.921 | 0.0839 | -10.852 | 0.0684 | -10.995 | 0.0706 | -10.799 | 0.0830 | -7.842 | 0.3391 | -9.482 | 0.2127 | -10.254 | 0.1408 |
| 515.6 | -10.966 | 0.0766 | -11.037 | 0.0840 | -10.967 | 0.0688 | -11.108 | 0.0710 | -10.911 | 0.0836 | -7.930 | 0.3353 | -9.581 | 0.2140 | -10.361 | 0.1415 |
| 517.5 | -11.081 | 0.0771 | -11.152 | 0.0841 | -11.082 | 0.0692 | -11.221 | 0.0714 | -11.023 | 0.0842 | -8.017 | 0.3315 | -9.680 | 0.2154 | -10.469 | 0.1421 |
| 519.4 | -11.196 | 0.0776 | -11.269 | 0.0843 | -11.198 | 0.0696 | -11.334 | 0.0718 | -11.135 | 0.0849 | -8.106 | 0.3278 | -9.780 | 0.2168 | -10.577 | 0.1428 |
| 521.3 | -11.312 | 0.0782 | -11.385 | 0.0844 | -11.314 | 0.0701 | -11.448 | 0.0722 | -11.248 | 0.0855 | -8.194 | 0.3241 | -9.880 | 0.2182 | -10.685 | 0.1435 |
| 523.2 | -11.428 | 0.0787 | -11.502 | 0.0846 | -11.430 | 0.0705 | -11.562 | 0.0726 | -11.361 | 0.0862 | -8.283 | 0.3206 | -9.981 | 0.2196 | -10.793 | 0.1442 |
| 525.1 | -11.544 | 0.0792 | -11.619 | 0.0848 | -11.547 | 0.0710 | -11.676 | 0.0730 | -11.475 | 0.0868 | -8.372 | 0.3172 | -10.082 | 0.2210 | -10.902 | 0.1449 |
| 527.0 | -11.661 | 0.0798 | -11.736 | 0.0850 | -11.664 | 0.0714 | -11.791 | 0.0735 | -11.589 | 0.0875 | -8.462 | 0.3138 | -10.183 | 0.2225 | -11.011 | 0.1457 |
| 528.9 | -11.778 | 0.0804 | -11.854 | 0.0852 | -11.781 | 0.0719 | -11.906 | 0.0739 | -11.703 | 0.0882 | -8.552 | 0.3106 | -10.284 | 0.2240 | -11.120 | 0.1464 |
| 530.8 | -11.896 | 0.0810 | -11.972 | 0.0854 | -11.899 | 0.0724 | -12.022 | 0.0744 | -11.817 | 0.0889 | -8.642 | 0.3074 | -10.385 | 0.2254 | -11.230 | 0.1472 |
| 532.7 | -12.013 | 0.0816 | -12.091 | 0.0857 | -12.017 | 0.0729 | -12.137 | 0.0748 | -11.932 | 0.0896 | -8.732 | 0.3044 | -10.487 | 0.2270 | -11.340 | 0.1480 |
| 534.6 | -12.132 | 0.0822 | -12.210 | 0.0859 | -12.135 | 0.0734 | -12.253 | 0.0753 | -12.047 | 0.0903 | -8.823 | 0.3015 | -10.590 | 0.2285 | -11.450 | 0.1488 |
| 536.5 | -12.250 | 0.0828 | -12.329 | 0.0861 | -12.254 | 0.0739 | -12.370 | 0.0758 | -12.163 | 0.0910 | -8.915 | 0.2986 | -10.692 | 0.2300 | -11.561 | 0.1497 |
| 538.4 | -12.369 | 0.0834 | -12.448 | 0.0864 | -12.374 | 0.0745 | -12.487 | 0.0763 | -12.279 | 0.0917 | -9.007 | 0.2959 | -10.795 | 0.2316 | -11.672 | 0.1505 |
| 540.4 | -12.489 | 0.0841 | -12.568 | 0.0867 | -12.493 | 0.0750 | -12.604 | 0.0768 | -12.395 | 0.0925 | -9.099 | 0.2933 | -10.898 | 0.2331 | -11.783 | 0.1514 |
| 542.3 | -12.608 | 0.0847 | -12.689 | 0.0870 | -12.613 | 0.0756 | -12.721 | 0.0773 | -12.511 | 0.0932 | -9.191 | 0.2908 | -11.002 | 0.2347 | -11.895 | 0.1523 |
| 544.2 | -12.728 | 0.0854 | -12.809 | 0.0872 | -12.734 | 0.0761 | -12.839 | 0.0778 | -12.628 | 0.0939 | -9.284 | 0.2884 | -11.105 | 0.2364 | -12.007 | 0.1532 |
| 546.1 | -12.849 | 0.0861 | -12.930 | 0.0876 | -12.854 | 0.0767 | -12.958 | 0.0784 | -12.745 | 0.0947 | -9.377 | 0.2861 | -11.210 | 0.2380 | -12.119 | 0.1541 |
| 548.0 | -12.970 | 0.0868 | -13.052 | 0.0879 | -12.976 | 0.0773 | -13.076 | 0.0789 | -12.863 | 0.0955 | -9.470 | 0.2839 | -11.314 | 0.2396 | -12.232 | 0.1550 |
| 549.9 | -13.091 | 0.0875 | -13.173 | 0.0882 | -13.097 | 0.0779 | -13.195 | 0.0795 | -12.981 | 0.0962 | -9.564 | 0.2818 | -11.419 | 0.2413 | -12.345 | 0.1559 |
| 551.8 | -13.213 | 0.0882 | -13.296 | 0.0885 | -13.219 | 0.0785 | -13.315 | 0.0800 | -13.099 | 0.0970 | -9.658 | 0.2799 | -11.524 | 0.2430 | -12.458 | 0.1569 |
| 553.7 | -13.335 | 0.0889 | -13.418 | 0.0889 | -13.341 | 0.0791 | -13.434 | 0.0806 | -13.218 | 0.0978 | -9.752 | 0.2780 | -11.629 | 0.2447 | -12.572 | 0.1578 |
| 555.6 | -13.457 | 0.0897 | -13.541 | 0.0892 | -13.464 | 0.0797 | -13.554 | 0.0812 | -13.337 | 0.0986 | -9.847 | 0.2762 | -11.735 | 0.2464 | -12.686 | 0.1588 |
| 557.5 | -13.580 | 0.0904 | -13.664 | 0.0896 | -13.587 | 0.0804 | -13.675 | 0.0818 | -13.456 | 0.0994 | -9.942 | 0.2746 | -11.841 | 0.2481 | -12.800 | 0.1598 |
| 559.4 | -13.703 | 0.0912 | -13.788 | 0.0899 | -13.711 | 0.0810 | -13.796 | 0.0824 | -13.576 | 0.1002 | -10.037 | 0.2730 | -11.947 | 0.2499 | -12.915 | 0.1608 |
| 561.3 | -13.826 | 0.0920 | -13.912 | 0.0903 | -13.834 | 0.0816 | -13.917 | 0.0830 | -13.696 | 0.1010 | -10.133 | 0.2716 | -12.054 | 0.2516 | -13.030 | 0.1619 |
| 563.2 | -13.950 | 0.0928 | -14.036 | 0.0907 | -13.959 | 0.0823 | -14.038 | 0.0836 | -13.816 | 0.1019 | -10.229 | 0.2702 | -12.161 | 0.2534 | -13.145 | 0.1629 |
| 565.1 | -14.075 | 0.0936 | -14.161 | 0.0911 | -14.083 | 0.0830 | -14.160 | 0.0842 | -13.937 | 0.1027 | -10.325 | 0.2689 | -12.268 | 0.2552 | -13.261 | 0.1639 |
| 567.0 | -14.200 | 0.0944 | -14.286 | 0.0915 | -14.208 | 0.0837 | -14.283 | 0.0849 | -14.058 | 0.1035 | -10.421 | 0.2677 | -12.376 | 0.2571 | -13.377 | 0.1650 |
| 568.9 | -14.325 | 0.0952 | -14.411 | 0.0919 | -14.334 | 0.0844 | -14.405 | 0.0855 | -14.179 | 0.1044 | -10.518 | 0.2666 | -12.484 | 0.2589 | -13.494 | 0.1661 |
| 570.8 | -14.450 | 0.0961 | -14.537 | 0.0924 | -14.460 | 0.0851 | -14.528 | 0.0862 | -14.301 | 0.1052 | -10.615 | 0.2656 | -12.592 | 0.2608 | -13.610 | 0.1672 |
| 572.7 | -14.576 | 0.0970 | -14.663 | 0.0928 | -14.586 | 0.0858 | -14.652 | 0.0869 | -14.423 | 0.1061 | -10.713 | 0.2647 | -12.701 | 0.2626 | -13.728 | 0.1683 |
| 574.6 | -14.703 | 0.0978 | -14.790 | 0.0932 | -14.712 | 0.0865 | -14.776 | 0.0876 | -14.546 | 0.1070 | -10.810 | 0.2639 | -12.810 | 0.2645 | -13.845 | 0.1694 |
| 576.5 | -14.829 | 0.0987 | -14.917 | 0.0937 | -14.839 | 0.0873 | -14.900 | 0.0883 | -14.669 | 0.1079 | -10.908 | 0.2631 | -12.919 | 0.2664 | -13.963 | 0.1705 |
| 578.4 | -14.956 | 0.0996 | -15.044 | 0.0942 | -14.967 | 0.0880 | -15.025 | 0.0890 | -14.792 | 0.1088 | -11.007 | 0.2624 | -13.029 | 0.2684 | -14.081 | 0.1716 |
| 580.3 | -15.084 | 0.1005 | -15.172 | 0.0946 | -15.095 | 0.0888 | -15.150 | 0.0897 | -14.915 | 0.1097 | -11.105 | 0.2618 | -13.139 | 0.2703 | -14.200 | 0.1728 |
| 582.2 | -15.212 | 0.1015 | -15.300 | 0.0951 | -15.223 | 0.0895 | -15.275 | 0.0904 | -15.039 | 0.1106 | -11.204 | 0.2613 | -13.249 | 0.2723 | -14.319 | 0.1740 |

| | | | | | | | | | | | | | | | |
|---|---|---|---|---|---|---|---|---|---|---|---|---|---|---|---|
| 584.1 | -15.340 | 0.1024 | -15.429 | 0.0956 | -15.351 | 0.0903 | -15.401 | 0.0911 | -15.164 | 0.1115 | -11.303 | 0.2609 | -13.360 | 0.2743 | -14.438 | 0.1751 |
| 586.0 | -15.469 | 0.1033 | -15.558 | 0.0961 | -15.480 | 0.0911 | -15.527 | 0.0919 | -15.289 | 0.1124 | -11.403 | 0.2605 | -13.471 | 0.2763 | -14.558 | 0.1763 |
| 587.9 | -15.598 | 0.1043 | -15.687 | 0.0966 | -15.610 | 0.0919 | -15.653 | 0.0926 | -15.414 | 0.1134 | -11.502 | 0.2601 | -13.582 | 0.2783 | -14.678 | 0.1775 |
| 589.8 | -15.728 | 0.1053 | -15.817 | 0.0971 | -15.740 | 0.0927 | -15.780 | 0.0934 | -15.539 | 0.1143 | -11.602 | 0.2599 | -13.694 | 0.2803 | -14.798 | 0.1787 |
| 591.7 | -15.858 | 0.1063 | -15.947 | 0.0977 | -15.870 | 0.0935 | -15.908 | 0.0942 | -15.665 | 0.1153 | -11.703 | 0.2597 | -13.806 | 0.2824 | -14.919 | 0.1800 |
| 593.6 | -15.988 | 0.1073 | -16.077 | 0.0982 | -16.001 | 0.0944 | -16.035 | 0.0950 | -15.791 | 0.1162 | -11.803 | 0.2596 | -13.918 | 0.2845 | -15.040 | 0.1812 |
| 595.5 | -16.119 | 0.1083 | -16.208 | 0.0987 | -16.132 | 0.0952 | -16.163 | 0.0957 | -15.918 | 0.1172 | -11.904 | 0.2595 | -14.031 | 0.2866 | -15.161 | 0.1825 |
| 597.4 | -16.250 | 0.1094 | -16.339 | 0.0993 | -16.263 | 0.0961 | -16.292 | 0.0966 | -16.045 | 0.1182 | -12.006 | 0.2595 | -14.144 | 0.2887 | -15.283 | 0.1837 |
| 599.3 | -16.382 | 0.1104 | -16.471 | 0.0999 | -16.395 | 0.0969 | -16.421 | 0.0974 | -16.172 | 0.1191 | -12.107 | 0.2596 | -14.257 | 0.2908 | -15.405 | 0.1850 |
| 601.2 | -16.514 | 0.1115 | -16.603 | 0.1004 | -16.527 | 0.0978 | -16.550 | 0.0982 | -16.300 | 0.1201 | -12.209 | 0.2597 | -14.371 | 0.2930 | -15.528 | 0.1863 |
| 603.1 | -16.646 | 0.1126 | -16.735 | 0.1010 | -16.660 | 0.0987 | -16.680 | 0.0990 | -16.428 | 0.1211 | -12.311 | 0.2598 | -14.485 | 0.2951 | -15.651 | 0.1876 |
| 605.0 | -16.779 | 0.1136 | -16.868 | 0.1016 | -16.793 | 0.0996 | -16.810 | 0.0999 | -16.556 | 0.1222 | -12.413 | 0.2600 | -14.599 | 0.2973 | -15.774 | 0.1889 |
| 606.9 | -16.913 | 0.1148 | -17.002 | 0.1022 | -16.926 | 0.1005 | -16.940 | 0.1007 | -16.685 | 0.1232 | -12.516 | 0.2603 | -14.714 | 0.2995 | -15.898 | 0.1903 |
| 608.8 | -17.046 | 0.1159 | -17.135 | 0.1028 | -17.060 | 0.1015 | -17.071 | 0.1016 | -16.814 | 0.1242 | -12.619 | 0.2606 | -14.829 | 0.3018 | -16.022 | 0.1916 |
| 610.7 | -17.181 | 0.1170 | -17.269 | 0.1034 | -17.195 | 0.1024 | -17.203 | 0.1025 | -16.944 | 0.1253 | -12.722 | 0.2610 | -14.945 | 0.3040 | -16.146 | 0.1930 |
| 612.6 | -17.315 | 0.1182 | -17.404 | 0.1041 | -17.330 | 0.1033 | -17.334 | 0.1034 | -17.074 | 0.1263 | -12.826 | 0.2614 | -15.061 | 0.3063 | -16.271 | 0.1943 |
| 614.5 | -17.450 | 0.1193 | -17.539 | 0.1047 | -17.465 | 0.1043 | -17.466 | 0.1043 | -17.204 | 0.1274 | -12.930 | 0.2618 | -15.177 | 0.3086 | -16.396 | 0.1957 |
| 616.4 | -17.586 | 0.1205 | -17.674 | 0.1054 | -17.600 | 0.1053 | -17.599 | 0.1052 | -17.335 | 0.1284 | -13.034 | 0.2623 | -15.293 | 0.3109 | -16.522 | 0.1971 |
| 618.3 | -17.721 | 0.1217 | -17.809 | 0.1060 | -17.736 | 0.1063 | -17.732 | 0.1061 | -17.466 | 0.1295 | -13.138 | 0.2629 | -15.410 | 0.3132 | -16.648 | 0.1985 |
| 620.2 | -17.858 | 0.1229 | -17.946 | 0.1067 | -17.873 | 0.1073 | -17.865 | 0.1071 | -17.598 | 0.1306 | -13.243 | 0.2634 | -15.527 | 0.3155 | -16.774 | 0.1999 |
| 622.1 | -17.994 | 0.1242 | -18.082 | 0.1074 | -18.010 | 0.1083 | -17.999 | 0.1080 | -17.730 | 0.1317 | -13.348 | 0.2641 | -15.645 | 0.3179 | -16.901 | 0.2014 |
| 624.0 | -18.131 | 0.1254 | -18.219 | 0.1080 | -18.147 | 0.1093 | -18.133 | 0.1090 | -17.862 | 0.1328 | -13.453 | 0.2647 | -15.763 | 0.3203 | -17.028 | 0.2028 |
| 625.9 | -18.269 | 0.1267 | -18.356 | 0.1087 | -18.285 | 0.1103 | -18.267 | 0.1100 | -17.995 | 0.1339 | -13.559 | 0.2654 | -15.881 | 0.3227 | -17.155 | 0.2043 |
| 627.8 | -18.407 | 0.1279 | -18.494 | 0.1094 | -18.423 | 0.1114 | -18.402 | 0.1110 | -18.128 | 0.1350 | -13.665 | 0.2662 | -16.000 | 0.3251 | -17.283 | 0.2058 |
| 629.7 | -18.545 | 0.1292 | -18.632 | 0.1102 | -18.561 | 0.1124 | -18.538 | 0.1119 | -18.261 | 0.1361 | -13.771 | 0.2669 | -16.119 | 0.3276 | -17.411 | 0.2073 |
| 631.6 | -18.684 | 0.1305 | -18.770 | 0.1109 | -18.700 | 0.1135 | -18.673 | 0.1130 | -18.395 | 0.1373 | -13.878 | 0.2677 | -16.238 | 0.3300 | -17.539 | 0.2088 |
| 633.5 | -18.823 | 0.1319 | -18.909 | 0.1116 | -18.840 | 0.1146 | -18.809 | 0.1140 | -18.529 | 0.1384 | -13.985 | 0.2686 | -16.358 | 0.3325 | -17.668 | 0.2103 |
| 635.4 | -18.963 | 0.1332 | -19.049 | 0.1124 | -18.979 | 0.1157 | -18.946 | 0.1150 | -18.664 | 0.1396 | -14.092 | 0.2695 | -16.478 | 0.3350 | -17.798 | 0.2118 |
| 637.3 | -19.103 | 0.1346 | -19.188 | 0.1131 | -19.120 | 0.1168 | -19.083 | 0.1161 | -18.799 | 0.1408 | -14.199 | 0.2704 | -16.599 | 0.3376 | -17.927 | 0.2133 |
| 639.2 | -19.244 | 0.1359 | -19.329 | 0.1139 | -19.260 | 0.1179 | -19.220 | 0.1171 | -18.934 | 0.1419 | -14.307 | 0.2713 | -16.719 | 0.3401 | -18.058 | 0.2149 |
| 641.1 | -19.385 | 0.1373 | -19.469 | 0.1147 | -19.401 | 0.1191 | -19.358 | 0.1182 | -19.070 | 0.1431 | -14.415 | 0.2723 | -16.841 | 0.3427 | -18.188 | 0.2165 |
| 643.0 | -19.526 | 0.1387 | -19.610 | 0.1154 | -19.543 | 0.1202 | -19.496 | 0.1193 | -19.206 | 0.1443 | -14.524 | 0.2733 | -16.962 | 0.3453 | -18.319 | 0.2181 |
| 644.9 | -19.668 | 0.1402 | -19.751 | 0.1162 | -19.685 | 0.1214 | -19.635 | 0.1204 | -19.343 | 0.1455 | -14.632 | 0.2744 | -17.084 | 0.3479 | -18.450 | 0.2196 |
| 646.8 | -19.810 | 0.1416 | -19.893 | 0.1170 | -19.827 | 0.1225 | -19.774 | 0.1215 | -19.480 | 0.1467 | -14.741 | 0.2754 | -17.206 | 0.3505 | -18.582 | 0.2213 |
| 648.7 | -19.953 | 0.1431 | -20.036 | 0.1179 | -19.970 | 0.1237 | -19.913 | 0.1226 | -19.617 | 0.1480 | -14.851 | 0.2766 | -17.329 | 0.3532 | -18.714 | 0.2229 |
| 650.6 | -20.096 | 0.1445 | -20.178 | 0.1187 | -20.113 | 0.1249 | -20.053 | 0.1237 | -19.755 | 0.1492 | -14.960 | 0.2777 | -17.452 | 0.3559 | -18.846 | 0.2245 |
| 652.5 | -20.240 | 0.1460 | -20.321 | 0.1195 | -20.257 | 0.1262 | -20.194 | 0.1249 | -19.893 | 0.1504 | -15.070 | 0.2789 | -17.575 | 0.3585 | -18.979 | 0.2262 |
| 654.4 | -20.384 | 0.1476 | -20.465 | 0.1204 | -20.401 | 0.1274 | -20.334 | 0.1260 | -20.032 | 0.1517 | -15.181 | 0.2801 | -17.699 | 0.3613 | -19.112 | 0.2278 |
| 656.3 | -20.528 | 0.1491 | -20.609 | 0.1212 | -20.546 | 0.1286 | -20.475 | 0.1272 | -20.171 | 0.1530 | -15.291 | 0.2813 | -17.823 | 0.3640 | -19.246 | 0.2295 |
| 658.2 | -20.673 | 0.1506 | -20.753 | 0.1221 | -20.691 | 0.1299 | -20.617 | 0.1284 | -20.310 | 0.1542 | -15.402 | 0.2825 | -17.948 | 0.3668 | -19.380 | 0.2312 |
| 660.1 | -20.819 | 0.1522 | -20.898 | 0.1229 | -20.836 | 0.1312 | -20.759 | 0.1296 | -20.450 | 0.1555 | -15.513 | 0.2838 | -18.072 | 0.3696 | -19.514 | 0.2329 |
| 662.0 | -20.964 | 0.1538 | -21.043 | 0.1238 | -20.982 | 0.1325 | -20.901 | 0.1308 | -20.590 | 0.1568 | -15.625 | 0.2851 | -18.198 | 0.3724 | -19.649 | 0.2347 |
| 663.9 | -21.111 | 0.1554 | -21.188 | 0.1247 | -21.128 | 0.1338 | -21.044 | 0.1320 | -20.730 | 0.1581 | -15.737 | 0.2865 | -18.323 | 0.3752 | -19.784 | 0.2364 |
| 665.8 | -21.257 | 0.1570 | -21.334 | 0.1256 | -21.275 | 0.1351 | -21.187 | 0.1332 | -20.871 | 0.1594 | -15.849 | 0.2878 | -18.449 | 0.3780 | -19.919 | 0.2381 |
| 667.7 | -21.404 | 0.1586 | -21.481 | 0.1266 | -21.422 | 0.1364 | -21.331 | 0.1345 | -21.013 | 0.1608 | -15.961 | 0.2892 | -18.575 | 0.3809 | -20.055 | 0.2399 |
| 669.6 | -21.552 | 0.1603 | -21.627 | 0.1275 | -21.570 | 0.1377 | -21.475 | 0.1358 | -21.154 | 0.1621 | -16.074 | 0.2906 | -18.702 | 0.3838 | -20.192 | 0.2417 |

| | | | | | | | | | | | | | | | |
|---|---|---|---|---|---|---|---|---|---|---|---|---|---|---|---|
| 671.5 | -21.700 | 0.1620 | -21.775 | 0.1284 | -21.718 | 0.1391 | -21.619 | 0.1370 | -21.296 | 0.1635 | -16.187 | 0.2921 | -18.829 | 0.3867 | -20.328 | 0.2435 |
| 673.4 | -21.849 | 0.1637 | -21.922 | 0.1294 | -21.866 | 0.1405 | -21.764 | 0.1383 | -21.439 | 0.1648 | -16.301 | 0.2935 | -18.956 | 0.3897 | -20.465 | 0.2453 |
| 675.3 | -21.997 | 0.1654 | -22.070 | 0.1303 | -22.015 | 0.1419 | -21.910 | 0.1396 | -21.582 | 0.1662 | -16.415 | 0.2950 | -19.084 | 0.3926 | -20.603 | 0.2471 |
| 677.2 | -22.147 | 0.1671 | -22.219 | 0.1313 | -22.164 | 0.1433 | -22.055 | 0.1410 | -21.725 | 0.1676 | -16.529 | 0.2966 | -19.212 | 0.3956 | -20.741 | 0.2490 |
| 679.1 | -22.297 | 0.1689 | -22.368 | 0.1323 | -22.314 | 0.1447 | -22.202 | 0.1423 | -21.869 | 0.1689 | -16.643 | 0.2981 | -19.341 | 0.3986 | -20.879 | 0.2508 |
| 681.0 | -22.447 | 0.1707 | -22.517 | 0.1333 | -22.464 | 0.1461 | -22.348 | 0.1436 | -22.013 | 0.1703 | -16.758 | 0.2997 | -19.470 | 0.4017 | -21.018 | 0.2527 |
| 682.9 | -22.598 | 0.1725 | -22.667 | 0.1343 | -22.615 | 0.1476 | -22.495 | 0.1450 | -22.158 | 0.1718 | -16.873 | 0.3013 | -19.599 | 0.4047 | -21.157 | 0.2546 |
| 684.8 | -22.749 | 0.1743 | -22.817 | 0.1353 | -22.766 | 0.1491 | -22.643 | 0.1464 | -22.303 | 0.1732 | -16.988 | 0.3029 | -19.729 | 0.4078 | -21.296 | 0.2565 |
| 686.7 | -22.900 | 0.1761 | -22.968 | 0.1363 | -22.918 | 0.1505 | -22.791 | 0.1478 | -22.448 | 0.1746 | -17.104 | 0.3045 | -19.859 | 0.4109 | -21.436 | 0.2584 |
| 688.6 | -23.052 | 0.1780 | -23.119 | 0.1374 | -23.070 | 0.1520 | -22.939 | 0.1492 | -22.594 | 0.1761 | -17.220 | 0.3062 | -19.989 | 0.4141 | -21.576 | 0.2604 |
| 690.5 | -23.205 | 0.1799 | -23.270 | 0.1384 | -23.222 | 0.1536 | -23.088 | 0.1506 | -22.740 | 0.1775 | -17.337 | 0.3079 | -20.120 | 0.4172 | -21.717 | 0.2623 |
| 692.4 | -23.358 | 0.1818 | -23.422 | 0.1395 | -23.375 | 0.1551 | -23.237 | 0.1521 | -22.886 | 0.1790 | -17.453 | 0.3096 | -20.251 | 0.4204 | -21.858 | 0.2643 |
| 694.3 | -23.511 | 0.1837 | -23.575 | 0.1405 | -23.528 | 0.1566 | -23.387 | 0.1535 | -23.033 | 0.1805 | -17.570 | 0.3114 | -20.382 | 0.4236 | -22.000 | 0.2663 |
| 696.2 | -23.665 | 0.1857 | -23.728 | 0.1416 | -23.682 | 0.1582 | -23.537 | 0.1550 | -23.181 | 0.1819 | -17.688 | 0.3131 | -20.514 | 0.4268 | -22.142 | 0.2683 |
| 698.1 | -23.820 | 0.1876 | -23.881 | 0.1427 | -23.836 | 0.1598 | -23.687 | 0.1565 | -23.329 | 0.1834 | -17.806 | 0.3149 | -20.647 | 0.4301 | -22.284 | 0.2703 |
| 700.0 | -23.974 | 0.1896 | -24.034 | 0.1438 | -23.991 | 0.1614 | -23.838 | 0.1580 | -23.477 | 0.1850 | -17.924 | 0.3167 | -20.779 | 0.4333 | -22.427 | 0.2723 |
| 701.9 | -24.130 | 0.1916 | -24.188 | 0.1450 | -24.146 | 0.1630 | -23.989 | 0.1595 | -23.625 | 0.1865 | -18.042 | 0.3186 | -20.912 | 0.4366 | -22.570 | 0.2744 |
| 703.8 | -24.285 | 0.1937 | -24.343 | 0.1461 | -24.302 | 0.1646 | -24.141 | 0.1610 | -23.774 | 0.1880 | -18.161 | 0.3204 | -21.046 | 0.4400 | -22.713 | 0.2764 |
| 705.7 | -24.442 | 0.1957 | -24.498 | 0.1472 | -24.458 | 0.1663 | -24.293 | 0.1626 | -23.924 | 0.1896 | -18.280 | 0.3223 | -21.180 | 0.4433 | -22.857 | 0.2785 |
| 707.6 | -24.598 | 0.1978 | -24.653 | 0.1484 | -24.614 | 0.1680 | -24.446 | 0.1641 | -24.074 | 0.1911 | -18.399 | 0.3242 | -21.314 | 0.4467 | -23.001 | 0.2806 |
| 709.5 | -24.755 | 0.1999 | -24.809 | 0.1496 | -24.771 | 0.1696 | -24.599 | 0.1657 | -24.224 | 0.1927 | -18.519 | 0.3262 | -21.448 | 0.4501 | -23.146 | 0.2827 |
| 711.4 | -24.913 | 0.2020 | -24.965 | 0.1507 | -24.929 | 0.1713 | -24.753 | 0.1673 | -24.374 | 0.1943 | -18.639 | 0.3281 | -21.583 | 0.4535 | -23.291 | 0.2849 |
| 713.3 | -25.071 | 0.2042 | -25.122 | 0.1519 | -25.087 | 0.1731 | -24.907 | 0.1689 | -24.525 | 0.1958 | -18.759 | 0.3301 | -21.719 | 0.4570 | -23.437 | 0.2870 |
| 715.2 | -25.229 | 0.2063 | -25.279 | 0.1531 | -25.245 | 0.1748 | -25.061 | 0.1706 | -24.677 | 0.1975 | -18.880 | 0.3321 | -21.854 | 0.4605 | -23.583 | 0.2892 |
| 717.1 | -25.388 | 0.2085 | -25.437 | 0.1544 | -25.404 | 0.1766 | -25.216 | 0.1722 | -24.829 | 0.1991 | -19.001 | 0.3342 | -21.990 | 0.4640 | -23.729 | 0.2914 |
| 719.0 | -25.548 | 0.2108 | -25.595 | 0.1556 | -25.563 | 0.1783 | -25.371 | 0.1739 | -24.981 | 0.2007 | -19.123 | 0.3362 | -22.127 | 0.4675 | -23.876 | 0.2936 |
| 720.9 | -25.707 | 0.2130 | -25.753 | 0.1568 | -25.722 | 0.1801 | -25.527 | 0.1756 | -25.134 | 0.2023 | -19.244 | 0.3383 | -22.264 | 0.4711 | -24.023 | 0.2958 |
| 722.8 | -25.868 | 0.2153 | -25.912 | 0.1581 | -25.883 | 0.1819 | -25.683 | 0.1773 | -25.287 | 0.2040 | -19.366 | 0.3404 | -22.401 | 0.4747 | -24.170 | 0.2980 |
| 724.7 | -26.029 | 0.2176 | -26.071 | 0.1593 | -26.043 | 0.1838 | -25.840 | 0.1790 | -25.440 | 0.2056 | -19.489 | 0.3425 | -22.539 | 0.4783 | -24.318 | 0.3003 |
| 726.6 | -26.190 | 0.2199 | -26.231 | 0.1606 | -26.204 | 0.1856 | -25.997 | 0.1807 | -25.594 | 0.2073 | -19.612 | 0.3447 | -22.677 | 0.4819 | -24.467 | 0.3025 |
| 728.5 | -26.352 | 0.2222 | -26.391 | 0.1619 | -26.366 | 0.1875 | -26.154 | 0.1825 | -25.749 | 0.2090 | -19.735 | 0.3469 | -22.815 | 0.4856 | -24.616 | 0.3048 |
| 730.4 | -26.514 | 0.2246 | -26.552 | 0.1632 | -26.528 | 0.1894 | -26.312 | 0.1842 | -25.903 | 0.2107 | -19.858 | 0.3491 | -22.954 | 0.4893 | -24.765 | 0.3071 |
| 732.3 | -26.676 | 0.2270 | -26.713 | 0.1646 | -26.690 | 0.1913 | -26.471 | 0.1860 | -26.059 | 0.2124 | -19.982 | 0.3513 | -23.093 | 0.4930 | -24.914 | 0.3095 |
| 734.2 | -26.840 | 0.2294 | -26.874 | 0.1659 | -26.853 | 0.1932 | -26.629 | 0.1878 | -26.214 | 0.2142 | -20.106 | 0.3535 | -23.233 | 0.4967 | -25.064 | 0.3118 |
| 736.1 | -27.003 | 0.2319 | -27.036 | 0.1672 | -27.016 | 0.1952 | -26.789 | 0.1896 | -26.370 | 0.2159 | -20.231 | 0.3558 | -23.373 | 0.5005 | -25.215 | 0.3142 |
| 738.0 | -27.167 | 0.2343 | -27.199 | 0.1686 | -27.180 | 0.1971 | -26.948 | 0.1915 | -26.527 | 0.2176 | -20.356 | 0.3581 | -23.513 | 0.5043 | -25.366 | 0.3165 |
| 739.9 | -27.332 | 0.2368 | -27.361 | 0.1700 | -27.344 | 0.1991 | -27.108 | 0.1933 | -26.684 | 0.2194 | -20.481 | 0.3604 | -23.654 | 0.5082 | -25.517 | 0.3189 |
| 741.9 | -27.497 | 0.2394 | -27.525 | 0.1714 | -27.509 | 0.2011 | -27.269 | 0.1952 | -26.841 | 0.2212 | -20.606 | 0.3628 | -23.795 | 0.5120 | -25.669 | 0.3214 |
| 743.8 | -27.662 | 0.2419 | -27.688 | 0.1728 | -27.674 | 0.2032 | -27.430 | 0.1971 | -26.998 | 0.2230 | -20.732 | 0.3651 | -23.936 | 0.5159 | -25.821 | 0.3238 |
| 745.7 | -27.828 | 0.2445 | -27.852 | 0.1742 | -27.840 | 0.2052 | -27.592 | 0.1990 | -27.157 | 0.2248 | -20.858 | 0.3675 | -24.078 | 0.5198 | -25.973 | 0.3262 |
| 747.6 | -27.995 | 0.2471 | -28.017 | 0.1756 | -28.006 | 0.2073 | -27.754 | 0.2010 | -27.315 | 0.2266 | -20.985 | 0.3700 | -24.221 | 0.5238 | -26.126 | 0.3287 |
| 749.5 | -28.162 | 0.2498 | -28.182 | 0.1771 | -28.172 | 0.2094 | -27.916 | 0.2029 | -27.474 | 0.2284 | -21.112 | 0.3724 | -24.363 | 0.5278 | -26.279 | 0.3312 |
| 751.4 | -28.329 | 0.2524 | -28.347 | 0.1785 | -28.339 | 0.2115 | -28.079 | 0.2049 | -27.633 | 0.2303 | -21.239 | 0.3749 | -24.506 | 0.5318 | -26.433 | 0.3337 |
| 753.3 | -28.497 | 0.2551 | -28.513 | 0.1800 | -28.507 | 0.2136 | -28.242 | 0.2069 | -27.793 | 0.2321 | -21.367 | 0.3774 | -24.650 | 0.5358 | -26.587 | 0.3363 |
| 755.2 | -28.665 | 0.2578 | -28.680 | 0.1815 | -28.675 | 0.2158 | -28.406 | 0.2089 | -27.953 | 0.2340 | -21.495 | 0.3799 | -24.794 | 0.5399 | -26.742 | 0.3388 |
| 757.1 | -28.834 | 0.2606 | -28.847 | 0.1830 | -28.843 | 0.2180 | -28.570 | 0.2110 | -28.114 | 0.2359 | -21.623 | 0.3825 | -24.938 | 0.5440 | -26.897 | 0.3414 |

| | | | | | | | | | | | | | | | |
|---|---|---|---|---|---|---|---|---|---|---|---|---|---|---|---|
| 759.0 | -29.003 | 0.2634 | -29.014 | 0.1845 | -29.012 | 0.2202 | -28.735 | 0.2130 | -28.275 | 0.2378 | -21.752 | 0.3850 | -25.083 | 0.5481 | -27.052 | 0.3440 |
| 760.9 | -29.173 | 0.2662 | -29.182 | 0.1860 | -29.181 | 0.2224 | -28.900 | 0.2151 | -28.436 | 0.2397 | -21.881 | 0.3876 | -25.228 | 0.5522 | -27.208 | 0.3466 |
| 762.8 | -29.343 | 0.2690 | -29.350 | 0.1876 | -29.351 | 0.2246 | -29.065 | 0.2172 | -28.598 | 0.2416 | -22.010 | 0.3903 | -25.373 | 0.5564 | -27.364 | 0.3492 |
| 764.7 | -29.514 | 0.2719 | -29.518 | 0.1891 | -29.521 | 0.2269 | -29.231 | 0.2193 | -28.761 | 0.2435 | -22.140 | 0.3929 | -25.519 | 0.5606 | -27.521 | 0.3519 |
| 766.6 | -29.685 | 0.2748 | -29.687 | 0.1907 | -29.692 | 0.2292 | -29.398 | 0.2215 | -28.923 | 0.2455 | -22.270 | 0.3956 | -25.665 | 0.5649 | -27.678 | 0.3546 |
| 768.5 | -29.857 | 0.2777 | -29.857 | 0.1923 | -29.863 | 0.2315 | -29.565 | 0.2236 | -29.086 | 0.2475 | -22.401 | 0.3983 | -25.812 | 0.5692 | -27.836 | 0.3573 |
| 770.4 | -30.029 | 0.2807 | -30.027 | 0.1939 | -30.035 | 0.2338 | -29.732 | 0.2258 | -29.250 | 0.2494 | -22.531 | 0.4010 | -25.959 | 0.5735 | -27.994 | 0.3600 |
| 772.3 | -30.202 | 0.2837 | -30.197 | 0.1955 | -30.207 | 0.2362 | -29.900 | 0.2280 | -29.414 | 0.2514 | -22.663 | 0.4038 | -26.106 | 0.5778 | -28.152 | 0.3627 |
| 774.2 | -30.375 | 0.2867 | -30.368 | 0.1972 | -30.380 | 0.2386 | -30.068 | 0.2302 | -29.578 | 0.2534 | -22.794 | 0.4066 | -26.254 | 0.5822 | -28.311 | 0.3655 |
| 776.1 | -30.549 | 0.2898 | -30.539 | 0.1988 | -30.553 | 0.2410 | -30.237 | 0.2325 | -29.743 | 0.2555 | -22.926 | 0.4094 | -26.402 | 0.5866 | -28.470 | 0.3682 |
| 778.0 | -30.723 | 0.2929 | -30.711 | 0.2005 | -30.726 | 0.2434 | -30.406 | 0.2347 | -29.908 | 0.2575 | -23.058 | 0.4122 | -26.551 | 0.5910 | -28.630 | 0.3710 |
| 779.9 | -30.897 | 0.2960 | -30.883 | 0.2022 | -30.900 | 0.2459 | -30.576 | 0.2370 | -30.074 | 0.2595 | -23.191 | 0.4151 | -26.700 | 0.5955 | -28.790 | 0.3739 |
| 781.8 | -31.072 | 0.2991 | -31.056 | 0.2039 | -31.075 | 0.2483 | -30.746 | 0.2393 | -30.240 | 0.2616 | -23.324 | 0.4180 | -26.850 | 0.6000 | -28.950 | 0.3767 |
| 783.7 | -31.248 | 0.3023 | -31.229 | 0.2056 | -31.250 | 0.2508 | -30.916 | 0.2417 | -30.407 | 0.2637 | -23.457 | 0.4209 | -26.999 | 0.6045 | -29.111 | 0.3796 |
| 785.6 | -31.424 | 0.3055 | -31.403 | 0.2073 | -31.425 | 0.2534 | -31.087 | 0.2440 | -30.574 | 0.2658 | -23.591 | 0.4238 | -27.150 | 0.6091 | -29.272 | 0.3825 |
| 787.5 | -31.601 | 0.3088 | -31.577 | 0.2091 | -31.601 | 0.2559 | -31.259 | 0.2464 | -30.741 | 0.2679 | -23.725 | 0.4268 | -27.300 | 0.6137 | -29.434 | 0.3854 |
| 789.4 | -31.778 | 0.3121 | -31.751 | 0.2109 | -31.778 | 0.2585 | -31.431 | 0.2488 | -30.909 | 0.2700 | -23.859 | 0.4298 | -27.451 | 0.6183 | -29.596 | 0.3883 |
| 791.3 | -31.955 | 0.3154 | -31.926 | 0.2126 | -31.955 | 0.2611 | -31.603 | 0.2513 | -31.077 | 0.2722 | -23.994 | 0.4329 | -27.603 | 0.6230 | -29.759 | 0.3913 |
| 793.2 | -32.133 | 0.3187 | -32.101 | 0.2144 | -32.132 | 0.2637 | -31.776 | 0.2537 | -31.246 | 0.2743 | -24.129 | 0.4359 | -27.755 | 0.6277 | -29.922 | 0.3942 |
| 795.1 | -32.312 | 0.3221 | -32.277 | 0.2163 | -32.310 | 0.2664 | -31.949 | 0.2562 | -31.415 | 0.2765 | -24.265 | 0.4390 | -27.907 | 0.6324 | -30.086 | 0.3972 |
| 797.0 | -32.491 | 0.3255 | -32.453 | 0.2181 | -32.488 | 0.2691 | -32.123 | 0.2587 | -31.584 | 0.2787 | -24.401 | 0.4421 | -28.060 | 0.6372 | -30.249 | 0.4003 |
| 798.9 | -32.670 | 0.3290 | -32.630 | 0.2199 | -32.667 | 0.2718 | -32.297 | 0.2612 | -31.754 | 0.2809 | -24.537 | 0.4453 | -28.213 | 0.6420 | -30.414 | 0.4033 |
| 800.8 | -32.850 | 0.3325 | -32.807 | 0.2218 | -32.846 | 0.2745 | -32.472 | 0.2638 | -31.925 | 0.2831 | -24.673 | 0.4484 | -28.366 | 0.6468 | -30.578 | 0.4064 |
| 802.7 | -33.031 | 0.3360 | -32.985 | 0.2237 | -33.026 | 0.2773 | -32.647 | 0.2663 | -32.095 | 0.2853 | -24.810 | 0.4517 | -28.520 | 0.6517 | -30.744 | 0.4095 |
| 804.6 | -33.212 | 0.3396 | -33.163 | 0.2256 | -33.206 | 0.2801 | -32.823 | 0.2689 | -32.267 | 0.2876 | -24.948 | 0.4549 | -28.674 | 0.6566 | -30.909 | 0.4126 |
| 806.5 | -33.393 | 0.3432 | -33.342 | 0.2275 | -33.387 | 0.2829 | -32.999 | 0.2715 | -32.438 | 0.2898 | -25.085 | 0.4582 | -28.829 | 0.6616 | -31.075 | 0.4157 |
| 808.4 | -33.575 | 0.3468 | -33.521 | 0.2295 | -33.568 | 0.2857 | -33.175 | 0.2742 | -32.610 | 0.2921 | -25.223 | 0.4614 | -28.984 | 0.6665 | -31.242 | 0.4189 |
| 810.3 | -33.757 | 0.3505 | -33.700 | 0.2314 | -33.750 | 0.2886 | -33.352 | 0.2769 | -32.783 | 0.2944 | -25.362 | 0.4648 | -29.140 | 0.6716 | -31.409 | 0.4221 |
| 812.2 | -33.940 | 0.3542 | -33.880 | 0.2334 | -33.932 | 0.2915 | -33.530 | 0.2796 | -32.956 | 0.2967 | -25.501 | 0.4681 | -29.295 | 0.6766 | -31.576 | 0.4253 |
| 814.1 | -34.124 | 0.3580 | -34.061 | 0.2354 | -34.115 | 0.2944 | -33.708 | 0.2823 | -33.129 | 0.2991 | -25.640 | 0.4715 | -29.452 | 0.6817 | -31.744 | 0.4285 |
| 816.0 | -34.308 | 0.3618 | -34.241 | 0.2374 | -34.298 | 0.2974 | -33.886 | 0.2850 | -33.303 | 0.3014 | -25.779 | 0.4749 | -29.609 | 0.6868 | -31.912 | 0.4318 |
| 817.9 | -34.492 | 0.3656 | -34.423 | 0.2394 | -34.482 | 0.3003 | -34.065 | 0.2878 | -33.477 | 0.3038 | -25.919 | 0.4784 | -29.766 | 0.6920 | -32.081 | 0.4351 |
| 819.8 | -34.677 | 0.3695 | -34.605 | 0.2415 | -34.666 | 0.3034 | -34.245 | 0.2906 | -33.652 | 0.3061 | -26.060 | 0.4819 | -29.923 | 0.6972 | -32.250 | 0.4384 |
| 821.7 | -34.863 | 0.3734 | -34.787 | 0.2436 | -34.850 | 0.3064 | -34.424 | 0.2935 | -33.827 | 0.3085 | -26.200 | 0.4854 | -30.081 | 0.7024 | -32.419 | 0.4417 |
| 823.6 | -35.049 | 0.3773 | -34.970 | 0.2456 | -35.035 | 0.3095 | -34.605 | 0.2963 | -34.003 | 0.3109 | -26.341 | 0.4889 | -30.240 | 0.7077 | -32.589 | 0.4451 |
| 825.5 | -35.235 | 0.3813 | -35.153 | 0.2478 | -35.221 | 0.3126 | -34.786 | 0.2992 | -34.179 | 0.3134 | -26.483 | 0.4925 | -30.398 | 0.7130 | -32.759 | 0.4485 |
| 827.4 | -35.422 | 0.3853 | -35.336 | 0.2499 | -35.407 | 0.3157 | -34.967 | 0.3021 | -34.355 | 0.3158 | -26.624 | 0.4961 | -30.558 | 0.7183 | -32.930 | 0.4519 |
| 829.3 | -35.609 | 0.3894 | -35.520 | 0.2520 | -35.594 | 0.3189 | -35.149 | 0.3050 | -34.532 | 0.3183 | -26.767 | 0.4997 | -30.717 | 0.7237 | -33.101 | 0.4553 |
| 831.2 | -35.797 | 0.3935 | -35.705 | 0.2542 | -35.781 | 0.3220 | -35.331 | 0.3080 | -34.709 | 0.3208 | -26.909 | 0.5034 | -30.877 | 0.7291 | -33.273 | 0.4588 |
| 833.1 | -35.986 | 0.3977 | -35.890 | 0.2564 | -35.968 | 0.3253 | -35.513 | 0.3110 | -34.887 | 0.3233 | -27.052 | 0.5071 | -31.038 | 0.7346 | -33.445 | 0.4623 |
| 835.0 | -36.175 | 0.4019 | -36.076 | 0.2586 | -36.156 | 0.3285 | -35.697 | 0.3140 | -35.065 | 0.3258 | -27.195 | 0.5109 | -31.198 | 0.7401 | -33.618 | 0.4658 |
| 836.9 | -36.364 | 0.4061 | -36.262 | 0.2608 | -36.345 | 0.3318 | -35.880 | 0.3171 | -35.244 | 0.3283 | -27.339 | 0.5146 | -31.360 | 0.7456 | -33.791 | 0.4693 |
| 838.8 | -36.554 | 0.4104 | -36.448 | 0.2630 | -36.534 | 0.3351 | -36.064 | 0.3202 | -35.423 | 0.3308 | -27.483 | 0.5184 | -31.521 | 0.7512 | -33.964 | 0.4729 |
| 840.7 | -36.745 | 0.4147 | -36.635 | 0.2653 | -36.723 | 0.3385 | -36.249 | 0.3233 | -35.602 | 0.3334 | -27.627 | 0.5223 | -31.683 | 0.7568 | -34.138 | 0.4765 |
| 842.6 | -36.936 | 0.4191 | -36.822 | 0.2676 | -36.913 | 0.3418 | -36.434 | 0.3264 | -35.782 | 0.3360 | -27.772 | 0.5261 | -31.846 | 0.7624 | -34.312 | 0.4801 |
| 844.5 | -37.127 | 0.4235 | -37.010 | 0.2699 | -37.104 | 0.3453 | -36.619 | 0.3296 | -35.962 | 0.3386 | -27.917 | 0.5300 | -32.009 | 0.7681 | -34.487 | 0.4838 |

| | | | | | | | | | | | | | | | |
|---|---|---|---|---|---|---|---|---|---|---|---|---|---|---|---|
| 846.4 | -37.319 | 0.4279 | -37.198 | 0.2722 | -37.295 | 0.3487 | -36.805 | 0.3328 | -36.143 | 0.3412 | -28.063 | 0.5340 | -32.172 | 0.7739 | -34.662 | 0.4875 |
| 848.3 | -37.512 | 0.4324 | -37.387 | 0.2746 | -37.486 | 0.3522 | -36.992 | 0.3360 | -36.324 | 0.3439 | -28.209 | 0.5380 | -32.336 | 0.7796 | -34.838 | 0.4912 |
| 850.2 | -37.705 | 0.4369 | -37.576 | 0.2769 | -37.678 | 0.3557 | -37.179 | 0.3393 | -36.506 | 0.3465 | -28.355 | 0.5420 | -32.500 | 0.7855 | -35.014 | 0.4949 |
| 852.1 | -37.899 | 0.4415 | -37.766 | 0.2793 | -37.871 | 0.3592 | -37.366 | 0.3426 | -36.688 | 0.3492 | -28.502 | 0.5460 | -32.665 | 0.7913 | -35.191 | 0.4987 |
| 854.0 | -38.093 | 0.4462 | -37.956 | 0.2817 | -38.063 | 0.3628 | -37.554 | 0.3459 | -36.871 | 0.3519 | -28.649 | 0.5501 | -32.830 | 0.7972 | -35.367 | 0.5025 |
| 855.9 | -38.287 | 0.4508 | -38.147 | 0.2842 | -38.257 | 0.3664 | -37.743 | 0.3492 | -37.054 | 0.3546 | -28.796 | 0.5542 | -32.995 | 0.8032 | -35.545 | 0.5063 |
| 857.8 | -38.482 | 0.4556 | -38.338 | 0.2866 | -38.451 | 0.3701 | -37.931 | 0.3526 | -37.237 | 0.3573 | -28.944 | 0.5584 | -33.161 | 0.8091 | -35.723 | 0.5101 |
| 859.7 | -38.678 | 0.4603 | -38.530 | 0.2891 | -38.645 | 0.3737 | -38.121 | 0.3560 | -37.421 | 0.3601 | -29.092 | 0.5626 | -33.327 | 0.8152 | -35.901 | 0.5140 |
| 861.6 | -38.874 | 0.4651 | -38.722 | 0.2916 | -38.840 | 0.3775 | -38.311 | 0.3595 | -37.605 | 0.3628 | -29.241 | 0.5668 | -33.494 | 0.8212 | -36.080 | 0.5179 |
| 863.5 | -39.071 | 0.4700 | -38.914 | 0.2941 | -39.036 | 0.3812 | -38.501 | 0.3630 | -37.790 | 0.3656 | -29.390 | 0.5711 | -33.661 | 0.8273 | -36.259 | 0.5219 |
| 865.4 | -39.268 | 0.4749 | -39.107 | 0.2966 | -39.232 | 0.3850 | -38.692 | 0.3665 | -37.975 | 0.3684 | -29.539 | 0.5754 | -33.829 | 0.8335 | -36.438 | 0.5258 |
| 867.3 | -39.466 | 0.4799 | -39.301 | 0.2992 | -39.428 | 0.3888 | -38.883 | 0.3700 | -38.161 | 0.3712 | -29.689 | 0.5797 | -33.997 | 0.8397 | -36.619 | 0.5298 |
| 869.2 | -39.664 | 0.4849 | -39.495 | 0.3018 | -39.625 | 0.3927 | -39.075 | 0.3736 | -38.347 | 0.3741 | -29.839 | 0.5841 | -34.165 | 0.8459 | -36.799 | 0.5339 |
| 871.1 | -39.862 | 0.4899 | -39.689 | 0.3044 | -39.823 | 0.3966 | -39.267 | 0.3772 | -38.533 | 0.3770 | -29.989 | 0.5885 | -34.334 | 0.8522 | -36.980 | 0.5379 |
| 873.0 | -40.062 | 0.4951 | -39.884 | 0.3070 | -40.020 | 0.4005 | -39.460 | 0.3808 | -38.720 | 0.3798 | -30.140 | 0.5930 | -34.503 | 0.8585 | -37.161 | 0.5420 |
| 874.9 | -40.261 | 0.5002 | -40.080 | 0.3097 | -40.219 | 0.4044 | -39.653 | 0.3845 | -38.908 | 0.3827 | -30.292 | 0.5975 | -34.673 | 0.8649 | -37.343 | 0.5461 |
| 876.8 | -40.462 | 0.5054 | -40.275 | 0.3124 | -40.418 | 0.4085 | -39.847 | 0.3882 | -39.095 | 0.3857 | -30.443 | 0.6020 | -34.843 | 0.8713 | -37.526 | 0.5503 |
| 878.7 | -40.663 | 0.5107 | -40.472 | 0.3151 | -40.617 | 0.4125 | -40.041 | 0.3920 | -39.284 | 0.3886 | -30.595 | 0.6066 | -35.013 | 0.8778 | -37.708 | 0.5544 |
| 880.6 | -40.864 | 0.5160 | -40.669 | 0.3178 | -40.817 | 0.4166 | -40.236 | 0.3958 | -39.472 | 0.3916 | -30.748 | 0.6112 | -35.184 | 0.8843 | -37.891 | 0.5587 |
| 882.5 | -41.066 | 0.5214 | -40.866 | 0.3206 | -41.018 | 0.4207 | -40.431 | 0.3996 | -39.662 | 0.3945 | -30.901 | 0.6158 | -35.356 | 0.8908 | -38.075 | 0.5629 |
| 884.4 | -41.268 | 0.5268 | -41.064 | 0.3233 | -41.219 | 0.4248 | -40.627 | 0.4034 | -39.851 | 0.3976 | -31.054 | 0.6205 | -35.528 | 0.8974 | -38.259 | 0.5672 |
| 886.3 | -41.471 | 0.5322 | -41.262 | 0.3261 | -41.420 | 0.4290 | -40.823 | 0.4073 | -40.041 | 0.4006 | -31.207 | 0.6253 | -35.700 | 0.9041 | -38.444 | 0.5715 |
| 888.2 | -41.674 | 0.5378 | -41.461 | 0.3290 | -41.622 | 0.4333 | -41.019 | 0.4112 | -40.232 | 0.4036 | -31.361 | 0.6300 | -35.872 | 0.9108 | -38.629 | 0.5758 |
| 890.1 | -41.878 | 0.5433 | -41.660 | 0.3318 | -41.825 | 0.4375 | -41.217 | 0.4152 | -40.423 | 0.4067 | -31.516 | 0.6348 | -36.045 | 0.9175 | -38.814 | 0.5802 |
| 892.0 | -42.083 | 0.5490 | -41.859 | 0.3347 | -42.028 | 0.4419 | -41.414 | 0.4192 | -40.614 | 0.4098 | -31.671 | 0.6397 | -36.219 | 0.9243 | -39.000 | 0.5846 |
| 893.9 | -42.288 | 0.5546 | -42.060 | 0.3376 | -42.231 | 0.4462 | -41.612 | 0.4232 | -40.806 | 0.4129 | -31.826 | 0.6446 | -36.393 | 0.9311 | -39.186 | 0.5890 |
| 895.8 | -42.493 | 0.5604 | -42.260 | 0.3405 | -42.435 | 0.4506 | -41.811 | 0.4273 | -40.998 | 0.4160 | -31.981 | 0.6495 | -36.567 | 0.9380 | -39.373 | 0.5935 |
| 897.7 | -42.699 | 0.5662 | -42.461 | 0.3435 | -42.640 | 0.4550 | -42.010 | 0.4314 | -41.191 | 0.4192 | -32.137 | 0.6545 | -36.742 | 0.9449 | -39.560 | 0.5980 |
| 899.6 | -42.906 | 0.5720 | -42.663 | 0.3464 | -42.845 | 0.4595 | -42.210 | 0.4355 | -41.384 | 0.4223 | -32.294 | 0.6595 | -36.917 | 0.9519 | -39.748 | 0.6025 |
| 901.5 | -43.113 | 0.5779 | -42.865 | 0.3494 | -43.050 | 0.4640 | -42.410 | 0.4397 | -41.578 | 0.4255 | -32.450 | 0.6646 | -37.093 | 0.9589 | -39.936 | 0.6071 |
| 903.4 | -43.321 | 0.5839 | -43.068 | 0.3525 | -43.256 | 0.4686 | -42.610 | 0.4439 | -41.772 | 0.4287 | -32.607 | 0.6697 | -37.269 | 0.9660 | -40.125 | 0.6117 |
| 905.3 | -43.529 | 0.5899 | -43.271 | 0.3555 | -43.463 | 0.4732 | -42.811 | 0.4482 | -41.966 | 0.4320 | -32.765 | 0.6749 | -37.446 | 0.9731 | -40.314 | 0.6163 |
| 907.2 | -43.738 | 0.5960 | -43.474 | 0.3586 | -43.670 | 0.4778 | -43.013 | 0.4524 | -42.161 | 0.4352 | -32.923 | 0.6801 | -37.622 | 0.9803 | -40.504 | 0.6210 |
| 909.1 | -43.947 | 0.6021 | -43.678 | 0.3617 | -43.878 | 0.4825 | -43.215 | 0.4568 | -42.357 | 0.4385 | -33.081 | 0.6853 | -37.800 | 0.9875 | -40.694 | 0.6257 |
| 911.0 | -44.157 | 0.6083 | -43.883 | 0.3648 | -44.086 | 0.4872 | -43.418 | 0.4611 | -42.553 | 0.4418 | -33.240 | 0.6906 | -37.978 | 0.9948 | -40.884 | 0.6305 |
| 912.9 | -44.367 | 0.6146 | -44.088 | 0.3680 | -44.295 | 0.4920 | -43.621 | 0.4656 | -42.749 | 0.4451 | -33.399 | 0.6959 | -38.156 | 1.0021 | -41.075 | 0.6352 |
| 914.8 | -44.578 | 0.6209 | -44.293 | 0.3712 | -44.504 | 0.4968 | -43.824 | 0.4700 | -42.946 | 0.4485 | -33.559 | 0.7013 | -38.335 | 1.0095 | -41.266 | 0.6400 |
| 916.7 | -44.789 | 0.6273 | -44.499 | 0.3744 | -44.713 | 0.5017 | -44.028 | 0.4745 | -43.143 | 0.4518 | -33.719 | 0.7067 | -38.514 | 1.0169 | -41.458 | 0.6449 |
| 918.6 | -45.001 | 0.6337 | -44.706 | 0.3776 | -44.924 | 0.5066 | -44.233 | 0.4790 | -43.341 | 0.4552 | -33.879 | 0.7122 | -38.693 | 1.0244 | -41.650 | 0.6498 |
| 920.5 | -45.214 | 0.6402 | -44.913 | 0.3809 | -45.134 | 0.5115 | -44.438 | 0.4836 | -43.539 | 0.4586 | -34.040 | 0.7177 | -38.873 | 1.0319 | -41.843 | 0.6547 |
| 922.4 | -45.427 | 0.6468 | -45.120 | 0.3842 | -45.346 | 0.5165 | -44.644 | 0.4882 | -43.737 | 0.4621 | -34.201 | 0.7233 | -39.054 | 1.0395 | -42.036 | 0.6597 |
| 924.3 | -45.640 | 0.6534 | -45.328 | 0.3875 | -45.557 | 0.5216 | -44.850 | 0.4929 | -43.936 | 0.4655 | -34.362 | 0.7289 | -39.235 | 1.0471 | -42.230 | 0.6646 |
| 926.2 | -45.854 | 0.6601 | -45.537 | 0.3909 | -45.770 | 0.5266 | -45.056 | 0.4976 | -44.136 | 0.4690 | -34.524 | 0.7346 | -39.416 | 1.0548 | -42.424 | 0.6697 |
| 928.1 | -46.069 | 0.6668 | -45.745 | 0.3943 | -45.982 | 0.5318 | -45.263 | 0.5023 | -44.336 | 0.4725 | -34.687 | 0.7403 | -39.598 | 1.0625 | -42.619 | 0.6747 |
| 930.0 | -46.284 | 0.6737 | -45.955 | 0.3977 | -46.196 | 0.5370 | -45.471 | 0.5071 | -44.536 | 0.4761 | -34.849 | 0.7460 | -39.780 | 1.0703 | -42.814 | 0.6798 |
| 931.9 | -46.500 | 0.6805 | -46.165 | 0.4011 | -46.409 | 0.5422 | -45.679 | 0.5119 | -44.737 | 0.4796 | -35.012 | 0.7518 | -39.962 | 1.0781 | -43.009 | 0.6850 |

| | | | | | | | | | | | | | | | |
|---|---|---|---|---|---|---|---|---|---|---|---|---|---|---|---|
| 933.8 | -46.716 | 0.6875 | -46.375 | 0.4046 | -46.624 | 0.5474 | -45.888 | 0.5168 | -44.939 | 0.4832 | -35.176 | 0.7577 | -40.146 | 1.0860 | -43.205 | 0.6902 |
| 935.7 | -46.933 | 0.6945 | -46.586 | 0.4081 | -46.839 | 0.5528 | -46.097 | 0.5217 | -45.140 | 0.4868 | -35.340 | 0.7636 | -40.329 | 1.0940 | -43.401 | 0.6954 |
| 937.6 | -47.150 | 0.7016 | -46.797 | 0.4116 | -47.054 | 0.5581 | -46.306 | 0.5267 | -45.343 | 0.4904 | -35.504 | 0.7695 | -40.513 | 1.1020 | -43.598 | 0.7006 |
| 939.5 | -47.368 | 0.7087 | -47.009 | 0.4152 | -47.270 | 0.5636 | -46.516 | 0.5317 | -45.545 | 0.4941 | -35.669 | 0.7756 | -40.697 | 1.1101 | -43.796 | 0.7059 |
| 941.5 | -47.586 | 0.7159 | -47.221 | 0.4188 | -47.486 | 0.5690 | -46.727 | 0.5367 | -45.748 | 0.4977 | -35.834 | 0.7816 | -40.882 | 1.1182 | -43.993 | 0.7113 |
| 943.4 | -47.805 | 0.7232 | -47.434 | 0.4224 | -47.703 | 0.5745 | -46.938 | 0.5418 | -45.952 | 0.5014 | -36.000 | 0.7877 | -41.067 | 1.1264 | -44.192 | 0.7167 |
| 945.3 | -48.025 | 0.7305 | -47.648 | 0.4260 | -47.921 | 0.5801 | -47.150 | 0.5469 | -46.156 | 0.5052 | -36.166 | 0.7939 | -41.253 | 1.1346 | -44.391 | 0.7221 |
| 947.2 | -48.245 | 0.7380 | -47.861 | 0.4297 | -48.139 | 0.5857 | -47.362 | 0.5521 | -46.360 | 0.5089 | -36.332 | 0.8001 | -41.439 | 1.1429 | -44.590 | 0.7275 |
| 949.1 | -48.466 | 0.7454 | -48.076 | 0.4334 | -48.358 | 0.5914 | -47.575 | 0.5573 | -46.565 | 0.5127 | -36.499 | 0.8063 | -41.626 | 1.1512 | -44.789 | 0.7330 |
| 951.0 | -48.687 | 0.7530 | -48.291 | 0.4372 | -48.577 | 0.5971 | -47.788 | 0.5626 | -46.771 | 0.5165 | -36.666 | 0.8126 | -41.813 | 1.1596 | -44.990 | 0.7386 |
| 952.9 | -48.909 | 0.7606 | -48.506 | 0.4409 | -48.796 | 0.6029 | -48.001 | 0.5679 | -46.977 | 0.5203 | -36.834 | 0.8190 | -42.000 | 1.1681 | -45.190 | 0.7441 |
| 954.8 | -49.131 | 0.7683 | -48.722 | 0.4447 | -49.016 | 0.6087 | -48.216 | 0.5733 | -47.183 | 0.5242 | -37.002 | 0.8254 | -42.188 | 1.1766 | -45.391 | 0.7497 |
| 956.7 | -49.354 | 0.7761 | -48.938 | 0.4486 | -49.237 | 0.6146 | -48.430 | 0.5787 | -47.390 | 0.5280 | -37.170 | 0.8319 | -42.377 | 1.1852 | -45.593 | 0.7554 |
| 958.6 | -49.577 | 0.7839 | -49.155 | 0.4525 | -49.458 | 0.6205 | -48.645 | 0.5842 | -47.597 | 0.5319 | -37.339 | 0.8384 | -42.566 | 1.1938 | -45.795 | 0.7611 |
| 960.5 | -49.801 | 0.7918 | -49.372 | 0.4564 | -49.680 | 0.6265 | -48.861 | 0.5897 | -47.805 | 0.5359 | -37.508 | 0.8450 | -42.755 | 1.2025 | -45.997 | 0.7668 |
| 962.4 | -50.026 | 0.7998 | -49.590 | 0.4603 | -49.902 | 0.6325 | -49.077 | 0.5952 | -48.013 | 0.5398 | -37.678 | 0.8517 | -42.945 | 1.2113 | -46.200 | 0.7726 |
| 964.3 | -50.251 | 0.8079 | -49.808 | 0.4643 | -50.125 | 0.6386 | -49.294 | 0.6008 | -48.222 | 0.5438 | -37.848 | 0.8583 | -43.135 | 1.2201 | -46.404 | 0.7785 |
| 966.2 | -50.476 | 0.8160 | -50.027 | 0.4683 | -50.348 | 0.6447 | -49.511 | 0.6065 | -48.431 | 0.5478 | -38.018 | 0.8651 | -43.325 | 1.2289 | -46.608 | 0.7843 |
| 968.1 | -50.702 | 0.8242 | -50.246 | 0.4723 | -50.572 | 0.6509 | -49.729 | 0.6122 | -48.640 | 0.5519 | -38.189 | 0.8719 | -43.516 | 1.2379 | -46.812 | 0.7902 |
| 970.0 | -50.929 | 0.8325 | -50.466 | 0.4764 | -50.797 | 0.6572 | -49.947 | 0.6180 | -48.850 | 0.5559 | -38.361 | 0.8788 | -43.708 | 1.2469 | -47.017 | 0.7962 |
| 971.9 | -51.156 | 0.8409 | -50.686 | 0.4805 | -51.022 | 0.6635 | -50.166 | 0.6238 | -49.061 | 0.5600 | -38.532 | 0.8857 | -43.900 | 1.2559 | -47.222 | 0.8022 |
| 973.8 | -51.384 | 0.8493 | -50.907 | 0.4846 | -51.247 | 0.6699 | -50.386 | 0.6296 | -49.272 | 0.5641 | -38.705 | 0.8927 | -44.092 | 1.2651 | -47.428 | 0.8082 |
| 975.7 | -51.613 | 0.8579 | -51.128 | 0.4888 | -51.473 | 0.6763 | -50.605 | 0.6356 | -49.483 | 0.5683 | -38.877 | 0.8997 | -44.285 | 1.2742 | -47.634 | 0.8143 |
| 977.6 | -51.841 | 0.8665 | -51.350 | 0.4930 | -51.700 | 0.6828 | -50.826 | 0.6415 | -49.695 | 0.5724 | -39.050 | 0.9068 | -44.479 | 1.2835 | -47.841 | 0.8204 |
| 979.5 | -52.071 | 0.8751 | -51.572 | 0.4973 | -51.927 | 0.6893 | -51.047 | 0.6475 | -49.907 | 0.5766 | -39.223 | 0.9139 | -44.672 | 1.2928 | -48.048 | 0.8266 |
| 981.4 | -52.301 | 0.8839 | -51.795 | 0.5015 | -52.154 | 0.6959 | -51.268 | 0.6536 | -50.120 | 0.5808 | -39.397 | 0.9211 | -44.867 | 1.3022 | -48.256 | 0.8328 |
| 983.3 | -52.532 | 0.8927 | -52.019 | 0.5058 | -52.383 | 0.7025 | -51.490 | 0.6597 | -50.334 | 0.5851 | -39.572 | 0.9284 | -45.061 | 1.3116 | -48.464 | 0.8391 |
| 985.2 | -52.763 | 0.9017 | -52.242 | 0.5102 | -52.611 | 0.7092 | -51.712 | 0.6659 | -50.547 | 0.5894 | -39.746 | 0.9357 | -45.256 | 1.3211 | -48.673 | 0.8454 |
| 987.1 | -52.995 | 0.9107 | -52.467 | 0.5146 | -52.840 | 0.7160 | -51.935 | 0.6721 | -50.762 | 0.5937 | -39.921 | 0.9431 | -45.452 | 1.3307 | -48.882 | 0.8518 |
| 989.0 | -53.227 | 0.9197 | -52.692 | 0.5190 | -53.070 | 0.7228 | -52.159 | 0.6784 | -50.976 | 0.5980 | -40.097 | 0.9506 | -45.648 | 1.3403 | -49.091 | 0.8582 |
| 990.9 | -53.460 | 0.9289 | -52.917 | 0.5235 | -53.301 | 0.7297 | -52.383 | 0.6847 | -51.191 | 0.6024 | -40.273 | 0.9581 | -45.845 | 1.3500 | -49.301 | 0.8646 |
| 992.8 | -53.693 | 0.9382 | -53.143 | 0.5280 | -53.531 | 0.7367 | -52.607 | 0.6911 | -51.407 | 0.6068 | -40.449 | 0.9657 | -46.042 | 1.3598 | -49.512 | 0.8711 |
| 994.7 | -53.927 | 0.9475 | -53.369 | 0.5325 | -53.763 | 0.7437 | -52.832 | 0.6976 | -51.623 | 0.6112 | -40.626 | 0.9733 | -46.239 | 1.3696 | -49.723 | 0.8777 |
| 996.6 | -54.162 | 0.9569 | -53.596 | 0.5371 | -53.995 | 0.7508 | -53.058 | 0.7041 | -51.840 | 0.6157 | -40.803 | 0.9810 | -46.437 | 1.3795 | -49.935 | 0.8843 |
| 998.5 | -54.397 | 0.9664 | -53.823 | 0.5417 | -54.227 | 0.7579 | -53.284 | 0.7106 | -52.057 | 0.6201 | -40.981 | 0.9888 | -46.635 | 1.3894 | -50.147 | 0.8909 |

**Supplementary Information References**

1. Fokin, D.A. *et al.* Electronic growth of Pb on the vicinal Si surface. *Phys. Status Solidi C* **7**, 165-168 (2010).
2. Su, W.B. *et al.* Correlation between quantized electronic states and oscillatory thickness relaxations of 2D Pb islands on Si (111)-(7× 7) surfaces. *Phys. Rev. Lett.* **86**, 5116 (2001).
3. Kern, R., Le Lay, G. and J.J. Metois. Current Topics in Materials Science 3, 131, ed. by E. Kaldis (1979).
4. Bromann, K. et al. Interlayer mass transport in homoepitaxial and heteroepitaxial metal growth. Phys. Rev. Lett. 75, 677 (1995).
5. Sette, F. et al. Coverage and chemical dependence of adsorbate-induced bond weakening in metal substrate surfaces. Phys. Rev. Lett. 61, 1384 (1988).
6. Baski, A.A. and Fuchs, H. Epitaxial growth of silver on mica as studied by AFM and STM. Surf. Sci. 313, 275-288 (1994).
7. Davey, W. P. Precision measurements of the lattice constants of twelve common metals. P*hys. Rev.* **25**, 753 (1925).
8. Zaumseil, P. High-resolution characterization of the forbidden Si 200 and Si 222 reflections. *J. Appl. Crystallogr.* **48**, 528 (2015).
9. Durand, O. et al. Interpretation of the two-components observed in high resolution X-ray diffraction ω scan peaks for mosaic ZnO thin films grown on c-sapphire substrates using pulsed laser deposition. *Thin Solid Films* **519**, 6369 (2011).
10. High, A.A. *et al.* Visible-frequency hyperbolic metasurface. *Nature* **522**, 192 (2015).
11. Park, J.H. *et al.* Single-Crystalline Silver Films for Plasmonics. *Adv. Mater.* **24**, 3988-3992 (2012).
12. Brendel R, Bormann D. An infrared dielectric function model for amorphous solids. *Journal of Appl. Phys.* **71**, 1–6 (1992).
13. Johnson, P.B. and Christy, R.W. Optical constants of the noble metals. *Phys. Rev. B* **6**, 4370 (1972).
14. Pinchuk, A., Kreibig, U., Hilger A. Optical properties of metallic nanoparticles: influence of interface effects and interband transitions. Surface Science 557, 269–280 (2004).
15. Rodionov, I.A. et al. Crystalline structure dependence on optical properties of silver thin film over time. 2017 Progress In Electromagnetics Research Symposium — Spring (PIERS), 1497-1502 (2017).
16. Kreibig, U. et al. Interfaces in nanostructures: optical investigations on cluster-matter. Nanostructured Materials, 11, 1335-1342 (1999).

17. Baburin, A.S. et al. Toward theoretically limited SPP propagation length above two hundred microns on ultra-smooth silver surface. eprint arXiv:1806.07606 (2018)
18. Wu, Y. *et al.* Intrinsic optical properties and enhanced plasmonic response of epitaxial silver. *Adv. Mater.* **26**, 6106-6110 (2014).